\documentclass[prd, superscriptaddress, tightenlines, longbibliography, nofootinbib, eqsecnum, amsfonts, amsmath, floatfix, twocolumn, notitlepage]{revtex4-2}

\usepackage{graphicx}
\usepackage{dcolumn}
\usepackage{bm}
\usepackage{color}
\usepackage{xcolor}
\usepackage{booktabs}

\def\gridline#1{\vskip6pt\hbox to\hsize{#1}\vskip6pt}

\def\fig#1#2#3{\hfill\vbox{\parskip=0pt\hsize=#2
\includegraphics[width=#2]{#1}\vskip2pt\vtop{\centering
\footnotesize
\hsize=#2
#3\vskip1pt
}}\hfill}

\usepackage{hyperref}

\usepackage[capitalize,noabbrev]{cleveref}
\usepackage{comment}
\usepackage{siunitx}

\crefname{proposition}{Prop.}{Props.}
\crefname{definition}{Def.}{Defs.}
\crefname{lemma}{Lemma}{Lemmas}
\crefname{example}{Ex.}{Exs.}
\crefname{equation}{}{}
\crefname{section}{\S\hspace{-0.2em}}{\S\S}
\crefname{appendix}{\S\hspace{-0.2em}}{\S\S}
\crefname{subsection}{\S\hspace{-0.2em}}{\S\S}
\crefname{subsubsection}{\S\hspace{-0.2em}}{\S\S}
\crefname{figure}{Fig.}{Figs.}
\crefname{wrapfigure}{Fig.}{Figs.}
\crefname{corollary}{Cor.}{Cors.}
\crefname{table}{Table}{Tables}

\newcommand{\udem}{Department of Physics, Universit\'{e} de Montr\'{e}al, Montr\'{e}al, 1375 Avenue Th\'{e}r\`{e}se-Lavoie-Roux, QC H2V 0B3, Canada}
\newcommand{\mila}{Mila - Quebec Artificial Intelligence Institute, Montr\'{e}al, 6666 Rue Saint-Urbain, QC H2S 3H1, Canada}
\newcommand{\ciela}{Ciela - Montreal Institute for Astrophysical Data Analysis and Machine Learning, Montréal, Canada}
\newcommand{\princeton}{Department of Physics, Princeton University, Princeton NJ 08544, USA}
\newcommand{\cca}{Center for Computational Astrophysics, Flatiron Institute, 162 5th Avenue, New York, NY 10010, USA}
\newcommand{\ccm}{Center for Computational Mathematics, Flatiron Institute, 162 5th Avenue, New York, NY 10010, USA}
\newcommand{\florida}{Department of Astronomy, University of Florida, 211 Bryant Space Science Center, Gainesville, FL 32611, USA}
\newcommand{\mpa}{Max-Planck-Institut f\"ur Extraterrestrische Physik, Postfach 1312, Giessenbachstrasse 1, 85748 Garching bei M\"unchen, Germany}
\newcommand{\waterloo}{Waterloo Centre for Astrophysics, University of Waterloo, 200 University Ave W, Waterloo, ON N2L 3G1, Canada}
\newcommand{\waterloob}{Department of Physics and Astronomy, University of Waterloo, 200 University Ave W, Waterloo, ON N2L 3G1, Canada}
\newcommand{\geneva}{D\'epartement de Physique Th\'eorique, Universit\'e de Gen\`eve, 24 quai Ernest Ansermet, 1211 Gen\`eve 4, Switzerland}
\newcommand{\nyu}{Center for Cosmology and Particle Physics, Department of Physics, New York University, New York, NY 10003, USA}
\newcommand{\cmu}{Department of Physics, Carnegie Mellon University, Pittsburgh, PA 15213, USA}

\begin{document}

\preprint{APS/123-QED}

\title{{\sc SimBIG}: Field-level Simulation-Based Inference of Galaxy Clustering}

\author{Pablo Lemos$^*$}
\affiliation{\udem}
\affiliation{\mila}
\affiliation{\ciela}
\affiliation{\cca}

\author{Liam Parker$^*$}
\affiliation{\princeton} 

\author{ChangHoon Hahn}
\affiliation{\princeton} 

\author{Shirley Ho}
\affiliation{\cca}
\affiliation{\princeton}
\affiliation{\nyu}
\affiliation{\cmu}

\author{Michael Eickenberg}
\affiliation{\ccm}

\author{Jiamin Hou}
\affiliation{\florida}
\affiliation{\mpa}

\author{Elena Massara}
\affiliation{\waterloo}
\affiliation{\waterloob}

\author{Chirag Modi}
\affiliation{\cca}
\affiliation{\ccm}

\author{Azadeh Moradinezhad Dizgah}
\affiliation{\geneva}

\author{Bruno R\'egaldo-Saint Blancard}
\affiliation{\ccm}

\author{David Spergel}
\affiliation{\cca}
\affiliation{\princeton}

\def\thefootnote{*}\footnotetext{These authors contributed equally to this work.}\def\thefootnote{\arabic{footnote}}

\date{\today}

\begin{abstract}
We present the first simulation-based inference (SBI) of cosmological parameters from field-level analysis of galaxy clustering. Standard galaxy clustering analyses rely on analyzing summary statistics, such as the power spectrum, $P_\ell$, with analytic models based on perturbation theory. Consequently, they do not fully exploit the non-linear and non-Gaussian features of the galaxy distribution. To address these limitations, we use the {\sc SimBIG} forward modelling framework to perform SBI using normalizing flows. We apply {\sc SimBIG} to a subset of the BOSS CMASS galaxy sample using a convolutional neural network with stochastic weight averaging to perform massive data compression of the galaxy field. We infer constraints on $\Omega_m = 0.267^{+0.033}_{-0.029}$ and $\sigma_8=0.762^{+0.036}_{-0.035}$. While our constraints on $\Omega_m$ are in-line with standard $P_\ell$ analyses, those on $\sigma_8$ are $2.65\times$ tighter. 
Our analysis also provides constraints on the Hubble constant $H_0=64.5 \pm 3.8 \ {\rm km / s / Mpc}$ 
from galaxy clustering alone. 
This higher constraining power comes from additional 
non-Gaussian cosmological information, inaccessible with $P_\ell$. 
We demonstrate the robustness of our analysis by showcasing our ability to infer unbiased cosmological constraints from a series of test simulations that are constructed using different forward models than the one used in our training dataset.
This work not only presents competitive cosmological constraints but 
also introduces novel methods for leveraging additional cosmological 
information in upcoming galaxy surveys like DESI, PFS, and {\em Euclid}.
\end{abstract}


\maketitle

\section{\label{sec:intro}Introduction:\protect}
Precision measurements of cosmological parameters, such as the matter density and the expansion rate of the Universe, play a crucial role in shaping our understanding of the evolution and structure of the cosmos. These parameters can be inferred from a variety of observational data, including measurements of the statistical properties of the large-scale structure (LSS) of the universe traced by the distribution of galaxies.

Traditionally, cosmological parameter inference has relied on 
analyzing the distribution of galaxies using summary 
statistics --- most often the power spectrum, $P_\ell(k)$~\citep[\emph{e.g.}][]{kaiser1987clustering, hamilton1997towards, peacock2001measurement, hawkins20032df, tegmark2006cosmological, guzzo2008test, beutler2017, ivanov2020cosmological, kobayashi2022}.
In addition, these analyses incorporate analytical modeling of galaxy clustering through perturbation theory~\citep[PT; see][for recent reviews]{bernardeau2002, desjacques2018}.
Consequently, these analyses have been limited to large, 
weakly non-linear scales 
where the deviation from PT is small.
By only considering the power spectrum, these analyses can not exploit the rich non-Gaussian information in the galaxy distribution, which is only weakly imprinted on the power spectrum.

Recent analyses of BOSS data have now established that there is in fact significant non-Gaussian cosmological information on non-linear scales in galaxy clustering. 
Furthermore, previous galaxy clustering analyses using higher-order clustering statistics have produced significantly tighter constraints than with $P_\ell$ alone~\citep[\emph{e.g.}][]{gil-marin2017, damico2022, philcox2022boss, ivanov2023}.
Furthermore, forecasts that employ various summary statistics beyond $P_\ell$ \citep[\emph{e.g.}][]{hahn2020, hahn2021a, Massara2022, wang2022, hou2022, eickenberg2022} 
have been shown to produce even tighter constraints by including non-linear scales. 
Nonetheless, these applications remain limited by the inability of PT to model galaxy clustering at scales beyond the quasi-linear, especially for higher-order statistics.

Another major challenge of galaxy clustering analyses is 
their inability to fully account for observational systematics. 
For example, fiber collisions have been shown to significantly bias $P_\ell$ on scales smaller than $k \sim 0.1\,h/\textrm{Mpc}$~\citep{hahn2017a, bianchi2018}.
Observational effects in targeting, imaging, and completeness also significantly impact clustering measurements \citep{ross2012clustering, ross2017}. 
Finally, these analyses assume a Gaussian functional form of the likelihood function used in their Bayesian framework. 
This assumption does not necessarily hold in general~\citep{scoccimarro2000gravitational, sellentin2018, hahn2019}. 

To overcome these limitations, we instead use Simulation-Based Inference\footnote{The terms `likelihood-free inference' and 
`implicit likelihood inference' have also been used to refer
to the same method} 
(SBI). 
SBI uses forward models of the observables, instead of analytic models, and then infers a posterior distribution over the parameters (or a likelihood, that can then be converted into the posterior with Bayes' theorem). 
This method enables us to leverage high-fidelity simulations that accurately 
model complex physical processes, leading to more robust inferences than methods 
based on analytical models.

There have already been multiple applications of SBI in astronomy \citep[\emph{e.g.}][]{hahn2017b, PerreaultLevasseur:2017ltk, PhysRevLett.127.241103,alsing2019fast,Wagner-Carena:2020yun,Legin:2021zup, Coogan:2020yux, Montel:2022fhv, Coogan:2022cky, Brehmer:2019jyt, Mishra-Sharma:2021oxe, Karchev:2022xyn, Hermans:2020skz, Karchev:2022ycy, lemos2021sum, hahn2022a}. In the specific context of galaxy clustering, \citet{hahn2022rm} introduced the SIMulation-Based Inference of Galaxies ({\sc SimBIG}) forward models, which produce realistic mock observations of the Sloan Digital Sky Survey III Baryon Oscillation Spectroscopic Survey~\citep[BOSS;][]{eisenstein2011sdss, dawson2012baryon} Southern Galactic Cap (SGC) at different cosmologies, and includes systematic effects such as survey geometry and fiber collisions. Using these models, we were able to robustly infer $\Lambda$CDM parameters from the BOSS CMASS-SGC sample at scales down to $k_{\rm max} \sim 0.5h/\textrm{Mpc}$. 
These works, however, focused on presenting the {\sc SimBIG} framework and relied 
on compressing the galaxy distribution to the power spectrum,
which does not capture the non-Gaussian information present.

\begin{figure*}
    \centering
    \includegraphics[width=.99\linewidth]{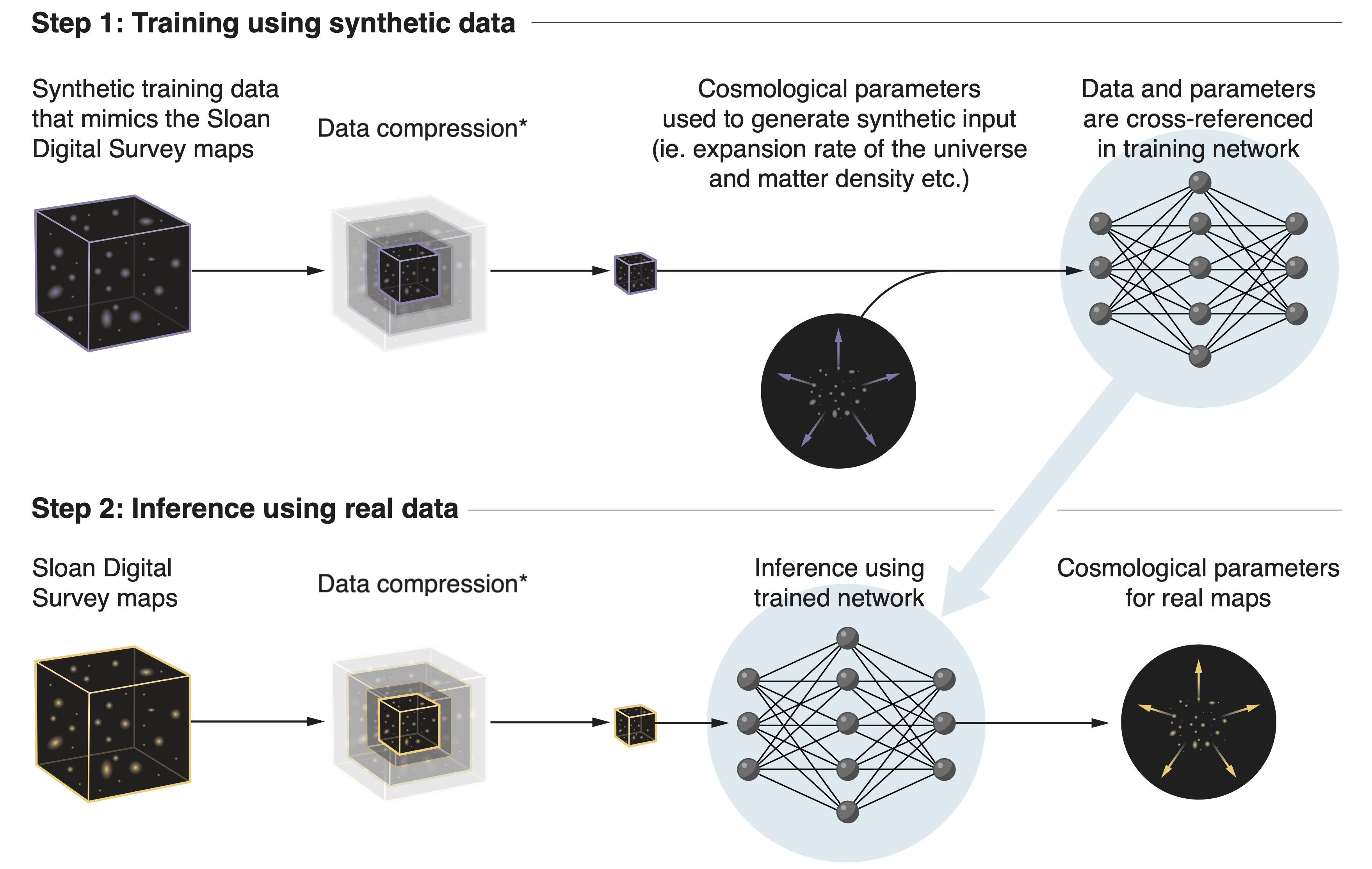}
    \caption{Schematic illustrating the various elements of the {\sc SimBIG} forward modelling pipeline. First, we generate synthetic galaxy catalogs that mimic the real BOSS observations. Then, we train a data compression step using our CNN to compress the catalog to its cosmological parameters. Next, we train a neural posterior estimator on the estimating parameters and the true parameters to estimate posteriors over the cosmological parameters. Once our data compression and neural posterior estimator are trained, we apply our pipeline to infer cosmological parameters from the real BOSS observations.}
    \label{fig:pipeline}
\end{figure*}

In this work, we extend {\sc SimBIG} to analyze the galaxy distribution directly at 
the field-level\footnote{
The term {\it field-level}, is used here to refer to the fact that our neural network takes as input the field, as opposed to perform some compression step on the field before feeding it to the neural network, such as power spectrum or bispectrum.}.
Specifically, we use convolutional neural networks (CNNs) to perform massive data 
compression and to extract maximally relevant features from the galaxy distribution. 
By learning the maximally relevant features with CNNs, our approach aims to extract 
even more cosmological information than summary statistics and establish a comprehensive 
framework for extracting {\em all} of the cosmological information in galaxy distributions. 

The rest of the paper is organized as follows. We describe the details of the observations and simulations in~\S\ref{sec:data}. Our methodology is explained in~\S\ref{sec:methods} and applied to observations in~\S\ref{sec:results}. Finally, we present our conclusions in~\S\ref{sec:conclusions}.

\section{\label{sec:data}Observations \& Simulations:\protect}

In this section, we describe the observational galaxy
sample as well as the forward-modeled training and 
test simulations. 

\subsection{Observations: BOSS CMASS SGC}
We use a sample of CMASS Luminous Red Galaxies from 
the BOSS Data Release 12 as our observational
data\footnote{\url{https://data.sdss.org/sas/dr12/boss/lss/}}.
We limit our analysis to the subsample of CMASS 
sample at the Southern Galactic Cap (SGC) within
the angular footprint, ${\rm DEC > - 6 \ deg.}$ and ${\rm -25 < RA < 28 \ deg.}$, and redshift range, 
$0.45<z<0.6$.
The reason for this is that the QUIJOTE simulation boxes are not big enough to include the full CMASS sample.
In total, our sample consists of 109,636 galaxies. 
Visual illustrations of the sample can be found in~\cite{hahn2022rm,simbig_mock_challenge}.

\subsection{{\sc SimBIG} Forward Model}
We use the {\sc SimBIG} forward modelling pipeline to generate field-level synthetic
observations that aim to be statistically indistinguishable from BOSS observations. 
This pipeline consists of four distinct steps: 
(1) $N$-body simulations, 
(2) a dark matter halo finder,
(3) a halo occupation distribution framework (HOD), 
and (4) application of survey realism. A schematic illustrating the pipeline is provided in \cref{fig:pipeline}. 

The $N$-body simulations are taken from the {\sc QUIJOTE} suite \citep{villaescusa2020quijote}. 
The simulations evolve $1024^3$ cold dark matter particles from $z=127$ to $z=0.5$ in a cosmological volume $1\,(h^{-1}\si{Gpc})^3$ using the TreePM {\sc Gadget-III}. 
These simulations accurately model matter clustering 
down to non-linear scales beyond $k=0.5h/\si{Mpc}$. 

From these $N$-body simulations, dark matter halos are identified using the {\sc ROCKSTAR} halo finder \citep{behroozi2012rockstar}, which has been shown to robustly and accurately track dark matter halo location and substructure using phase-space information.
Specifically, the standard \cite{zheng2007galaxy}
HOD model, which populates halos using $M_h$ and five free HOD parameters, is expanded by including assembly, concentration, and velocity biases. 
These biases add the necessary flexibility to account 
for recent evidence suggesting that galaxies occupy 
halos in ways that depend on halo properties beyond
$M_h$ \citep[\emph{e.g.} assembly history;][]{more2016detection, vakili2019galaxies, zentner2019constraints, hadzhiyska2021galaxy}. 

Finally, survey realism is applied to the HOD galaxy catalog to produce a CMASS-like galaxy catalog. 
First, the $1(h^{-1}\si{Gpc})^3$ box is remapped to a 
cuboid \citep{carlson2010embedding} and then cut to the BOSS survey geometry.
Then, the galaxy catalog is trimmed to $z \in (0.45, 0.6)$, 
and fiber collisions are applied. 
Ultimately, the forward models are determined by five $\Lambda$CDM cosmological parameters, $\Omega_m, \Omega_b, h, n_s, \sigma_8$, and nine HOD parameters.
We refer readers to \citet{hahn2022rm, simbig_mock_challenge} for further details.

To construct our training set, we use 2,518 high-resolution {\sc QUIJOTE} $N$-body simulations\footnote{We supplement the 2,000 QUIJOTE $N$-body simulations used in \cite{hahn2022rm} with 518 additionally constructed simulations.}
arranged in a Latin hypercube configuration (LHC),  which imposes priors on the cosmological parameters that conservatively encompass the 
\textit{Planck} cosmological constraints. 
For each simulation, we forward-model 10 CMASS-like
galaxy catalogs using unique HOD parameters randomly 
sampled from a conservative prior. While this is suboptimal, as it leads to samples that are not independent and identically distributed (i.i.d.), this factor 10 increase in the number of available simulations greatly improves our results, and we expect regularization to deal with any potential issues arising from not i.i.d. samples.
We split the resulting $25,180$ simulations into a 
$20,000$ and $5,180$ training and validation set.

\subsection{Test simulations}\label{sec:test}
In order to demonstrate that we can infer accurate 
and unbiased cosmological constraints, we test our 
analysis on three different sets of realistic test simulations that differ from the training dataset and have been developed within {\sc SimBIG} and introduced in \cite{simbig_mock_challenge}: {\sc TEST0, TEST1,} and {\sc TEST2}.

{\sc TEST0} uses {\sc QUIJOTE} $N$-body simulations that have the same specifications as those arranged in the LHC, but were run at a fiducial cosmology with $\Omega_m = 0.3175, \Omega_b = 0.049, h = 0.6711, n_s = 0.9624, \sigma_8 = 0.834$. The halo finder, HOD framework and survey realism are the same as those 
used
in the training set, but the HOD parameters span a narrower prior. 
This test dataset contains 500 synthetic galaxy catalogs.

{\sc TEST1} involves the same $N$-body simulations as 
{\sc TEST0}, but a different halo finder: the Friend-of-Friend
algorithm~\citep{davis1985evolution}. 
Assembly, concentration, and satellite velocity biases are also not considered in the HOD model. 
Central velocity bias is implemented, as the halo 
velocities in FoF halo catalogs correspond to the bulk velocity of the dark matter particles in the halo rather than the velocity of the central density peak of the halo. This test dataset contains 500 synthetic galaxy catalogs.

{\sc TEST2} uses 25 {\sc AbacusSummit} $N$-body simulations \citep{maksimova2021abacussummit} in the ``base'' configuration of the suite. 
The simulations contain $6912^3$ particles in a $(2h^{-1}\si{Gpc})^3$ volume box. 
Halo catalogs are constructed from these simulations 
using the {\sc CompaSO} halo finder 
\citep{hadzhiyska2022compaso} and each of them is divided into 8 boxes of volume $1\,(h^{-1}\si{Gpc})^3$. Halos are populated with galaxies using the
same HOD model implemented in the training set, with 
HOD parameters that sample the same narrower 
priors used in {\sc TEST0}. 
This test dataset contains 1,000 synthetic galaxy catalogs.

All three test datasets incorporate the same survey 
realism as the training dataset to produce CMASS-like 
galaxy catalogs.

It would be ideal to have a third set, that we only tested on after passing validation tests on {\sc TEST0}, {\sc TEST1} and {\sc TEST2}. However, due to the high computational cost of our simulations, this was unfeasible. 

\subsection{Galaxy Density Field},
To apply CNNs to our observational and simulated galaxy 
samples, we mesh the galaxy distribution into a box, with voxel size $64 \times 128 \times 128$. We choose this size because divisibility by two allows for easier downsampling in the CNN. 
First, we place the distribution into a 
$[707, 1414, 1414] \ {\rm Mpc} / h$ box and convert
it into a 3D density field using a 
cloud-in-cell mass assignment~\citep{birdsall1969clouds}.
For our observational sample, we include systematics 
weights for multiple effects ~\citep[redshift failures, stellar density, and seeing conditions;][]{ross2012clustering, anderson2012clustering} in the mass assignment. 

Since our data occupies a 
$[577.3, 1414, 1224] \ {\rm Mpc} / h$ box, we fill some 
of the box with zero-valued voxels.
Our voxels have size 
$\sim [11, 11, 11] \ {\rm Mpc} / h$, thus we  
impose an effective scale cut of $k < k_{\rm max} = 0.28 \ h {\rm / Mpc}$. 
While this is larger than the scale cut imposed 
in the {\sc SimBIG} $P_\ell$ analysis~\citep{hahnsimbig}, we find that it is sufficient to place significant cosmological constraints. Moreover, pushing to even smaller scale cuts presents its own set of challenges. For one, smaller scale cuts present significant computational challenges in terms of the required memory to train on larger forward model sizes. Additionally, we find that models trained on smaller scale cuts tend to overfit on the training dataset significantly, limiting the robustness of their inferred parameters.  

\section{\label{sec:methods}Methods}

Our approach to field-level inference of cosmological parameters consists of two main components: a massive data compression/feature extraction step performed by a CNN, followed by SBI. In the following section, we describe each step in more detail\footnote{We also attempted a one-step approach, where the CNN served as embedding to the SBI step, however, we found that constraints were significantly weaker when using this approach.}. We also describe two additional elements of our analysis, designed to ensure accurate posterior estimates: weight marginalization and validation with coverage probability tests.

\subsection{CNN-based Feature Extraction}
CNNs are flexible machine learning models that can be optimized to extract maximally relevant features from their inputs across a wide variety tasks. They consist of multiple layers of specialized kernels that are convolved across the input to extract features in a hierarchical scheme. These networks are particularly well-suited for image-based tasks due to their ability to (1) exploit local receptive fields, (2) recognize patterns regardless of their position in the input due to translational invariance, and (3) extract increasingly complex features by combining lower-level features from previous layers hierarchically \citep[for a review of CNNs, see][]{alzubaidi2021review}.

In this study, we train a three-dimensional CNN to compress the galaxy density fields produced by the {\sc SimBIG} forward models to the cosmological parameters of those models. Specifically, the CNN takes as input the three-dimensional tensor representing the discretized forward model, $x \in \mathbb{R}^{64 \times 128 \times 128}$, and outputs a prediction, $\bm{\hat{\theta}}$, of the $\Lambda$CDM cosmological parameters, $\{\Omega_m, \Omega_b, h, n_s, \sigma_8\}$, used to generate that forward model. 

The CNN architecture consists of 5 convolutional blocks. Each convolutional block begins with a convolutional layer that convolves its input with a number of $3 \times 3 \times 3$ kernels. This convolution is performed with 1-voxel zero-padding. This is followed by a rectified linear unit (ReLU). The output of the ReLU unit is then downsampled using max-pooling, which enables the network to learn features at increasing scales by reducing the size of its internal representations. Finally, batch normalization is applied, which typically speeds up training and has been shown to help with generalization \citep{santurkar2018does}. Following the convolutional blocks, the activation maps are flattened and fed into three fully-connected layers that output  $\bm{\hat{\theta}}$. These layers also use ReLU activation functions, but do not perform batch normalization.

In order to prevent overfitting on the training simulations, we include in the CNN's final architecture significant levels of dropout. This technique randomly sets to zero a percentage of neuron activations during training. Specifically, we use dropout percentages of $p=0.15$ for each convolutional block and $p=0.4$ for each fully connected blocks. Additionally, we introduce a large $\ell_2$ penalty term with normalization value $\lambda=0.0275$ on the network weights. In applying dropout in both the convolutional and fully connected layers, the network is forced to train on a smaller subset of active neurons, leading to underutilization of the network's capacity. Moreover, with the $\ell_2$ penalty term in the loss function, the network's flexibility, and subsequently its ability to learn specific features, is limited. While these measures ultimately limit the constraining power of the CNN, they ensure robustness and generalizability, and thus protect against the fact that the {\sc SimBIG} forward models, and in general any forward model, are approximate.

CNN training is performed using a supervised learning approach. We optimize the weights of the network to minimize the mean-squared-error (MSE) loss between 
$\bm{\hat{\theta}}_{\rm normed}$ and $\bm{\theta}^{\rm true}_{\rm normed}$, where 
we normalize both $\bm{\hat{\theta}}$ and $\bm{\theta}^{\rm true}$ to $(0, 1)$,
to prevent their varying ranges from affecting the loss differently. 
The optimization is performed using stochastic gradient descent with momentum $\beta=0.9$. The neural network is trained in mini-batches of 32 galaxy fields. We use the OneCycleLR learning rate scheduler, which involves gradually increasing and then decreasing the learning rate during a single training cycle, and has been shown to lead to faster convergence and improved generalization \citep{smith2019super}. We use a maximal learning rate of $r = 0.01$. During training, the input fields are also randomly flipped horizontally and vertically with $p=0.5$ to further improve network generalization. We train the CNN on a single A100 GPU core until the MSE computed on the validation set has not improved for 20 consecutive epochs. Training the CNN in this context takes roughly 8 hours. 

The CNN's architecture and hyperparameters are determined through experimentation and are roughly modelled off of previous successful image classifiers in \cite{szegedy2016rethinking, krizhevsky2017imagenet}. To determine the specifics of our network, we train 60 networks using the Optuna hyperparameter framework \citep{optuna}. Specifically, we vary the number of convolutional blocks between 3 and 6, the number of fully-connected layers between 1 and 6, the base number of channels of the convolutional blocks between 2 and 14, the width of the fully-connected layers between 128 and 1024, the dropout in both convolutional and fully-connected layers between $p=0$ and $p=0.5$, the $\ell_2$ penalty between $\lambda = 10^{-4}$ and $\lambda = 10^{-1}$, and the max learning rate between $r = 10^{-5}$ and $r = 10^{-2}$. Ultimately, we aim to maximize the network's ability to extract relevant features from the galaxy density field while maintaining its ability to generalize beyond the {\sc SimBIG} training simulations. To that end, we select the network configuration that maximizes the network's MSE on the held-out validation models while minimizing the ratio between training MSE and validation MSE. However, in order to pass the validation tests on the out-of-distribution {\sc TEST1} and {\sc TEST2}, we found that it was necessary to impose slightly stricter regularization on the network. Thus, the dropout and $\ell_2$ terms were increased through trial-and-error from the optuna output to their reported values. Ultimately, the significant amounts of regularization are included due to the model's tendency to overfit on the relatively small dataset size. 

\subsection{Weight Marginalization}

In order to further prevent the CNN from overfitting on the 
training set, we perform a weight marginalization step, converting our CNN into a Bayesian Neural Network (BNN). 
In contrast to other neural networks, BNNs train the model weights as a distribution rather than searching for an optimal value. This allows them to capture the uncertainty in the weights and outputs of the model. The ultimate goal of BNNs is to quantify the uncertainty introduced by the models in terms of outputs and weights so as to explain the trustworthiness of the prediction.

In this work, we use Stochastic Weight Averaging~\citep[SWA;][]{maddox2019simple, wilson2020bayesian}. SWA is predicated on the observation that the parameters of deep neural networks often converge to the edges of low-loss regions. This edge-type convergence is sub-optimal, as these solutions are more susceptible to the shift between train and test error surfaces. SWA approximates the posterior distribution of the weights of the CNN as a Normal distribution, whose mean and covariance are
given by 

\begin{equation}
    \bar{w} = \frac{1}{N_{\rm swa}}\sum_{n=1}^{N_{\rm swa}} w_n, \quad \Sigma = \frac{1}{N_{\rm swa}}\sum_{n=1}^{N_{\rm swa}} (w_n - \bar{w})(w_n - \bar{w})^T,
\end{equation}
respectively, where $w$ are the weights of the network, $n$ is the time step during network optimization/training, and $N_{{\rm swa}}$ are the total steps over which SWA is performed.

By adopting this scheme, SWA solutions tend to converge to the center of flat loss regions, thereby leading to more stable and generalizeable solutions. Indeed, SWA has already been shown to lead to better generalization to out-of-distribution data~\citep{wilson2020bayesian}, which is expected to improve the robustness of our analysis. 
Moreover, SWA has been shown to outperform competing methods in multiple tasks~\citep{maddox2019simple}, and has been previously applied to astrophysics~\citep{cranmer2021bayesian} and cosmology~\citep{lemos2023robust}. We use the publicly available {\sc cosmoSWAG} implementation\footnote{\url{https://github.com/Pablo-Lemos/cosmoSWAG}}. The compressed galaxy field that we feed as input to SBI is the output of the SWA network: a set of 10 samples of the posterior distribution weights of the CNN --- a 50-dimensional data vector.

\subsection{Simulation-based inference}

After training the CNN, we use the {\sc SimBIG} SBI framework to estimate posterior distributions of the cosmological parameters, $\bm{\theta}$, from the compressed representation of the observables obtained from the CNN, $\bm{\hat{\theta}}$. We represent this posterior as $p(\bm{\theta}|\bm{\hat{\theta}})$.

There are multiple existing frameworks for SBI, such as Approximate Bayesian Computation~\citep[\emph{e.g.}][]{10.1214/aos/1176346785, pritchard1999population, beaumont2002approximate, marjoram2003markov, fearnhead2012constructing}, 
Neural Ratio Estimation~\citep[\emph{e.g.}][]{cranmer2015approximating, thomas2022likelihood, hermans2020likelihood, durkan2020contrastive, Miller:2022haf}, 
Neural Likelihood Estimation~\citep[\emph{e.g.}][]{price2018bayesian, papamakarios2019sequential, frazier2022bayesian}, 
and Neural Posterior Score Estimation~\citep[\emph{e.g.}][]{sharrock2022sequential, geffner2022score}. 
We use Neural Posterior Estimation~\citep[NPE; \emph{e.g.}][]{rezende2015variational, papamakarios2016fast, 2018arXiv180509294L, lueckmann2017flexible, greenberg2019automatic}, which uses a neural density estimator (NDE) to estimate the posterior distribution from a training set. In this case, the training set consists of the ground-truth/CNN-compressed $\{\bm \theta, \bm{\hat{\theta}}\}$ parameter pairs of the {\sc SimBIG} forward models. We use the publicly available {\sc sbi} implementation from \citet{tejero-cantero2020sbi}. 

Previous {\sc SimBIG} analyses employed a Masked Autoregressive Flow \citep{Papamakarios2017} as the density estimator. 
For our density estimator, we instead use Neural Spline Flows~\citep[NSF; ][]{durkan2019neural}, a more expressive alternative. 
Denoting our NSF as $q_\phi(\bm{\theta}|\bm{\hat{\theta}})$, where $\phi$ represents its hyperparameters, we train $q_\phi$ by minimizing the KL divergence between $p(\bm{\theta}, \bm{\hat{\theta}})$ and $q_\phi(\bm{\theta}| \bm{\hat{\theta}})p(\hat{\bm{\theta}})$. 
This is equivalent to maximizing the log-likelihood over the training set of {\sc SimBIG} forward models.
In practice, we split the catalogs into a training and 
validation set with 90/10 split, and use an early stopping procedure to prevent overfitting by stopping training when the validation log-likelihood has failed to increase after 20 epochs. 
Additionally, to improve the robustness of our NDE, we use an ensemble of five NSFs, which has been shown to produce more reliable 
approximations~\citep{lakshminarayanan2017simple, hermans2022trust}.

\begin{figure}[th!]
    \centering
    \includegraphics[width=.8\linewidth]{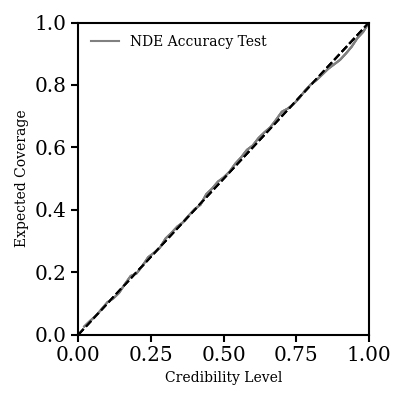}
    \caption{TARP expected coverage probability vs probability level. For an accurate posterior estimator, the line will follow the diagonal, while deviations from the diagonal are indicative of over or under confidence. We show the {\it NDE Accuracy test}, using $5,180$ of our base simulations that were not used in the training of our CNN.}
    \label{fig:tests_nde}
\end{figure}

\begin{figure*}
    \centering
    \centering
    \gridline{\fig{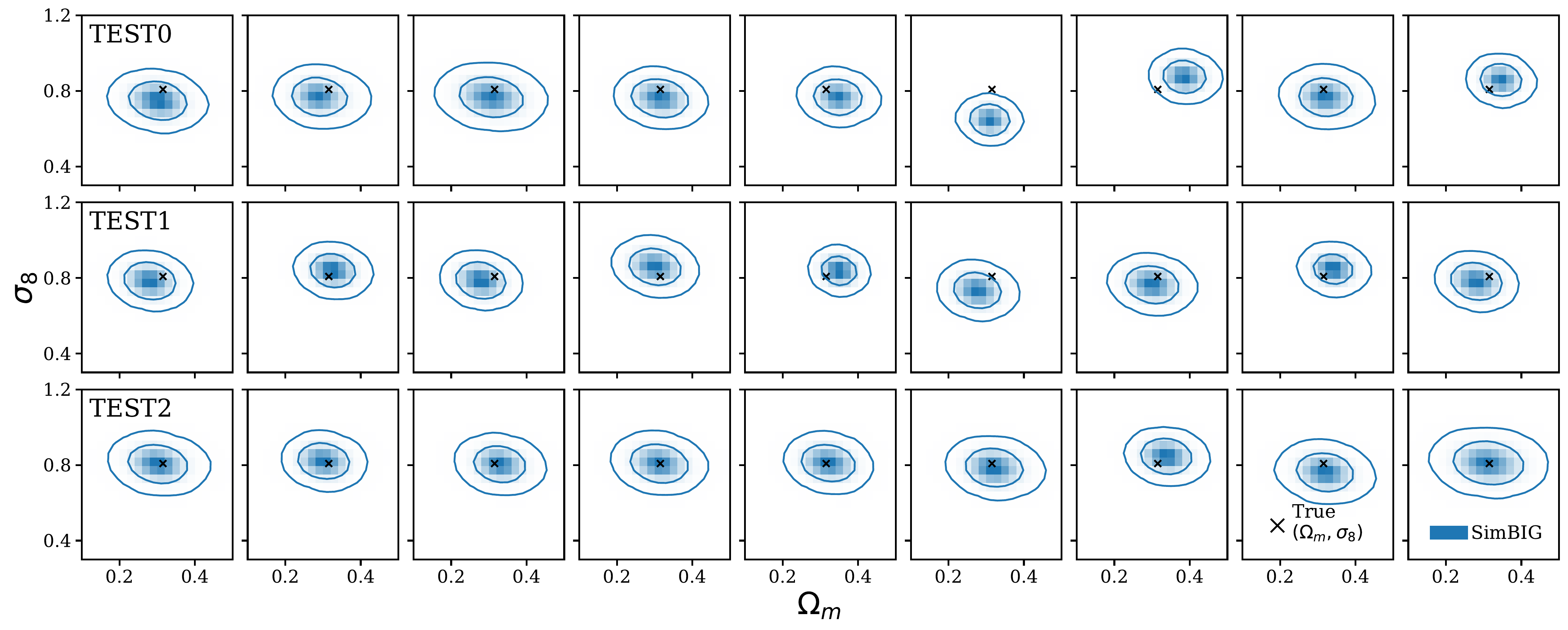}{1.0\hsize}{(a) Posterior distributions for $(\Omega_m, \sigma_8)$. Each row shows 9 randomly selected examples taken from each test set, as labeled. True parameters are marked in black and the contours the contours represent the 68 and 95 percentiles.}}
    \gridline{\fig{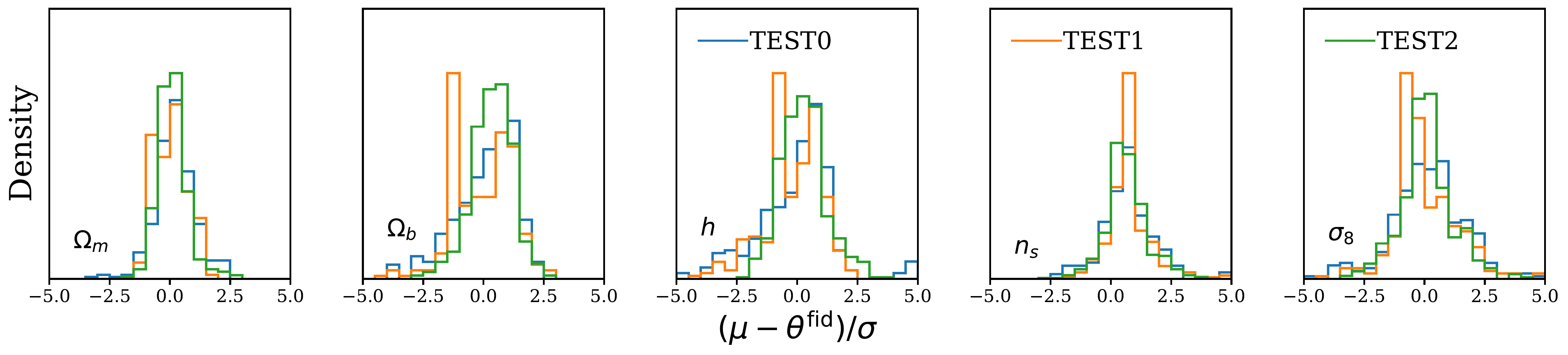}{1.0\hsize}{(b) Distributions of the differences between the posterior mean $\mu$ and the true parameter $\bm \theta^{\rm fid}$ normalized by the posterior standard deviation $\sigma$, for each cosmological parameter and test set. Differences between the distributions amongst the three datasets would be indicative of likelihood variations when we change our forward model. We find good agreement across all three forward models and all five parameters.}}
    \caption{Validation of our model on the {\sc SimBIG} mock challenge data.}
    \label{fig:tests_Om_s8}
\end{figure*}

\subsection{Validation}
Before analyzing observations, we first validate our posterior estimation in two stages. 
First, we validate on the $5,180$ simulations that were excluded from the training of our pipeline. 
We refer to this as the ``NDE accuracy test''. 
Second, we conduct the {\sc SimBIG} ``mock challenge'', where we validate our analysis on the suite of test simulations described in~\S\ref{sec:data}.
 
For the NDE accuracy test,  we use the Tests of Accuracy with Random 
Point (TARP) expected coverage probability (ECP) test as our metric. 
ECP is a necessary and sufficient test for the optimality of the estimated posterior, $q_\phi$~\citep{lemos2023sampling}\footnote{\url{https://github.com/Ciela-Institute/tarp}}. 
$p(\bm{\theta}\,|\,\bm{\hat{\theta}}) \equiv q_\phi(\bm{\theta}\,|\,\bm{\hat{\theta}})$
is only true in the limit of infinite data, and 
therefore we can only test for approximate equality, which 
is satisfied if and only if 
\begin{equation}\label{eq:ecp}
    {\rm ECP}(\hat{p}, \alpha) = 1 - \alpha \qquad \forall \alpha \in [0, 1],
\end{equation}
where ${\rm ECP}(\hat{p}, \alpha)$ is the expected coverage probability of the posterior estimate $\hat{p}$. 
TARP coverage probabilities are a robust method for estimating ECP that do not rely on evaluations of the posterior estimate. 
We can use it to calculate ECP for both the full-dimensional parameter space, or for
each parameter separately. 
The latter is equivalent to the Simulation-Based Calibration~\citep{talts2018validating} used in the 
other {\sc SimBIG} analyses.

We present the results of our NDE accuracy test using TARP in~\cref{fig:tests_nde}, 
where we plot the ECP versus the confidence level $1-\alpha$ \footnote{This figure is often referred to as a probability-probability (PP) plot.}.
We evaluate the TARP ECP over the full dimensionality of our parameter space.
If the ECP and confidence level are equal for every $\alpha \in [0, 1]$, 
\emph{i.e.} it follows a diagonal line, 
then the estimator is well calibrated since the probability of our posterior estimate 
containing the true parameter values matches the actual confidence level.
We find that the NDE accuracy test is perfectly calibrated, as 
the ECP line is perfectly in the diagonal. 

We then move on to the {\sc SimBIG} mock challenge.
~\cref{fig:tests_Om_s8} (a) shows the marginalized two-dimensional posterior distribution of $\{\Omega_m, \sigma_8\}$ for 9 randomly selected simulations from each of the test sets: {\sc TEST0} (top), {\sc TEST1} (center), and {\sc TEST2} (bottom).
We mark the true $\Omega_m$ and $\sigma_8$ values in each panel (black x).
For all three test sets the posteriors appear to be well calibrated and unbiased, 
which qualitatively demonstrate the robustness of our analysis.

\begin{figure*}[th!]
    \centering
    \includegraphics[width=.99\linewidth]{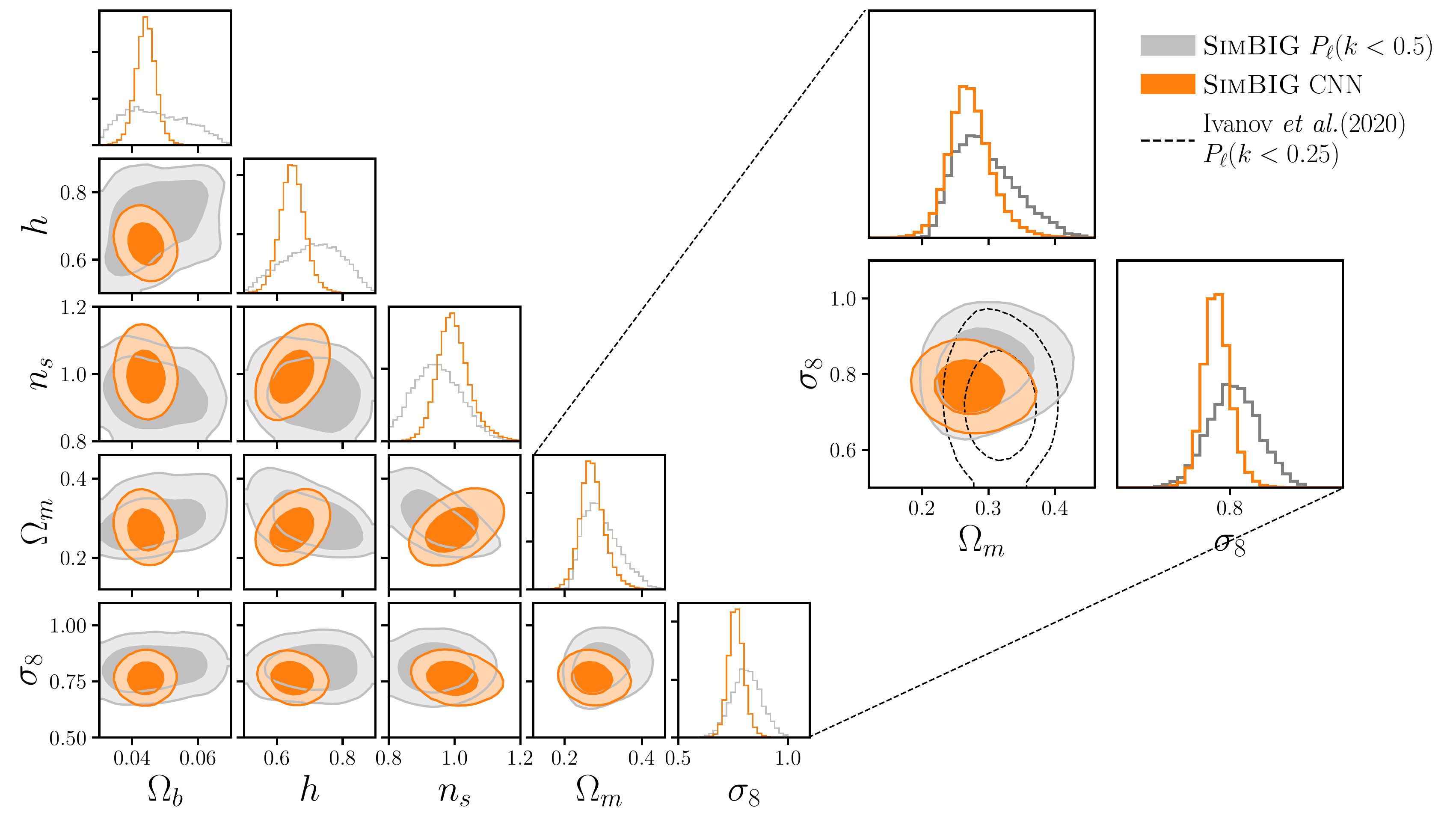}
    \caption{
    {\em Left}: 
    Posterior distributions for all $\Lambda$CDM 
    cosmological parameters from our CNN-based field 
    level inference of BOSS observations (orange). 
    For comparison, we include the {\sc SimBIG} 
    $P_\ell$  analysis (gray). 
    The contours represent the $68\%$ and $95\%$ confidence intervals. 
    Our CNN-based field-level inference produces 
    tighter, yet consistent, constraints to the 
    {\sc SimBIG} $P_\ell$ analysis.
    {\em Right}: Posterior distributions for 
    $\Omega_m$ and $\sigma_8$. 
    For comparison, we include posteriors from the 
    {\sc SimBIG} $P_\ell$ analysis (gray) and  
    the standard PT-based $P_\ell$
    analysis~\citep[black dashed;][]{ivanov2020cosmological}.
    Our analysis constrains $\Omega_m$ and $\sigma_8$ 
    $1.76$ and $1.92\times$ tighter than the {\sc SimBIG} 
    $P_\ell$ analysis. Moreover, our constraints on $\Omega_m$ are in-line with the PT-based $P_\ell$ analysis, and those on $\sigma_8$ are $2.65\times$ tighter}
    \label{fig:corner}
\end{figure*}

Next, we assess robustness more quantitatively using {\sc TEST0}, {\sc TEST1}, and {\sc TEST2}.
For the test simulations, we cannot use the same method as the NDE accuracy test due to the fact that the ECP relies on averaging over the prior distribution, but all of these simulations are run at fixed fiducial cosmologies. Therefore, we follow~\cite{simbig_mock_challenge} and we assess robustness by comparing the likelihoods over the three test sets. Specifically, we compute the posterior mean $\mu$ and standard deviation $\sigma$ for each $\Lambda$CDM parameter for each suite of test simulations. Then, we analyze the difference between $\mu$ and the true parameter value $\theta^{\rm fid}$ in units of $\sigma$. For a robust pipeline, we expect to find consistency of these estimates across all three datasets. 
On the other hand, variations between the distributions would be indicative of likelihood variations that come from changing the forward model and imply that 
our analysis is sensitive to model variations. 

In ~\cref{fig:tests_Om_s8} (b), we present the likelihoods of 
{\sc TEST0} (blue), {\sc TEST1} (orange), and {\sc TEST2} (green) for each of the
$\Lambda$CDM parameters. 
We find consistent distributions for all parameters across test sets. 
This indicates that our posterior inference is robust to variations in the forward model. 
It also suggests that our use of weight marginalization led to better generalization 
properties. 

These validation tests form a crucial part of our analysis. We note that it is possible to obtain significantly tighter constraints that pass only the NDE accuracy test. 
However, in doing so, we would need to assume that our forward model accurately models every aspect of the 
observations. 
Given the complexities of galaxy formation, {\em any} forward
model of galaxy clustering is an approximate model. 
Hence, validating that we can successfully infer unbiased 
cosmological constraints from simulated test galaxy catalogs
generated with different forward models ({\sc TEST1} and 
{\sc TEST2}) serves as a powerful test against model 
misspecification, even if it come at the expense of 
significant constraining power.
In future work, we will explore additional tests of model 
misspecification and ``blind challenges'' where we test our 
analysis on simulations without knowing the true cosmological 
parameters or the forward model used to generate them.

\section{\label{sec:results}Results:\protect}
In \cref{fig:corner}, we present the posterior distribution of all $\Lambda$CDM 
cosmological parameters inferred from our field-level analysis of the BOSS CMASS SGC
using {\sc SimBIG} (orange). 
In the right panels, we focus on the growth of structure parameters $\Omega_m$ and
$\sigma_8$.
The diagonal panels present the 1D marginalized posteriors; the rest of the
panels present marginalized 2D posteriors of different parameter pairs. 
The contours represent the 68 and 95 percentiles and the ranges of the panels 
match the prior.
For comparison, we include posteriors from the {\sc SimBIG} 
$P_\ell(k_{\rm max} < 0.5\,h/{\rm Mpc})$ 
analysis \citep[grey;][]{hahnsimbig} as well as the constraints from the  
PT based $P_\ell(k_{\rm max} < 0.25\,h/{\rm Mpc})$ analysis of the CMASS SGC sample~\citep[dashed;][]{ivanov2020cosmological}. 

Overall, our field-level analysis using the CNN provides tighter, yet consistent, cosmological constraints to the previous BOSS analyses. 
Specifically, our constraints on $\Omega_m$ and $\sigma_8$ are
$1.76\times$ and $1.92\times$ tighter than the {\sc SimBIG}
$P_\ell$ analysis. Moreover, our constraints on $\Omega_m$ are in-line with the PT-based $P_\ell$ analysis, and those on $\sigma_8$ are $2.65\times$ tighter. 
This higher constraining power is expected.
Indeed, by using the full galaxy field, we are able to exploit non-Gaussian cosmological information on non-linear scales that is inaccessible to $P_\ell$ analyses. 
Moreover, in using the {\sc SimBIG} SBI approach, we are able 
to more robustly account for observational systematics compared to the standard clustering analyses.

In fact, with the added constraining power of our field-level analysis, we also can place significant 
constraints on $H_0=63.1 \pm 4.1 \ {\rm km / s / Mpc}$, albeit weaker than those 
on $\Omega_m$ and $\sigma_8$. 
This is in contrast to standard $P_\ell$ analyses, which  
cannot independently constrain $H_0$  and typically 
rely on priors from Big Bang Nucleosynthesis or CMB 
experiments. 
Our constraints support a low value of $H_0$ in good agreement with {\it Planck} constraints~\citep{aghanim2020planck}. 
However, we do not have not have enough constraining power to make strong statements. 
We will further investigate the cosmological implications of this result 
and how they compare with other surveys and cosmological 
probes in an accompanying paper.


\section{\label{sec:conclusions}Conclusions:\protect}

In this paper, we present cosmological constraints from a field-level analysis of the CMASS galaxy catalogs using simulation-based inference. 
We demonstrate that our analysis passes a number of stringent validation tests, including a robustness test based on simulations constructed using different forward models. 
These test sets provide key validation against model misspecification and demonstrates some robustness against
discrepancies between observations and our forward model.

Furthermore, we show that our cosmological parameter 
constraints are consistent but significantly tighter 
than those from $P_\ell$ analyses. 
In particular, our constraints on $\Omega_m$ and $\sigma_8$ 
are in-line and $2.65\times$ tighter than the standard PT-based 
$P_\ell$ analyses. 
We are even able to produce significant constraints on
$H_0$, without any priors from external experiments.
These improvements demonstrate that our method successfully 
extracts additional non-Gaussian and non-linear cosmological
information from the galaxy distribution. 

As simulations become more realistic and efficient in the 
future, we will be able to extend our analyses to 
smaller scales an the larger volumes covered by upcoming
surveys such as the 
Dark Energy Spectroscopic Instrument~\citep[DESI;][]{desicollaboration2016, desicollaboration2016a, abareshi2022}, 
Subaru Prime Focus Spectrograph~\citep[PFS;][]{takada2014, tamura2016}, 
the ESA {\em Euclid} satellite mission~\citep{laureijs2011}, and the
Nancy Grace Roman Space Telescope~\citep[Roman;][]{spergel2015, wang2022a}. 
Our results demonstrate that these analyses will be able to 
produce leading cosmological constraints from galaxy 
clustering.
The methodology and tests presented in this paper lay the groundwork for such analyses. 

In accompanying papers 
we present the {\sc SimBIG} analysis of galaxy clustering using two summary statistics: the galaxy bispectrum and the wavelet scattering transform statistics. 
Furthermore, in \cite{simbig2023compare}, we present a comparison of the different
{\sc SimBIG} analyses, including the field-level constraints presented in this work. 
We also discuss their cosmological implications and present forecasts for extending
{\sc SimBIG} to upcoming galaxy surveys.

\section*{Acknowledgements}
It is a pleasure to thank 
Mikhail M. Ivanov 
for providing us with the posteriors used for comparison, and Ben Wandelt for discussions that greatly helped the papers. We thank the Learning the Universe Collaboration for helpful feedback and stimulating discussions.
PL acknowledges support from the Simons Foundation.
JH has received funding from the European Union's Horizon 2020 research and innovation program under the Marie Sk\l{}odowska-Curie grant agreement No 101025187. AMD acknowledges funding from Tomalla Foundation for Research in Gravity. 


\bibliography{main}

\providecommand{\aj}{Astron. J. }\providecommand{\apj}{ApJ
  }\providecommand{\apjl}{ApJ
  }\providecommand{\mnras}{MNRAS}\providecommand{\prl}{PRL}\providecommand{\prd}{PRD}\providecommand{\jcap}{JCAP}\providecommand{\aap}{A\&A}
\begin{thebibliography}{109}%
\makeatletter
\providecommand \@ifxundefined [1]{%
 \@ifx{#1\undefined}
}%
\providecommand \@ifnum [1]{%
 \ifnum #1\expandafter \@firstoftwo
 \else \expandafter \@secondoftwo
 \fi
}%
\providecommand \@ifx [1]{%
 \ifx #1\expandafter \@firstoftwo
 \else \expandafter \@secondoftwo
 \fi
}%
\providecommand \natexlab [1]{#1}%
\providecommand \enquote  [1]{``#1''}%
\providecommand \bibnamefont  [1]{#1}%
\providecommand \bibfnamefont [1]{#1}%
\providecommand \citenamefont [1]{#1}%
\providecommand \href@noop [0]{\@secondoftwo}%
\providecommand \href [0]{\begingroup \@sanitize@url \@href}%
\providecommand \@href[1]{\@@startlink{#1}\@@href}%
\providecommand \@@href[1]{\endgroup#1\@@endlink}%
\providecommand \@sanitize@url [0]{\catcode `\\12\catcode `\$12\catcode
  `\&12\catcode `\#12\catcode `\^12\catcode `\_12\catcode `\%12\relax}%
\providecommand \@@startlink[1]{}%
\providecommand \@@endlink[0]{}%
\providecommand \url  [0]{\begingroup\@sanitize@url \@url }%
\providecommand \@url [1]{\endgroup\@href {#1}{\urlprefix }}%
\providecommand \urlprefix  [0]{URL }%
\providecommand \Eprint [0]{\href }%
\providecommand \doibase [0]{http://dx.doi.org/}%
\providecommand \selectlanguage [0]{\@gobble}%
\providecommand \bibinfo  [0]{\@secondoftwo}%
\providecommand \bibfield  [0]{\@secondoftwo}%
\providecommand \translation [1]{[#1]}%
\providecommand \BibitemOpen [0]{}%
\providecommand \bibitemStop [0]{}%
\providecommand \bibitemNoStop [0]{.\EOS\space}%
\providecommand \EOS [0]{\spacefactor3000\relax}%
\providecommand \BibitemShut  [1]{\csname bibitem#1\endcsname}%
\let\auto@bib@innerbib\@empty
\bibitem [{\citenamefont {Kaiser}(1987)}]{kaiser1987clustering}%
  \BibitemOpen
  \bibfield  {author} {\bibinfo {author} {\bibfnamefont {N.}~\bibnamefont
  {Kaiser}},\ }\href@noop {} {\bibfield  {journal} {\bibinfo  {journal}
  {Monthly Notices of the Royal Astronomical Society}\ }\textbf {\bibinfo
  {volume} {227}},\ \bibinfo {pages} {1} (\bibinfo {year} {1987})}\BibitemShut
  {NoStop}%
\bibitem [{\citenamefont {Hamilton}(1997)}]{hamilton1997towards}%
  \BibitemOpen
  \bibfield  {author} {\bibinfo {author} {\bibfnamefont {A.}~\bibnamefont
  {Hamilton}},\ }\href@noop {} {\bibfield  {journal} {\bibinfo  {journal}
  {Monthly Notices of the Royal Astronomical Society}\ }\textbf {\bibinfo
  {volume} {289}},\ \bibinfo {pages} {285} (\bibinfo {year}
  {1997})}\BibitemShut {NoStop}%
\bibitem [{\citenamefont {Peacock}\ \emph {et~al.}(2001)\citenamefont
  {Peacock}, \citenamefont {Cole}, \citenamefont {Norberg}, \citenamefont
  {Baugh}, \citenamefont {Bland-Hawthorn}, \citenamefont {Bridges},
  \citenamefont {Cannon}, \citenamefont {Colless}, \citenamefont {Collins},
  \citenamefont {Couch} \emph {et~al.}}]{peacock2001measurement}%
  \BibitemOpen
  \bibfield  {author} {\bibinfo {author} {\bibfnamefont {J.~A.}\ \bibnamefont
  {Peacock}}, \bibinfo {author} {\bibfnamefont {S.}~\bibnamefont {Cole}},
  \bibinfo {author} {\bibfnamefont {P.}~\bibnamefont {Norberg}}, \bibinfo
  {author} {\bibfnamefont {C.~M.}\ \bibnamefont {Baugh}}, \bibinfo {author}
  {\bibfnamefont {J.}~\bibnamefont {Bland-Hawthorn}}, \bibinfo {author}
  {\bibfnamefont {T.}~\bibnamefont {Bridges}}, \bibinfo {author} {\bibfnamefont
  {R.~D.}\ \bibnamefont {Cannon}}, \bibinfo {author} {\bibfnamefont
  {M.}~\bibnamefont {Colless}}, \bibinfo {author} {\bibfnamefont
  {C.}~\bibnamefont {Collins}}, \bibinfo {author} {\bibfnamefont
  {W.}~\bibnamefont {Couch}},  \emph {et~al.},\ }\href@noop {} {\bibfield
  {journal} {\bibinfo  {journal} {Nature}\ }\textbf {\bibinfo {volume} {410}},\
  \bibinfo {pages} {169} (\bibinfo {year} {2001})}\BibitemShut {NoStop}%
\bibitem [{\citenamefont {Hawkins}\ \emph {et~al.}(2003)\citenamefont
  {Hawkins}, \citenamefont {Maddox}, \citenamefont {Cole}, \citenamefont
  {Lahav}, \citenamefont {Madgwick}, \citenamefont {Norberg}, \citenamefont
  {Peacock}, \citenamefont {Baldry}, \citenamefont {Baugh}, \citenamefont
  {Bland-Hawthorn} \emph {et~al.}}]{hawkins20032df}%
  \BibitemOpen
  \bibfield  {author} {\bibinfo {author} {\bibfnamefont {E.}~\bibnamefont
  {Hawkins}}, \bibinfo {author} {\bibfnamefont {S.}~\bibnamefont {Maddox}},
  \bibinfo {author} {\bibfnamefont {S.}~\bibnamefont {Cole}}, \bibinfo {author}
  {\bibfnamefont {O.}~\bibnamefont {Lahav}}, \bibinfo {author} {\bibfnamefont
  {D.~S.}\ \bibnamefont {Madgwick}}, \bibinfo {author} {\bibfnamefont
  {P.}~\bibnamefont {Norberg}}, \bibinfo {author} {\bibfnamefont {J.~A.}\
  \bibnamefont {Peacock}}, \bibinfo {author} {\bibfnamefont {I.~K.}\
  \bibnamefont {Baldry}}, \bibinfo {author} {\bibfnamefont {C.~M.}\
  \bibnamefont {Baugh}}, \bibinfo {author} {\bibfnamefont {J.}~\bibnamefont
  {Bland-Hawthorn}},  \emph {et~al.},\ }\href@noop {} {\bibfield  {journal}
  {\bibinfo  {journal} {Monthly Notices of the Royal Astronomical Society}\
  }\textbf {\bibinfo {volume} {346}},\ \bibinfo {pages} {78} (\bibinfo {year}
  {2003})}\BibitemShut {NoStop}%
\bibitem [{\citenamefont {Tegmark}\ \emph {et~al.}(2006)\citenamefont
  {Tegmark}, \citenamefont {Eisenstein}, \citenamefont {Strauss}, \citenamefont
  {Weinberg}, \citenamefont {Blanton}, \citenamefont {Frieman}, \citenamefont
  {Fukugita}, \citenamefont {Gunn}, \citenamefont {Hamilton}, \citenamefont
  {Knapp} \emph {et~al.}}]{tegmark2006cosmological}%
  \BibitemOpen
  \bibfield  {author} {\bibinfo {author} {\bibfnamefont {M.}~\bibnamefont
  {Tegmark}}, \bibinfo {author} {\bibfnamefont {D.~J.}\ \bibnamefont
  {Eisenstein}}, \bibinfo {author} {\bibfnamefont {M.~A.}\ \bibnamefont
  {Strauss}}, \bibinfo {author} {\bibfnamefont {D.~H.}\ \bibnamefont
  {Weinberg}}, \bibinfo {author} {\bibfnamefont {M.~R.}\ \bibnamefont
  {Blanton}}, \bibinfo {author} {\bibfnamefont {J.~A.}\ \bibnamefont
  {Frieman}}, \bibinfo {author} {\bibfnamefont {M.}~\bibnamefont {Fukugita}},
  \bibinfo {author} {\bibfnamefont {J.~E.}\ \bibnamefont {Gunn}}, \bibinfo
  {author} {\bibfnamefont {A.~J.}\ \bibnamefont {Hamilton}}, \bibinfo {author}
  {\bibfnamefont {G.~R.}\ \bibnamefont {Knapp}},  \emph {et~al.},\ }\href@noop
  {} {\bibfield  {journal} {\bibinfo  {journal} {Physical Review D}\ }\textbf
  {\bibinfo {volume} {74}},\ \bibinfo {pages} {123507} (\bibinfo {year}
  {2006})}\BibitemShut {NoStop}%
\bibitem [{\citenamefont {Guzzo}\ \emph {et~al.}(2008)\citenamefont {Guzzo},
  \citenamefont {Pierleoni}, \citenamefont {Meneux}, \citenamefont {Branchini},
  \citenamefont {Le~F{\`e}vre}, \citenamefont {Marinoni}, \citenamefont
  {Garilli}, \citenamefont {Blaizot}, \citenamefont {De~Lucia}, \citenamefont
  {Pollo} \emph {et~al.}}]{guzzo2008test}%
  \BibitemOpen
  \bibfield  {author} {\bibinfo {author} {\bibfnamefont {L.}~\bibnamefont
  {Guzzo}}, \bibinfo {author} {\bibfnamefont {M.}~\bibnamefont {Pierleoni}},
  \bibinfo {author} {\bibfnamefont {B.}~\bibnamefont {Meneux}}, \bibinfo
  {author} {\bibfnamefont {E.}~\bibnamefont {Branchini}}, \bibinfo {author}
  {\bibfnamefont {O.}~\bibnamefont {Le~F{\`e}vre}}, \bibinfo {author}
  {\bibfnamefont {C.}~\bibnamefont {Marinoni}}, \bibinfo {author}
  {\bibfnamefont {B.}~\bibnamefont {Garilli}}, \bibinfo {author} {\bibfnamefont
  {J.}~\bibnamefont {Blaizot}}, \bibinfo {author} {\bibfnamefont
  {G.}~\bibnamefont {De~Lucia}}, \bibinfo {author} {\bibfnamefont
  {A.}~\bibnamefont {Pollo}},  \emph {et~al.},\ }\href@noop {} {\bibfield
  {journal} {\bibinfo  {journal} {Nature}\ }\textbf {\bibinfo {volume} {451}},\
  \bibinfo {pages} {541} (\bibinfo {year} {2008})}\BibitemShut {NoStop}%
\bibitem [{\citenamefont {Beutler}\ \emph {et~al.}(2017)\citenamefont
  {Beutler}, \citenamefont {Seo}, \citenamefont {Saito}, \citenamefont
  {Chuang}, \citenamefont {Cuesta}, \citenamefont {Eisenstein}, \citenamefont
  {{Gil-Mar{\'i}n}}, \citenamefont {Grieb}, \citenamefont {Hand}, \citenamefont
  {Kitaura}, \citenamefont {Modi}, \citenamefont {Nichol}, \citenamefont
  {Olmstead}, \citenamefont {Percival}, \citenamefont {Prada}, \citenamefont
  {S{\'a}nchez}, \citenamefont {{Rodriguez-Torres}}, \citenamefont {Ross},
  \citenamefont {Ross}, \citenamefont {Schneider}, \citenamefont {Tinker},
  \citenamefont {Tojeiro},\ and\ \citenamefont
  {{Vargas-Maga{\~n}a}}}]{beutler2017}%
  \BibitemOpen
  \bibfield  {author} {\bibinfo {author} {\bibfnamefont {F.}~\bibnamefont
  {Beutler}}, \bibinfo {author} {\bibfnamefont {H.-J.}\ \bibnamefont {Seo}},
  \bibinfo {author} {\bibfnamefont {S.}~\bibnamefont {Saito}}, \bibinfo
  {author} {\bibfnamefont {C.-H.}\ \bibnamefont {Chuang}}, \bibinfo {author}
  {\bibfnamefont {A.~J.}\ \bibnamefont {Cuesta}}, \bibinfo {author}
  {\bibfnamefont {D.~J.}\ \bibnamefont {Eisenstein}}, \bibinfo {author}
  {\bibfnamefont {H.}~\bibnamefont {{Gil-Mar{\'i}n}}}, \bibinfo {author}
  {\bibfnamefont {J.~N.}\ \bibnamefont {Grieb}}, \bibinfo {author}
  {\bibfnamefont {N.}~\bibnamefont {Hand}}, \bibinfo {author} {\bibfnamefont
  {F.-S.}\ \bibnamefont {Kitaura}}, \bibinfo {author} {\bibfnamefont
  {C.}~\bibnamefont {Modi}}, \bibinfo {author} {\bibfnamefont {R.~C.}\
  \bibnamefont {Nichol}}, \bibinfo {author} {\bibfnamefont {M.~D.}\
  \bibnamefont {Olmstead}}, \bibinfo {author} {\bibfnamefont {W.~J.}\
  \bibnamefont {Percival}}, \bibinfo {author} {\bibfnamefont {F.}~\bibnamefont
  {Prada}}, \bibinfo {author} {\bibfnamefont {A.~G.}\ \bibnamefont
  {S{\'a}nchez}}, \bibinfo {author} {\bibfnamefont {S.}~\bibnamefont
  {{Rodriguez-Torres}}}, \bibinfo {author} {\bibfnamefont {A.~J.}\ \bibnamefont
  {Ross}}, \bibinfo {author} {\bibfnamefont {N.~P.}\ \bibnamefont {Ross}},
  \bibinfo {author} {\bibfnamefont {D.~P.}\ \bibnamefont {Schneider}}, \bibinfo
  {author} {\bibfnamefont {J.}~\bibnamefont {Tinker}}, \bibinfo {author}
  {\bibfnamefont {R.}~\bibnamefont {Tojeiro}}, \ and\ \bibinfo {author}
  {\bibfnamefont {M.}~\bibnamefont {{Vargas-Maga{\~n}a}}},\ }\href {\doibase
  10.1093/mnras/stw3298} {\bibfield  {journal} {\bibinfo  {journal} {Monthly
  Notices of the Royal Astronomical Society}\ }\textbf {\bibinfo {volume}
  {466}},\ \bibinfo {pages} {2242} (\bibinfo {year} {2017})}\BibitemShut
  {NoStop}%
\bibitem [{\citenamefont {Ivanov}\ \emph {et~al.}(2020)\citenamefont {Ivanov},
  \citenamefont {Simonovi{\'c}},\ and\ \citenamefont
  {Zaldarriaga}}]{ivanov2020cosmological}%
  \BibitemOpen
  \bibfield  {author} {\bibinfo {author} {\bibfnamefont {M.~M.}\ \bibnamefont
  {Ivanov}}, \bibinfo {author} {\bibfnamefont {M.}~\bibnamefont
  {Simonovi{\'c}}}, \ and\ \bibinfo {author} {\bibfnamefont {M.}~\bibnamefont
  {Zaldarriaga}},\ }\href@noop {} {\bibfield  {journal} {\bibinfo  {journal}
  {Physical Review D}\ }\textbf {\bibinfo {volume} {101}},\ \bibinfo {pages}
  {083504} (\bibinfo {year} {2020})}\BibitemShut {NoStop}%
\bibitem [{\citenamefont {{Kobayashi}}\ \emph {et~al.}(2022)\citenamefont
  {{Kobayashi}}, \citenamefont {{Nishimichi}}, \citenamefont {{Takada}},\ and\
  \citenamefont {{Miyatake}}}]{kobayashi2022}%
  \BibitemOpen
  \bibfield  {author} {\bibinfo {author} {\bibfnamefont {Y.}~\bibnamefont
  {{Kobayashi}}}, \bibinfo {author} {\bibfnamefont {T.}~\bibnamefont
  {{Nishimichi}}}, \bibinfo {author} {\bibfnamefont {M.}~\bibnamefont
  {{Takada}}}, \ and\ \bibinfo {author} {\bibfnamefont {H.}~\bibnamefont
  {{Miyatake}}},\ }\href {\doibase 10.1103/PhysRevD.105.083517} {\bibfield
  {journal} {\bibinfo  {journal} {\prd}\ }\textbf {\bibinfo {volume} {105}},\
  \bibinfo {eid} {083517} (\bibinfo {year} {2022})},\ \Eprint
  {http://arxiv.org/abs/2110.06969} {arXiv:2110.06969 [astro-ph.CO]}
  \BibitemShut {NoStop}%
\bibitem [{\citenamefont {Bernardeau}\ \emph {et~al.}(2002)\citenamefont
  {Bernardeau}, \citenamefont {Colombi}, \citenamefont {Gaztanaga},\ and\
  \citenamefont {Scoccimarro}}]{bernardeau2002}%
  \BibitemOpen
  \bibfield  {author} {\bibinfo {author} {\bibfnamefont {F.}~\bibnamefont
  {Bernardeau}}, \bibinfo {author} {\bibfnamefont {S.}~\bibnamefont {Colombi}},
  \bibinfo {author} {\bibfnamefont {E.}~\bibnamefont {Gaztanaga}}, \ and\
  \bibinfo {author} {\bibfnamefont {R.}~\bibnamefont {Scoccimarro}},\ }\href
  {\doibase 10.1016/S0370-1573(02)00135-7} {\bibfield  {journal} {\bibinfo
  {journal} {Physics Reports}\ }\textbf {\bibinfo {volume} {367}},\ \bibinfo
  {pages} {1} (\bibinfo {year} {2002})},\ \Eprint
  {http://arxiv.org/abs/astro-ph/0112551} {arXiv:astro-ph/0112551} \BibitemShut
  {NoStop}%
\bibitem [{\citenamefont {Desjacques}\ \emph {et~al.}(2018)\citenamefont
  {Desjacques}, \citenamefont {Jeong},\ and\ \citenamefont
  {Schmidt}}]{desjacques2018}%
  \BibitemOpen
  \bibfield  {author} {\bibinfo {author} {\bibfnamefont {V.}~\bibnamefont
  {Desjacques}}, \bibinfo {author} {\bibfnamefont {D.}~\bibnamefont {Jeong}}, \
  and\ \bibinfo {author} {\bibfnamefont {F.}~\bibnamefont {Schmidt}},\ }\href
  {\doibase 10.1016/j.physrep.2017.12.002} {\bibfield  {journal} {\bibinfo
  {journal} {Physics Reports}\ }\textbf {\bibinfo {volume} {733}},\ \bibinfo
  {pages} {1} (\bibinfo {year} {2018})}\BibitemShut {NoStop}%
\bibitem [{\citenamefont {{Gil-Mar{\'i}n}}\ \emph {et~al.}(2017)\citenamefont
  {{Gil-Mar{\'i}n}}, \citenamefont {Percival}, \citenamefont {Verde},
  \citenamefont {Brownstein}, \citenamefont {Chuang}, \citenamefont {Kitaura},
  \citenamefont {{Rodr{\'i}guez-Torres}},\ and\ \citenamefont
  {Olmstead}}]{gil-marin2017}%
  \BibitemOpen
  \bibfield  {author} {\bibinfo {author} {\bibfnamefont {H.}~\bibnamefont
  {{Gil-Mar{\'i}n}}}, \bibinfo {author} {\bibfnamefont {W.~J.}\ \bibnamefont
  {Percival}}, \bibinfo {author} {\bibfnamefont {L.}~\bibnamefont {Verde}},
  \bibinfo {author} {\bibfnamefont {J.~R.}\ \bibnamefont {Brownstein}},
  \bibinfo {author} {\bibfnamefont {C.-H.}\ \bibnamefont {Chuang}}, \bibinfo
  {author} {\bibfnamefont {F.-S.}\ \bibnamefont {Kitaura}}, \bibinfo {author}
  {\bibfnamefont {S.~A.}\ \bibnamefont {{Rodr{\'i}guez-Torres}}}, \ and\
  \bibinfo {author} {\bibfnamefont {M.~D.}\ \bibnamefont {Olmstead}},\ }\href
  {\doibase 10.1093/mnras/stw2679} {\bibfield  {journal} {\bibinfo  {journal}
  {Monthly Notices of the Royal Astronomical Society}\ }\textbf {\bibinfo
  {volume} {465}},\ \bibinfo {pages} {1757} (\bibinfo {year}
  {2017})}\BibitemShut {NoStop}%
\bibitem [{\citenamefont {D'Amico}\ \emph {et~al.}(2022)\citenamefont
  {D'Amico}, \citenamefont {Donath}, \citenamefont {Lewandowski}, \citenamefont
  {Senatore},\ and\ \citenamefont {Zhang}}]{damico2022}%
  \BibitemOpen
  \bibfield  {author} {\bibinfo {author} {\bibfnamefont {G.}~\bibnamefont
  {D'Amico}}, \bibinfo {author} {\bibfnamefont {Y.}~\bibnamefont {Donath}},
  \bibinfo {author} {\bibfnamefont {M.}~\bibnamefont {Lewandowski}}, \bibinfo
  {author} {\bibfnamefont {L.}~\bibnamefont {Senatore}}, \ and\ \bibinfo
  {author} {\bibfnamefont {P.}~\bibnamefont {Zhang}},\ }\href@noop {} {\enquote
  {\bibinfo {title} {The {{BOSS}} bispectrum analysis at one loop from the
  {{Effective Field Theory}} of {{Large-Scale Structure}}},}\ } (\bibinfo
  {year} {2022})\BibitemShut {NoStop}%
\bibitem [{\citenamefont {Philcox}\ and\ \citenamefont
  {Ivanov}(2022)}]{philcox2022boss}%
  \BibitemOpen
  \bibfield  {author} {\bibinfo {author} {\bibfnamefont {O.~H.}\ \bibnamefont
  {Philcox}}\ and\ \bibinfo {author} {\bibfnamefont {M.~M.}\ \bibnamefont
  {Ivanov}},\ }\href@noop {} {\bibfield  {journal} {\bibinfo  {journal}
  {Physical Review D}\ }\textbf {\bibinfo {volume} {105}},\ \bibinfo {pages}
  {043517} (\bibinfo {year} {2022})}\BibitemShut {NoStop}%
\bibitem [{\citenamefont {{Ivanov}}\ \emph {et~al.}(2023)\citenamefont
  {{Ivanov}}, \citenamefont {{Philcox}}, \citenamefont {{Cabass}},
  \citenamefont {{Nishimichi}}, \citenamefont {{Simonovi{\'c}}},\ and\
  \citenamefont {{Zaldarriaga}}}]{ivanov2023}%
  \BibitemOpen
  \bibfield  {author} {\bibinfo {author} {\bibfnamefont {M.~M.}\ \bibnamefont
  {{Ivanov}}}, \bibinfo {author} {\bibfnamefont {O.~H.~E.}\ \bibnamefont
  {{Philcox}}}, \bibinfo {author} {\bibfnamefont {G.}~\bibnamefont {{Cabass}}},
  \bibinfo {author} {\bibfnamefont {T.}~\bibnamefont {{Nishimichi}}}, \bibinfo
  {author} {\bibfnamefont {M.}~\bibnamefont {{Simonovi{\'c}}}}, \ and\ \bibinfo
  {author} {\bibfnamefont {M.}~\bibnamefont {{Zaldarriaga}}},\ }\href {\doibase
  10.1103/PhysRevD.107.083515} {\bibfield  {journal} {\bibinfo  {journal}
  {\prd}\ }\textbf {\bibinfo {volume} {107}},\ \bibinfo {eid} {083515}
  (\bibinfo {year} {2023})},\ \Eprint {http://arxiv.org/abs/2302.04414}
  {arXiv:2302.04414 [astro-ph.CO]} \BibitemShut {NoStop}%
\bibitem [{\citenamefont {Hahn}\ \emph {et~al.}(2020)\citenamefont {Hahn},
  \citenamefont {{Villaescusa-Navarro}}, \citenamefont {Castorina},\ and\
  \citenamefont {Scoccimarro}}]{hahn2020}%
  \BibitemOpen
  \bibfield  {author} {\bibinfo {author} {\bibfnamefont {C.}~\bibnamefont
  {Hahn}}, \bibinfo {author} {\bibfnamefont {F.}~\bibnamefont
  {{Villaescusa-Navarro}}}, \bibinfo {author} {\bibfnamefont {E.}~\bibnamefont
  {Castorina}}, \ and\ \bibinfo {author} {\bibfnamefont {R.}~\bibnamefont
  {Scoccimarro}},\ }\href {\doibase 10.1088/1475-7516/2020/03/040} {\bibfield
  {journal} {\bibinfo  {journal} {Journal of Cosmology and Astroparticle
  Physics}\ }\textbf {\bibinfo {volume} {03}},\ \bibinfo {pages} {040}
  (\bibinfo {year} {2020})}\BibitemShut {NoStop}%
\bibitem [{\citenamefont {Hahn}\ and\ \citenamefont
  {{Villaescusa-Navarro}}(2021)}]{hahn2021a}%
  \BibitemOpen
  \bibfield  {author} {\bibinfo {author} {\bibfnamefont {C.}~\bibnamefont
  {Hahn}}\ and\ \bibinfo {author} {\bibfnamefont {F.}~\bibnamefont
  {{Villaescusa-Navarro}}},\ }\href {\doibase 10.1088/1475-7516/2021/04/029}
  {\bibfield  {journal} {\bibinfo  {journal} {Journal of Cosmology and
  Astroparticle Physics}\ }\textbf {\bibinfo {volume} {2021}},\ \bibinfo
  {pages} {029} (\bibinfo {year} {2021})}\BibitemShut {NoStop}%
\bibitem [{\citenamefont {{Massara}}\ \emph {et~al.}(2022)\citenamefont
  {{Massara}}, \citenamefont {{Villaescusa-Navarro}}, \citenamefont {{Hahn}},
  \citenamefont {{Abidi}}, \citenamefont {{Eickenberg}}, \citenamefont {{Ho}},
  \citenamefont {{Lemos}}, \citenamefont {{Moradinezhad Dizgah}},\ and\
  \citenamefont {{R{\'e}galdo-Saint Blancard}}}]{Massara2022}%
  \BibitemOpen
  \bibfield  {author} {\bibinfo {author} {\bibfnamefont {E.}~\bibnamefont
  {{Massara}}}, \bibinfo {author} {\bibfnamefont {F.}~\bibnamefont
  {{Villaescusa-Navarro}}}, \bibinfo {author} {\bibfnamefont {C.}~\bibnamefont
  {{Hahn}}}, \bibinfo {author} {\bibfnamefont {M.~M.}\ \bibnamefont {{Abidi}}},
  \bibinfo {author} {\bibfnamefont {M.}~\bibnamefont {{Eickenberg}}}, \bibinfo
  {author} {\bibfnamefont {S.}~\bibnamefont {{Ho}}}, \bibinfo {author}
  {\bibfnamefont {P.}~\bibnamefont {{Lemos}}}, \bibinfo {author} {\bibfnamefont
  {A.}~\bibnamefont {{Moradinezhad Dizgah}}}, \ and\ \bibinfo {author}
  {\bibfnamefont {B.}~\bibnamefont {{R{\'e}galdo-Saint Blancard}}},\
  }\href@noop {} {\bibfield  {journal} {\bibinfo  {journal} {arXiv e-prints}\
  ,\ \bibinfo {eid} {arXiv:2206.01709}} (\bibinfo {year} {2022})},\ \Eprint
  {http://arxiv.org/abs/2206.01709} {arXiv:2206.01709 [astro-ph.CO]}
  \BibitemShut {NoStop}%
\bibitem [{\citenamefont {Wang}\ \emph
  {et~al.}(2022{\natexlab{a}})\citenamefont {Wang}, \citenamefont {Zhao},
  \citenamefont {Koyama}, \citenamefont {Percival}, \citenamefont {Takahashi},
  \citenamefont {Hikage}, \citenamefont {{Gil-Mar{\'i}n}}, \citenamefont
  {Hahn}, \citenamefont {Zhao}, \citenamefont {Zhang}, \citenamefont {Mu},
  \citenamefont {Yu}, \citenamefont {Zhu},\ and\ \citenamefont
  {Ge}}]{wang2022}%
  \BibitemOpen
  \bibfield  {author} {\bibinfo {author} {\bibfnamefont {Y.}~\bibnamefont
  {Wang}}, \bibinfo {author} {\bibfnamefont {G.-B.}\ \bibnamefont {Zhao}},
  \bibinfo {author} {\bibfnamefont {K.}~\bibnamefont {Koyama}}, \bibinfo
  {author} {\bibfnamefont {W.~J.}\ \bibnamefont {Percival}}, \bibinfo {author}
  {\bibfnamefont {R.}~\bibnamefont {Takahashi}}, \bibinfo {author}
  {\bibfnamefont {C.}~\bibnamefont {Hikage}}, \bibinfo {author} {\bibfnamefont
  {H.}~\bibnamefont {{Gil-Mar{\'i}n}}}, \bibinfo {author} {\bibfnamefont
  {C.}~\bibnamefont {Hahn}}, \bibinfo {author} {\bibfnamefont {R.}~\bibnamefont
  {Zhao}}, \bibinfo {author} {\bibfnamefont {W.}~\bibnamefont {Zhang}},
  \bibinfo {author} {\bibfnamefont {X.}~\bibnamefont {Mu}}, \bibinfo {author}
  {\bibfnamefont {Y.}~\bibnamefont {Yu}}, \bibinfo {author} {\bibfnamefont
  {H.-M.}\ \bibnamefont {Zhu}}, \ and\ \bibinfo {author} {\bibfnamefont
  {F.}~\bibnamefont {Ge}},\ }\href {\doibase 10.48550/arXiv.2202.05248}
  {\enquote {\bibinfo {title} {Extracting high-order cosmological information
  in galaxy surveys with power spectra},}\ } (\bibinfo {year}
  {2022}{\natexlab{a}}),\ \Eprint {http://arxiv.org/abs/2202.05248}
  {arXiv:2202.05248 [astro-ph]} \BibitemShut {NoStop}%
\bibitem [{\citenamefont {{Hou}}\ \emph {et~al.}(2022)\citenamefont {{Hou}},
  \citenamefont {{Moradinezhad Dizgah}}, \citenamefont {{Hahn}},\ and\
  \citenamefont {{Massara}}}]{hou2022}%
  \BibitemOpen
  \bibfield  {author} {\bibinfo {author} {\bibfnamefont {J.}~\bibnamefont
  {{Hou}}}, \bibinfo {author} {\bibfnamefont {A.}~\bibnamefont {{Moradinezhad
  Dizgah}}}, \bibinfo {author} {\bibfnamefont {C.}~\bibnamefont {{Hahn}}}, \
  and\ \bibinfo {author} {\bibfnamefont {E.}~\bibnamefont {{Massara}}},\
  }\href@noop {} {\bibfield  {journal} {\bibinfo  {journal} {arXiv e-prints}\
  ,\ \bibinfo {eid} {arXiv:2210.12743}} (\bibinfo {year} {2022})},\ \Eprint
  {http://arxiv.org/abs/2210.12743} {arXiv:2210.12743 [astro-ph.CO]}
  \BibitemShut {NoStop}%
\bibitem [{\citenamefont {Eickenberg}\ \emph {et~al.}(2022)\citenamefont
  {Eickenberg}, \citenamefont {Allys}, \citenamefont {Moradinezhad~Dizgah},
  \citenamefont {Lemos}, \citenamefont {Massara}, \citenamefont {Abidi},
  \citenamefont {Hahn}, \citenamefont {Hassan}, \citenamefont {{Regaldo-Saint
  Blancard}}, \citenamefont {Ho}, \citenamefont {Mallat}, \citenamefont
  {And{\'e}n},\ and\ \citenamefont {{Villaescusa-Navarro}}}]{eickenberg2022}%
  \BibitemOpen
  \bibfield  {author} {\bibinfo {author} {\bibfnamefont {M.}~\bibnamefont
  {Eickenberg}}, \bibinfo {author} {\bibfnamefont {E.}~\bibnamefont {Allys}},
  \bibinfo {author} {\bibfnamefont {A.}~\bibnamefont {Moradinezhad~Dizgah}},
  \bibinfo {author} {\bibfnamefont {P.}~\bibnamefont {Lemos}}, \bibinfo
  {author} {\bibfnamefont {E.}~\bibnamefont {Massara}}, \bibinfo {author}
  {\bibfnamefont {M.}~\bibnamefont {Abidi}}, \bibinfo {author} {\bibfnamefont
  {C.}~\bibnamefont {Hahn}}, \bibinfo {author} {\bibfnamefont {S.}~\bibnamefont
  {Hassan}}, \bibinfo {author} {\bibfnamefont {B.}~\bibnamefont {{Regaldo-Saint
  Blancard}}}, \bibinfo {author} {\bibfnamefont {S.}~\bibnamefont {Ho}},
  \bibinfo {author} {\bibfnamefont {S.}~\bibnamefont {Mallat}}, \bibinfo
  {author} {\bibfnamefont {J.}~\bibnamefont {And{\'e}n}}, \ and\ \bibinfo
  {author} {\bibfnamefont {F.}~\bibnamefont {{Villaescusa-Navarro}}},\
  }\href@noop {} {\enquote {\bibinfo {title} {Wavelet {{Moments}} for
  {{Cosmological Parameter Estimation}}},}\ } (\bibinfo {year}
  {2022})\BibitemShut {NoStop}%
\bibitem [{\citenamefont {Hahn}\ \emph
  {et~al.}(2017{\natexlab{a}})\citenamefont {Hahn}, \citenamefont
  {Scoccimarro}, \citenamefont {Blanton}, \citenamefont {Tinker},\ and\
  \citenamefont {{Rodr{\'i}guez-Torres}}}]{hahn2017a}%
  \BibitemOpen
  \bibfield  {author} {\bibinfo {author} {\bibfnamefont {C.}~\bibnamefont
  {Hahn}}, \bibinfo {author} {\bibfnamefont {R.}~\bibnamefont {Scoccimarro}},
  \bibinfo {author} {\bibfnamefont {M.~R.}\ \bibnamefont {Blanton}}, \bibinfo
  {author} {\bibfnamefont {J.~L.}\ \bibnamefont {Tinker}}, \ and\ \bibinfo
  {author} {\bibfnamefont {S.~A.}\ \bibnamefont {{Rodr{\'i}guez-Torres}}},\
  }\href {\doibase 10.1093/mnras/stx185} {\bibfield  {journal} {\bibinfo
  {journal} {Monthly Notices of the Royal Astronomical Society}\ }\textbf
  {\bibinfo {volume} {467}},\ \bibinfo {pages} {1940} (\bibinfo {year}
  {2017}{\natexlab{a}})}\BibitemShut {NoStop}%
\bibitem [{\citenamefont {Bianchi}\ \emph {et~al.}(2018)\citenamefont
  {Bianchi}, \citenamefont {Burden}, \citenamefont {Percival}, \citenamefont
  {Brooks}, \citenamefont {Cahn}, \citenamefont {{Forero-Romero}},
  \citenamefont {Levi}, \citenamefont {Ross},\ and\ \citenamefont
  {Tarle}}]{bianchi2018}%
  \BibitemOpen
  \bibfield  {author} {\bibinfo {author} {\bibfnamefont {D.}~\bibnamefont
  {Bianchi}}, \bibinfo {author} {\bibfnamefont {A.}~\bibnamefont {Burden}},
  \bibinfo {author} {\bibfnamefont {W.~J.}\ \bibnamefont {Percival}}, \bibinfo
  {author} {\bibfnamefont {D.}~\bibnamefont {Brooks}}, \bibinfo {author}
  {\bibfnamefont {R.~N.}\ \bibnamefont {Cahn}}, \bibinfo {author}
  {\bibfnamefont {J.~E.}\ \bibnamefont {{Forero-Romero}}}, \bibinfo {author}
  {\bibfnamefont {M.}~\bibnamefont {Levi}}, \bibinfo {author} {\bibfnamefont
  {A.~J.}\ \bibnamefont {Ross}}, \ and\ \bibinfo {author} {\bibfnamefont
  {G.}~\bibnamefont {Tarle}},\ }\href {\doibase 10.1093/mnras/sty2377}
  {\bibfield  {journal} {\bibinfo  {journal} {Monthly Notices of the Royal
  Astronomical Society}\ }\textbf {\bibinfo {volume} {481}},\ \bibinfo {pages}
  {2338} (\bibinfo {year} {2018})}\BibitemShut {NoStop}%
\bibitem [{\citenamefont {Ross}\ \emph {et~al.}(2012)\citenamefont {Ross},
  \citenamefont {Percival}, \citenamefont {S{\'a}nchez}, \citenamefont
  {Samushia}, \citenamefont {Ho}, \citenamefont {Kazin}, \citenamefont
  {Manera}, \citenamefont {Reid}, \citenamefont {White}, \citenamefont
  {Tojeiro} \emph {et~al.}}]{ross2012clustering}%
  \BibitemOpen
  \bibfield  {author} {\bibinfo {author} {\bibfnamefont {A.~J.}\ \bibnamefont
  {Ross}}, \bibinfo {author} {\bibfnamefont {W.~J.}\ \bibnamefont {Percival}},
  \bibinfo {author} {\bibfnamefont {A.~G.}\ \bibnamefont {S{\'a}nchez}},
  \bibinfo {author} {\bibfnamefont {L.}~\bibnamefont {Samushia}}, \bibinfo
  {author} {\bibfnamefont {S.}~\bibnamefont {Ho}}, \bibinfo {author}
  {\bibfnamefont {E.}~\bibnamefont {Kazin}}, \bibinfo {author} {\bibfnamefont
  {M.}~\bibnamefont {Manera}}, \bibinfo {author} {\bibfnamefont
  {B.}~\bibnamefont {Reid}}, \bibinfo {author} {\bibfnamefont {M.}~\bibnamefont
  {White}}, \bibinfo {author} {\bibfnamefont {R.}~\bibnamefont {Tojeiro}},
  \emph {et~al.},\ }\href@noop {} {\bibfield  {journal} {\bibinfo  {journal}
  {Monthly Notices of the Royal Astronomical Society}\ }\textbf {\bibinfo
  {volume} {424}},\ \bibinfo {pages} {564} (\bibinfo {year}
  {2012})}\BibitemShut {NoStop}%
\bibitem [{\citenamefont {Ross}\ \emph {et~al.}(2017)\citenamefont {Ross},
  \citenamefont {Beutler}, \citenamefont {Chuang}, \citenamefont
  {{Pellejero-Ibanez}}, \citenamefont {Seo}, \citenamefont
  {{Vargas-Maga{\~n}a}}, \citenamefont {Cuesta}, \citenamefont {Percival},
  \citenamefont {Burden}, \citenamefont {S{\'a}nchez}, \citenamefont {Grieb},
  \citenamefont {Reid}, \citenamefont {Brownstein}, \citenamefont {Dawson},
  \citenamefont {Eisenstein}, \citenamefont {Ho}, \citenamefont {Kitaura},
  \citenamefont {Nichol}, \citenamefont {Olmstead}, \citenamefont {Prada},
  \citenamefont {{Rodr{\'i}guez-Torres}}, \citenamefont {Saito}, \citenamefont
  {{Salazar-Albornoz}}, \citenamefont {Schneider}, \citenamefont {Thomas},
  \citenamefont {Tinker}, \citenamefont {Tojeiro}, \citenamefont {Wang},
  \citenamefont {White},\ and\ \citenamefont {Zhao}}]{ross2017}%
  \BibitemOpen
  \bibfield  {author} {\bibinfo {author} {\bibfnamefont {A.~J.}\ \bibnamefont
  {Ross}}, \bibinfo {author} {\bibfnamefont {F.}~\bibnamefont {Beutler}},
  \bibinfo {author} {\bibfnamefont {C.-H.}\ \bibnamefont {Chuang}}, \bibinfo
  {author} {\bibfnamefont {M.}~\bibnamefont {{Pellejero-Ibanez}}}, \bibinfo
  {author} {\bibfnamefont {H.-J.}\ \bibnamefont {Seo}}, \bibinfo {author}
  {\bibfnamefont {M.}~\bibnamefont {{Vargas-Maga{\~n}a}}}, \bibinfo {author}
  {\bibfnamefont {A.~J.}\ \bibnamefont {Cuesta}}, \bibinfo {author}
  {\bibfnamefont {W.~J.}\ \bibnamefont {Percival}}, \bibinfo {author}
  {\bibfnamefont {A.}~\bibnamefont {Burden}}, \bibinfo {author} {\bibfnamefont
  {A.~G.}\ \bibnamefont {S{\'a}nchez}}, \bibinfo {author} {\bibfnamefont
  {J.~N.}\ \bibnamefont {Grieb}}, \bibinfo {author} {\bibfnamefont
  {B.}~\bibnamefont {Reid}}, \bibinfo {author} {\bibfnamefont {J.~R.}\
  \bibnamefont {Brownstein}}, \bibinfo {author} {\bibfnamefont {K.~S.}\
  \bibnamefont {Dawson}}, \bibinfo {author} {\bibfnamefont {D.~J.}\
  \bibnamefont {Eisenstein}}, \bibinfo {author} {\bibfnamefont
  {S.}~\bibnamefont {Ho}}, \bibinfo {author} {\bibfnamefont {F.-S.}\
  \bibnamefont {Kitaura}}, \bibinfo {author} {\bibfnamefont {R.~C.}\
  \bibnamefont {Nichol}}, \bibinfo {author} {\bibfnamefont {M.~D.}\
  \bibnamefont {Olmstead}}, \bibinfo {author} {\bibfnamefont {F.}~\bibnamefont
  {Prada}}, \bibinfo {author} {\bibfnamefont {S.~A.}\ \bibnamefont
  {{Rodr{\'i}guez-Torres}}}, \bibinfo {author} {\bibfnamefont {S.}~\bibnamefont
  {Saito}}, \bibinfo {author} {\bibfnamefont {S.}~\bibnamefont
  {{Salazar-Albornoz}}}, \bibinfo {author} {\bibfnamefont {D.~P.}\ \bibnamefont
  {Schneider}}, \bibinfo {author} {\bibfnamefont {D.}~\bibnamefont {Thomas}},
  \bibinfo {author} {\bibfnamefont {J.}~\bibnamefont {Tinker}}, \bibinfo
  {author} {\bibfnamefont {R.}~\bibnamefont {Tojeiro}}, \bibinfo {author}
  {\bibfnamefont {Y.}~\bibnamefont {Wang}}, \bibinfo {author} {\bibfnamefont
  {M.}~\bibnamefont {White}}, \ and\ \bibinfo {author} {\bibfnamefont {G.-b.}\
  \bibnamefont {Zhao}},\ }\href {\doibase 10.1093/mnras/stw2372} {\bibfield
  {journal} {\bibinfo  {journal} {Monthly Notices of the Royal Astronomical
  Society}\ }\textbf {\bibinfo {volume} {464}},\ \bibinfo {pages} {1168}
  (\bibinfo {year} {2017})}\BibitemShut {NoStop}%
\bibitem [{\citenamefont {Scoccimarro}(2000)}]{scoccimarro2000gravitational}%
  \BibitemOpen
  \bibfield  {author} {\bibinfo {author} {\bibfnamefont {R.}~\bibnamefont
  {Scoccimarro}},\ }\href@noop {} {\bibfield  {journal} {\bibinfo  {journal}
  {The Astrophysical Journal}\ }\textbf {\bibinfo {volume} {542}},\ \bibinfo
  {pages} {1} (\bibinfo {year} {2000})}\BibitemShut {NoStop}%
\bibitem [{\citenamefont {{Sellentin}}\ and\ \citenamefont
  {{Heavens}}(2018)}]{sellentin2018}%
  \BibitemOpen
  \bibfield  {author} {\bibinfo {author} {\bibfnamefont {E.}~\bibnamefont
  {{Sellentin}}}\ and\ \bibinfo {author} {\bibfnamefont {A.~F.}\ \bibnamefont
  {{Heavens}}},\ }\href {\doibase 10.1093/mnras/stx2491} {\bibfield  {journal}
  {\bibinfo  {journal} {\mnras}\ }\textbf {\bibinfo {volume} {473}},\ \bibinfo
  {pages} {2355} (\bibinfo {year} {2018})},\ \Eprint
  {http://arxiv.org/abs/1707.04488} {arXiv:1707.04488 [astro-ph.CO]}
  \BibitemShut {NoStop}%
\bibitem [{\citenamefont {{Hahn}}\ \emph {et~al.}(2019)\citenamefont {{Hahn}},
  \citenamefont {{Beutler}}, \citenamefont {{Sinha}}, \citenamefont
  {{Berlind}}, \citenamefont {{Ho}},\ and\ \citenamefont {{Hogg}}}]{hahn2019}%
  \BibitemOpen
  \bibfield  {author} {\bibinfo {author} {\bibfnamefont {C.}~\bibnamefont
  {{Hahn}}}, \bibinfo {author} {\bibfnamefont {F.}~\bibnamefont {{Beutler}}},
  \bibinfo {author} {\bibfnamefont {M.}~\bibnamefont {{Sinha}}}, \bibinfo
  {author} {\bibfnamefont {A.}~\bibnamefont {{Berlind}}}, \bibinfo {author}
  {\bibfnamefont {S.}~\bibnamefont {{Ho}}}, \ and\ \bibinfo {author}
  {\bibfnamefont {D.~W.}\ \bibnamefont {{Hogg}}},\ }\href {\doibase
  10.1093/mnras/stz558} {\bibfield  {journal} {\bibinfo  {journal} {\mnras}\
  }\textbf {\bibinfo {volume} {485}},\ \bibinfo {pages} {2956} (\bibinfo {year}
  {2019})},\ \Eprint {http://arxiv.org/abs/1803.06348} {arXiv:1803.06348
  [astro-ph.CO]} \BibitemShut {NoStop}%
\bibitem [{\citenamefont {Hahn}\ \emph
  {et~al.}(2017{\natexlab{b}})\citenamefont {Hahn}, \citenamefont {Vakili},
  \citenamefont {Walsh}, \citenamefont {Hearin}, \citenamefont {Hogg},\ and\
  \citenamefont {Campbell}}]{hahn2017b}%
  \BibitemOpen
  \bibfield  {author} {\bibinfo {author} {\bibfnamefont {C.}~\bibnamefont
  {Hahn}}, \bibinfo {author} {\bibfnamefont {M.}~\bibnamefont {Vakili}},
  \bibinfo {author} {\bibfnamefont {K.}~\bibnamefont {Walsh}}, \bibinfo
  {author} {\bibfnamefont {A.~P.}\ \bibnamefont {Hearin}}, \bibinfo {author}
  {\bibfnamefont {D.~W.}\ \bibnamefont {Hogg}}, \ and\ \bibinfo {author}
  {\bibfnamefont {D.}~\bibnamefont {Campbell}},\ }\href {\doibase
  10.1093/mnras/stx894} {\bibfield  {journal} {\bibinfo  {journal} {Monthly
  Notices of the Royal Astronomical Society}\ }\textbf {\bibinfo {volume}
  {469}},\ \bibinfo {pages} {2791} (\bibinfo {year} {2017}{\natexlab{b}})},\
  \Eprint {http://arxiv.org/abs/1607.01782} {arXiv:1607.01782} \BibitemShut
  {NoStop}%
\bibitem [{\citenamefont {Perreault~Levasseur}\ \emph
  {et~al.}(2017)\citenamefont {Perreault~Levasseur}, \citenamefont {Hezaveh},\
  and\ \citenamefont {Wechsler}}]{PerreaultLevasseur:2017ltk}%
  \BibitemOpen
  \bibfield  {author} {\bibinfo {author} {\bibfnamefont {L.}~\bibnamefont
  {Perreault~Levasseur}}, \bibinfo {author} {\bibfnamefont {Y.~D.}\
  \bibnamefont {Hezaveh}}, \ and\ \bibinfo {author} {\bibfnamefont {R.~H.}\
  \bibnamefont {Wechsler}},\ }\href {\doibase 10.3847/2041-8213/aa9704}
  {\bibfield  {journal} {\bibinfo  {journal} {Astrophys. J. Lett.}\ }\textbf
  {\bibinfo {volume} {850}},\ \bibinfo {pages} {L7} (\bibinfo {year} {2017})},\
  \Eprint {http://arxiv.org/abs/1708.08843} {arXiv:1708.08843 [astro-ph.CO]}
  \BibitemShut {NoStop}%
\bibitem [{\citenamefont {Dax}\ \emph {et~al.}(2021)\citenamefont {Dax},
  \citenamefont {Green}, \citenamefont {Gair}, \citenamefont {Macke},
  \citenamefont {Buonanno},\ and\ \citenamefont
  {Sch\"olkopf}}]{PhysRevLett.127.241103}%
  \BibitemOpen
  \bibfield  {author} {\bibinfo {author} {\bibfnamefont {M.}~\bibnamefont
  {Dax}}, \bibinfo {author} {\bibfnamefont {S.~R.}\ \bibnamefont {Green}},
  \bibinfo {author} {\bibfnamefont {J.}~\bibnamefont {Gair}}, \bibinfo {author}
  {\bibfnamefont {J.~H.}\ \bibnamefont {Macke}}, \bibinfo {author}
  {\bibfnamefont {A.}~\bibnamefont {Buonanno}}, \ and\ \bibinfo {author}
  {\bibfnamefont {B.}~\bibnamefont {Sch\"olkopf}},\ }\href {\doibase
  10.1103/PhysRevLett.127.241103} {\bibfield  {journal} {\bibinfo  {journal}
  {Phys. Rev. Lett.}\ }\textbf {\bibinfo {volume} {127}},\ \bibinfo {pages}
  {241103} (\bibinfo {year} {2021})}\BibitemShut {NoStop}%
\bibitem [{\citenamefont {Alsing}\ \emph {et~al.}(2019)\citenamefont {Alsing},
  \citenamefont {Charnock}, \citenamefont {Feeney},\ and\ \citenamefont
  {Wandelt}}]{alsing2019fast}%
  \BibitemOpen
  \bibfield  {author} {\bibinfo {author} {\bibfnamefont {J.}~\bibnamefont
  {Alsing}}, \bibinfo {author} {\bibfnamefont {T.}~\bibnamefont {Charnock}},
  \bibinfo {author} {\bibfnamefont {S.}~\bibnamefont {Feeney}}, \ and\ \bibinfo
  {author} {\bibfnamefont {B.}~\bibnamefont {Wandelt}},\ }\href@noop {}
  {\bibfield  {journal} {\bibinfo  {journal} {Monthly Notices of the Royal
  Astronomical Society}\ }\textbf {\bibinfo {volume} {488}},\ \bibinfo {pages}
  {4440} (\bibinfo {year} {2019})}\BibitemShut {NoStop}%
\bibitem [{\citenamefont {Wagner-Carena}\ \emph {et~al.}(2021)\citenamefont
  {Wagner-Carena}, \citenamefont {Park}, \citenamefont {Birrer}, \citenamefont
  {Marshall}, \citenamefont {Roodman},\ and\ \citenamefont
  {Wechsler}}]{Wagner-Carena:2020yun}%
  \BibitemOpen
  \bibfield  {author} {\bibinfo {author} {\bibfnamefont {S.}~\bibnamefont
  {Wagner-Carena}}, \bibinfo {author} {\bibfnamefont {J.~W.}\ \bibnamefont
  {Park}}, \bibinfo {author} {\bibfnamefont {S.}~\bibnamefont {Birrer}},
  \bibinfo {author} {\bibfnamefont {P.~J.}\ \bibnamefont {Marshall}}, \bibinfo
  {author} {\bibfnamefont {A.}~\bibnamefont {Roodman}}, \ and\ \bibinfo
  {author} {\bibfnamefont {R.~H.}\ \bibnamefont {Wechsler}} (\bibinfo
  {collaboration} {LSST Dark Energy Science}),\ }\href {\doibase
  10.3847/1538-4357/abdf59} {\bibfield  {journal} {\bibinfo  {journal}
  {Astrophys. J.}\ }\textbf {\bibinfo {volume} {909}},\ \bibinfo {pages} {187}
  (\bibinfo {year} {2021})},\ \Eprint {http://arxiv.org/abs/2010.13787}
  {arXiv:2010.13787 [astro-ph.CO]} \BibitemShut {NoStop}%
\bibitem [{\citenamefont {Legin}\ \emph {et~al.}(2021)\citenamefont {Legin},
  \citenamefont {Hezaveh}, \citenamefont {Levasseur},\ and\ \citenamefont
  {Wandelt}}]{Legin:2021zup}%
  \BibitemOpen
  \bibfield  {author} {\bibinfo {author} {\bibfnamefont {R.}~\bibnamefont
  {Legin}}, \bibinfo {author} {\bibfnamefont {Y.}~\bibnamefont {Hezaveh}},
  \bibinfo {author} {\bibfnamefont {L.~P.}\ \bibnamefont {Levasseur}}, \ and\
  \bibinfo {author} {\bibfnamefont {B.}~\bibnamefont {Wandelt}},\ }\href@noop
  {} {\  (\bibinfo {year} {2021})},\ \Eprint {http://arxiv.org/abs/2112.05278}
  {arXiv:2112.05278 [astro-ph.CO]} \BibitemShut {NoStop}%
\bibitem [{\citenamefont {Coogan}\ \emph {et~al.}(2020)\citenamefont {Coogan},
  \citenamefont {Karchev},\ and\ \citenamefont {Weniger}}]{Coogan:2020yux}%
  \BibitemOpen
  \bibfield  {author} {\bibinfo {author} {\bibfnamefont {A.}~\bibnamefont
  {Coogan}}, \bibinfo {author} {\bibfnamefont {K.}~\bibnamefont {Karchev}}, \
  and\ \bibinfo {author} {\bibfnamefont {C.}~\bibnamefont {Weniger}},\ }in\
  \href@noop {} {\emph {\bibinfo {booktitle} {{34th Conference on Neural
  Information Processing Systems}}}}\ (\bibinfo {year} {2020})\ \Eprint
  {http://arxiv.org/abs/2010.07032} {arXiv:2010.07032 [astro-ph.CO]}
  \BibitemShut {NoStop}%
\bibitem [{\citenamefont {Montel}\ \emph {et~al.}(2022)\citenamefont {Montel},
  \citenamefont {Coogan}, \citenamefont {Correa}, \citenamefont {Karchev},\
  and\ \citenamefont {Weniger}}]{Montel:2022fhv}%
  \BibitemOpen
  \bibfield  {author} {\bibinfo {author} {\bibfnamefont {N.~A.}\ \bibnamefont
  {Montel}}, \bibinfo {author} {\bibfnamefont {A.}~\bibnamefont {Coogan}},
  \bibinfo {author} {\bibfnamefont {C.}~\bibnamefont {Correa}}, \bibinfo
  {author} {\bibfnamefont {K.}~\bibnamefont {Karchev}}, \ and\ \bibinfo
  {author} {\bibfnamefont {C.}~\bibnamefont {Weniger}},\ }\href {\doibase
  10.1093/mnras/stac3215} {\bibfield  {journal} {\bibinfo  {journal} {Mon. Not.
  Roy. Astron. Soc.}\ }\textbf {\bibinfo {volume} {518}},\ \bibinfo {pages}
  {2746} (\bibinfo {year} {2022})},\ \Eprint {http://arxiv.org/abs/2205.09126}
  {arXiv:2205.09126 [astro-ph.CO]} \BibitemShut {NoStop}%
\bibitem [{\citenamefont {Coogan}\ \emph {et~al.}(2022)\citenamefont {Coogan},
  \citenamefont {Anau~Montel}, \citenamefont {Karchev}, \citenamefont
  {Grootes}, \citenamefont {Nattino},\ and\ \citenamefont
  {Weniger}}]{Coogan:2022cky}%
  \BibitemOpen
  \bibfield  {author} {\bibinfo {author} {\bibfnamefont {A.}~\bibnamefont
  {Coogan}}, \bibinfo {author} {\bibfnamefont {N.}~\bibnamefont {Anau~Montel}},
  \bibinfo {author} {\bibfnamefont {K.}~\bibnamefont {Karchev}}, \bibinfo
  {author} {\bibfnamefont {M.~W.}\ \bibnamefont {Grootes}}, \bibinfo {author}
  {\bibfnamefont {F.}~\bibnamefont {Nattino}}, \ and\ \bibinfo {author}
  {\bibfnamefont {C.}~\bibnamefont {Weniger}},\ }\href@noop {} {\  (\bibinfo
  {year} {2022})},\ \Eprint {http://arxiv.org/abs/2209.09918} {arXiv:2209.09918
  [astro-ph.CO]} \BibitemShut {NoStop}%
\bibitem [{\citenamefont {Brehmer}\ \emph {et~al.}(2019)\citenamefont
  {Brehmer}, \citenamefont {Mishra-Sharma}, \citenamefont {Hermans},
  \citenamefont {Louppe},\ and\ \citenamefont {Cranmer}}]{Brehmer:2019jyt}%
  \BibitemOpen
  \bibfield  {author} {\bibinfo {author} {\bibfnamefont {J.}~\bibnamefont
  {Brehmer}}, \bibinfo {author} {\bibfnamefont {S.}~\bibnamefont
  {Mishra-Sharma}}, \bibinfo {author} {\bibfnamefont {J.}~\bibnamefont
  {Hermans}}, \bibinfo {author} {\bibfnamefont {G.}~\bibnamefont {Louppe}}, \
  and\ \bibinfo {author} {\bibfnamefont {K.}~\bibnamefont {Cranmer}},\ }\href
  {\doibase 10.3847/1538-4357/ab4c41} {\bibfield  {journal} {\bibinfo
  {journal} {Astrophys. J.}\ }\textbf {\bibinfo {volume} {886}},\ \bibinfo
  {pages} {49} (\bibinfo {year} {2019})},\ \Eprint
  {http://arxiv.org/abs/1909.02005} {arXiv:1909.02005 [astro-ph.CO]}
  \BibitemShut {NoStop}%
\bibitem [{\citenamefont {Mishra-Sharma}\ and\ \citenamefont
  {Cranmer}(2022)}]{Mishra-Sharma:2021oxe}%
  \BibitemOpen
  \bibfield  {author} {\bibinfo {author} {\bibfnamefont {S.}~\bibnamefont
  {Mishra-Sharma}}\ and\ \bibinfo {author} {\bibfnamefont {K.}~\bibnamefont
  {Cranmer}},\ }\href {\doibase 10.1103/PhysRevD.105.063017} {\bibfield
  {journal} {\bibinfo  {journal} {Phys. Rev. D}\ }\textbf {\bibinfo {volume}
  {105}},\ \bibinfo {pages} {063017} (\bibinfo {year} {2022})},\ \Eprint
  {http://arxiv.org/abs/2110.06931} {arXiv:2110.06931 [astro-ph.HE]}
  \BibitemShut {NoStop}%
\bibitem [{\citenamefont {Karchev}\ \emph
  {et~al.}(2022{\natexlab{a}})\citenamefont {Karchev}, \citenamefont {Trotta},\
  and\ \citenamefont {Weniger}}]{Karchev:2022xyn}%
  \BibitemOpen
  \bibfield  {author} {\bibinfo {author} {\bibfnamefont {K.}~\bibnamefont
  {Karchev}}, \bibinfo {author} {\bibfnamefont {R.}~\bibnamefont {Trotta}}, \
  and\ \bibinfo {author} {\bibfnamefont {C.}~\bibnamefont {Weniger}},\ }\href
  {\doibase 10.1093/mnras/stac3785} {\  (\bibinfo {year}
  {2022}{\natexlab{a}}),\ 10.1093/mnras/stac3785},\ \Eprint
  {http://arxiv.org/abs/2209.06733} {arXiv:2209.06733 [astro-ph.CO]}
  \BibitemShut {NoStop}%
\bibitem [{\citenamefont {Hermans}\ \emph {et~al.}(2021)\citenamefont
  {Hermans}, \citenamefont {Banik}, \citenamefont {Weniger}, \citenamefont
  {Bertone},\ and\ \citenamefont {Louppe}}]{Hermans:2020skz}%
  \BibitemOpen
  \bibfield  {author} {\bibinfo {author} {\bibfnamefont {J.}~\bibnamefont
  {Hermans}}, \bibinfo {author} {\bibfnamefont {N.}~\bibnamefont {Banik}},
  \bibinfo {author} {\bibfnamefont {C.}~\bibnamefont {Weniger}}, \bibinfo
  {author} {\bibfnamefont {G.}~\bibnamefont {Bertone}}, \ and\ \bibinfo
  {author} {\bibfnamefont {G.}~\bibnamefont {Louppe}},\ }\href {\doibase
  10.1093/mnras/stab2181} {\bibfield  {journal} {\bibinfo  {journal} {Mon. Not.
  Roy. Astron. Soc.}\ }\textbf {\bibinfo {volume} {507}},\ \bibinfo {pages}
  {1999} (\bibinfo {year} {2021})},\ \Eprint {http://arxiv.org/abs/2011.14923}
  {arXiv:2011.14923 [astro-ph.GA]} \BibitemShut {NoStop}%
\bibitem [{\citenamefont {Karchev}\ \emph
  {et~al.}(2022{\natexlab{b}})\citenamefont {Karchev}, \citenamefont
  {Anau~Montel}, \citenamefont {Coogan},\ and\ \citenamefont
  {Weniger}}]{Karchev:2022ycy}%
  \BibitemOpen
  \bibfield  {author} {\bibinfo {author} {\bibfnamefont {K.}~\bibnamefont
  {Karchev}}, \bibinfo {author} {\bibfnamefont {N.}~\bibnamefont
  {Anau~Montel}}, \bibinfo {author} {\bibfnamefont {A.}~\bibnamefont {Coogan}},
  \ and\ \bibinfo {author} {\bibfnamefont {C.}~\bibnamefont {Weniger}},\ }in\
  \href@noop {} {\emph {\bibinfo {booktitle} {{36th Conference on Neural
  Information Processing Systems}}}}\ (\bibinfo {year} {2022})\ \Eprint
  {http://arxiv.org/abs/2211.04365} {arXiv:2211.04365 [astro-ph.IM]}
  \BibitemShut {NoStop}%
\bibitem [{\citenamefont {Lemos}\ \emph {et~al.}(2021)\citenamefont {Lemos},
  \citenamefont {Jeffrey}, \citenamefont {Whiteway}, \citenamefont {Lahav},
  \citenamefont {Noam~Libeskind},\ and\ \citenamefont
  {Hoffman}}]{lemos2021sum}%
  \BibitemOpen
  \bibfield  {author} {\bibinfo {author} {\bibfnamefont {P.}~\bibnamefont
  {Lemos}}, \bibinfo {author} {\bibfnamefont {N.}~\bibnamefont {Jeffrey}},
  \bibinfo {author} {\bibfnamefont {L.}~\bibnamefont {Whiteway}}, \bibinfo
  {author} {\bibfnamefont {O.}~\bibnamefont {Lahav}}, \bibinfo {author}
  {\bibfnamefont {I.}~\bibnamefont {Noam~Libeskind}}, \ and\ \bibinfo {author}
  {\bibfnamefont {Y.}~\bibnamefont {Hoffman}},\ }\href@noop {} {\bibfield
  {journal} {\bibinfo  {journal} {Physical Review D}\ }\textbf {\bibinfo
  {volume} {103}},\ \bibinfo {pages} {023009} (\bibinfo {year}
  {2021})}\BibitemShut {NoStop}%
\bibitem [{\citenamefont {Hahn}\ and\ \citenamefont
  {Melchior}(2022)}]{hahn2022a}%
  \BibitemOpen
  \bibfield  {author} {\bibinfo {author} {\bibfnamefont {C.}~\bibnamefont
  {Hahn}}\ and\ \bibinfo {author} {\bibfnamefont {P.}~\bibnamefont
  {Melchior}},\ }\href@noop {} {\enquote {\bibinfo {title} {Accelerated
  {{Bayesian SED Modeling}} using {{Amortized Neural Posterior Estimation}}},}\
  } (\bibinfo {year} {2022})\BibitemShut {NoStop}%
\bibitem [{\citenamefont {Hahn}\ \emph
  {et~al.}(2022{\natexlab{a}})\citenamefont {Hahn}, \citenamefont {Eickenberg},
  \citenamefont {Ho}, \citenamefont {Hou}, \citenamefont {Lemos}, \citenamefont
  {Massara}, \citenamefont {Modi}, \citenamefont {Dizgah}, \citenamefont
  {Blancard},\ and\ \citenamefont {Abidi}}]{hahn2022rm}%
  \BibitemOpen
  \bibfield  {author} {\bibinfo {author} {\bibfnamefont {C.}~\bibnamefont
  {Hahn}}, \bibinfo {author} {\bibfnamefont {M.}~\bibnamefont {Eickenberg}},
  \bibinfo {author} {\bibfnamefont {S.}~\bibnamefont {Ho}}, \bibinfo {author}
  {\bibfnamefont {J.}~\bibnamefont {Hou}}, \bibinfo {author} {\bibfnamefont
  {P.}~\bibnamefont {Lemos}}, \bibinfo {author} {\bibfnamefont
  {E.}~\bibnamefont {Massara}}, \bibinfo {author} {\bibfnamefont
  {C.}~\bibnamefont {Modi}}, \bibinfo {author} {\bibfnamefont {A.~M.}\
  \bibnamefont {Dizgah}}, \bibinfo {author} {\bibfnamefont {B.~R.-S.}\
  \bibnamefont {Blancard}}, \ and\ \bibinfo {author} {\bibfnamefont {M.~M.}\
  \bibnamefont {Abidi}},\ }\href@noop {} {\bibfield  {journal} {\bibinfo
  {journal} {arXiv preprint arXiv:2211.00660}\ } (\bibinfo {year}
  {2022}{\natexlab{a}})}\BibitemShut {NoStop}%
\bibitem [{\citenamefont {Eisenstein}\ \emph {et~al.}(2011)\citenamefont
  {Eisenstein}, \citenamefont {Weinberg}, \citenamefont {Agol}, \citenamefont
  {Aihara}, \citenamefont {Prieto}, \citenamefont {Anderson}, \citenamefont
  {Arns}, \citenamefont {Aubourg}, \citenamefont {Bailey}, \citenamefont
  {Balbinot} \emph {et~al.}}]{eisenstein2011sdss}%
  \BibitemOpen
  \bibfield  {author} {\bibinfo {author} {\bibfnamefont {D.~J.}\ \bibnamefont
  {Eisenstein}}, \bibinfo {author} {\bibfnamefont {D.~H.}\ \bibnamefont
  {Weinberg}}, \bibinfo {author} {\bibfnamefont {E.}~\bibnamefont {Agol}},
  \bibinfo {author} {\bibfnamefont {H.}~\bibnamefont {Aihara}}, \bibinfo
  {author} {\bibfnamefont {C.~A.}\ \bibnamefont {Prieto}}, \bibinfo {author}
  {\bibfnamefont {S.~F.}\ \bibnamefont {Anderson}}, \bibinfo {author}
  {\bibfnamefont {J.~A.}\ \bibnamefont {Arns}}, \bibinfo {author}
  {\bibfnamefont {{\'E}.}~\bibnamefont {Aubourg}}, \bibinfo {author}
  {\bibfnamefont {S.}~\bibnamefont {Bailey}}, \bibinfo {author} {\bibfnamefont
  {E.}~\bibnamefont {Balbinot}},  \emph {et~al.},\ }\href@noop {} {\bibfield
  {journal} {\bibinfo  {journal} {The Astronomical Journal}\ }\textbf {\bibinfo
  {volume} {142}},\ \bibinfo {pages} {72} (\bibinfo {year} {2011})}\BibitemShut
  {NoStop}%
\bibitem [{\citenamefont {Dawson}\ \emph {et~al.}(2012)\citenamefont {Dawson},
  \citenamefont {Schlegel}, \citenamefont {Ahn}, \citenamefont {Anderson},
  \citenamefont {Aubourg}, \citenamefont {Bailey}, \citenamefont {Barkhouser},
  \citenamefont {Bautista}, \citenamefont {Beifiori}, \citenamefont {Berlind}
  \emph {et~al.}}]{dawson2012baryon}%
  \BibitemOpen
  \bibfield  {author} {\bibinfo {author} {\bibfnamefont {K.~S.}\ \bibnamefont
  {Dawson}}, \bibinfo {author} {\bibfnamefont {D.~J.}\ \bibnamefont
  {Schlegel}}, \bibinfo {author} {\bibfnamefont {C.~P.}\ \bibnamefont {Ahn}},
  \bibinfo {author} {\bibfnamefont {S.~F.}\ \bibnamefont {Anderson}}, \bibinfo
  {author} {\bibfnamefont {{\'E}.}~\bibnamefont {Aubourg}}, \bibinfo {author}
  {\bibfnamefont {S.}~\bibnamefont {Bailey}}, \bibinfo {author} {\bibfnamefont
  {R.~H.}\ \bibnamefont {Barkhouser}}, \bibinfo {author} {\bibfnamefont
  {J.~E.}\ \bibnamefont {Bautista}}, \bibinfo {author} {\bibfnamefont
  {A.}~\bibnamefont {Beifiori}}, \bibinfo {author} {\bibfnamefont {A.~A.}\
  \bibnamefont {Berlind}},  \emph {et~al.},\ }\href@noop {} {\bibfield
  {journal} {\bibinfo  {journal} {The Astronomical Journal}\ }\textbf {\bibinfo
  {volume} {145}},\ \bibinfo {pages} {10} (\bibinfo {year} {2012})}\BibitemShut
  {NoStop}%
\bibitem [{\citenamefont {{Hahn}}\ \emph {et~al.}(2023)\citenamefont {{Hahn}},
  \citenamefont {{Eickenberg}}, \citenamefont {{Ho}}, \citenamefont {{Hou}},
  \citenamefont {{Lemos}}, \citenamefont {{Massara}}, \citenamefont {{Modi}},
  \citenamefont {{Moradinezhad Dizgah}}, \citenamefont {{R{\'e}galdo-Saint
  Blancard}},\ and\ \citenamefont {{Abidi}}}]{simbig_mock_challenge}%
  \BibitemOpen
  \bibfield  {author} {\bibinfo {author} {\bibfnamefont {C.}~\bibnamefont
  {{Hahn}}}, \bibinfo {author} {\bibfnamefont {M.}~\bibnamefont
  {{Eickenberg}}}, \bibinfo {author} {\bibfnamefont {S.}~\bibnamefont {{Ho}}},
  \bibinfo {author} {\bibfnamefont {J.}~\bibnamefont {{Hou}}}, \bibinfo
  {author} {\bibfnamefont {P.}~\bibnamefont {{Lemos}}}, \bibinfo {author}
  {\bibfnamefont {E.}~\bibnamefont {{Massara}}}, \bibinfo {author}
  {\bibfnamefont {C.}~\bibnamefont {{Modi}}}, \bibinfo {author} {\bibfnamefont
  {A.}~\bibnamefont {{Moradinezhad Dizgah}}}, \bibinfo {author} {\bibfnamefont
  {B.}~\bibnamefont {{R{\'e}galdo-Saint Blancard}}}, \ and\ \bibinfo {author}
  {\bibfnamefont {M.~M.}\ \bibnamefont {{Abidi}}},\ }\href {\doibase
  10.1088/1475-7516/2023/04/010} {\bibfield  {journal} {\bibinfo  {journal}
  {\jcap}\ }\textbf {\bibinfo {volume} {2023}},\ \bibinfo {eid} {010} (\bibinfo
  {year} {2023})},\ \Eprint {http://arxiv.org/abs/2211.00660} {arXiv:2211.00660
  [astro-ph.CO]} \BibitemShut {NoStop}%
\bibitem [{\citenamefont {Villaescusa-Navarro}\ \emph
  {et~al.}(2020)\citenamefont {Villaescusa-Navarro}, \citenamefont {Hahn},
  \citenamefont {Massara}, \citenamefont {Banerjee}, \citenamefont {Delgado},
  \citenamefont {Ramanah}, \citenamefont {Charnock}, \citenamefont {Giusarma},
  \citenamefont {Li}, \citenamefont {Allys} \emph
  {et~al.}}]{villaescusa2020quijote}%
  \BibitemOpen
  \bibfield  {author} {\bibinfo {author} {\bibfnamefont {F.}~\bibnamefont
  {Villaescusa-Navarro}}, \bibinfo {author} {\bibfnamefont {C.}~\bibnamefont
  {Hahn}}, \bibinfo {author} {\bibfnamefont {E.}~\bibnamefont {Massara}},
  \bibinfo {author} {\bibfnamefont {A.}~\bibnamefont {Banerjee}}, \bibinfo
  {author} {\bibfnamefont {A.~M.}\ \bibnamefont {Delgado}}, \bibinfo {author}
  {\bibfnamefont {D.~K.}\ \bibnamefont {Ramanah}}, \bibinfo {author}
  {\bibfnamefont {T.}~\bibnamefont {Charnock}}, \bibinfo {author}
  {\bibfnamefont {E.}~\bibnamefont {Giusarma}}, \bibinfo {author}
  {\bibfnamefont {Y.}~\bibnamefont {Li}}, \bibinfo {author} {\bibfnamefont
  {E.}~\bibnamefont {Allys}},  \emph {et~al.},\ }\href@noop {} {\bibfield
  {journal} {\bibinfo  {journal} {The Astrophysical Journal Supplement Series}\
  }\textbf {\bibinfo {volume} {250}},\ \bibinfo {pages} {2} (\bibinfo {year}
  {2020})}\BibitemShut {NoStop}%
\bibitem [{\citenamefont {Behroozi}\ \emph {et~al.}(2012)\citenamefont
  {Behroozi}, \citenamefont {Wechsler},\ and\ \citenamefont
  {Wu}}]{behroozi2012rockstar}%
  \BibitemOpen
  \bibfield  {author} {\bibinfo {author} {\bibfnamefont {P.~S.}\ \bibnamefont
  {Behroozi}}, \bibinfo {author} {\bibfnamefont {R.~H.}\ \bibnamefont
  {Wechsler}}, \ and\ \bibinfo {author} {\bibfnamefont {H.-Y.}\ \bibnamefont
  {Wu}},\ }\href@noop {} {\bibfield  {journal} {\bibinfo  {journal} {The
  Astrophysical Journal}\ }\textbf {\bibinfo {volume} {762}},\ \bibinfo {pages}
  {109} (\bibinfo {year} {2012})}\BibitemShut {NoStop}%
\bibitem [{\citenamefont {Zheng}\ \emph {et~al.}(2007)\citenamefont {Zheng},
  \citenamefont {Coil},\ and\ \citenamefont {Zehavi}}]{zheng2007galaxy}%
  \BibitemOpen
  \bibfield  {author} {\bibinfo {author} {\bibfnamefont {Z.}~\bibnamefont
  {Zheng}}, \bibinfo {author} {\bibfnamefont {A.~L.}\ \bibnamefont {Coil}}, \
  and\ \bibinfo {author} {\bibfnamefont {I.}~\bibnamefont {Zehavi}},\
  }\href@noop {} {\bibfield  {journal} {\bibinfo  {journal} {The Astrophysical
  Journal}\ }\textbf {\bibinfo {volume} {667}},\ \bibinfo {pages} {760}
  (\bibinfo {year} {2007})}\BibitemShut {NoStop}%
\bibitem [{\citenamefont {More}\ \emph {et~al.}(2016)\citenamefont {More},
  \citenamefont {Miyatake}, \citenamefont {Takada}, \citenamefont {Diemer},
  \citenamefont {Kravtsov}, \citenamefont {Dalal}, \citenamefont {More},
  \citenamefont {Murata}, \citenamefont {Mandelbaum}, \citenamefont {Rozo}
  \emph {et~al.}}]{more2016detection}%
  \BibitemOpen
  \bibfield  {author} {\bibinfo {author} {\bibfnamefont {S.}~\bibnamefont
  {More}}, \bibinfo {author} {\bibfnamefont {H.}~\bibnamefont {Miyatake}},
  \bibinfo {author} {\bibfnamefont {M.}~\bibnamefont {Takada}}, \bibinfo
  {author} {\bibfnamefont {B.}~\bibnamefont {Diemer}}, \bibinfo {author}
  {\bibfnamefont {A.~V.}\ \bibnamefont {Kravtsov}}, \bibinfo {author}
  {\bibfnamefont {N.~K.}\ \bibnamefont {Dalal}}, \bibinfo {author}
  {\bibfnamefont {A.}~\bibnamefont {More}}, \bibinfo {author} {\bibfnamefont
  {R.}~\bibnamefont {Murata}}, \bibinfo {author} {\bibfnamefont
  {R.}~\bibnamefont {Mandelbaum}}, \bibinfo {author} {\bibfnamefont
  {E.}~\bibnamefont {Rozo}},  \emph {et~al.},\ }\href@noop {} {\bibfield
  {journal} {\bibinfo  {journal} {The Astrophysical Journal}\ }\textbf
  {\bibinfo {volume} {825}},\ \bibinfo {pages} {39} (\bibinfo {year}
  {2016})}\BibitemShut {NoStop}%
\bibitem [{\citenamefont {Vakili}\ and\ \citenamefont
  {Hahn}(2019)}]{vakili2019galaxies}%
  \BibitemOpen
  \bibfield  {author} {\bibinfo {author} {\bibfnamefont {M.}~\bibnamefont
  {Vakili}}\ and\ \bibinfo {author} {\bibfnamefont {C.}~\bibnamefont {Hahn}},\
  }\href@noop {} {\bibfield  {journal} {\bibinfo  {journal} {The Astrophysical
  Journal}\ }\textbf {\bibinfo {volume} {872}},\ \bibinfo {pages} {115}
  (\bibinfo {year} {2019})}\BibitemShut {NoStop}%
\bibitem [{\citenamefont {Zentner}\ \emph {et~al.}(2019)\citenamefont
  {Zentner}, \citenamefont {Hearin}, \citenamefont {van~den Bosch},
  \citenamefont {Lange},\ and\ \citenamefont
  {Villarreal}}]{zentner2019constraints}%
  \BibitemOpen
  \bibfield  {author} {\bibinfo {author} {\bibfnamefont {A.~R.}\ \bibnamefont
  {Zentner}}, \bibinfo {author} {\bibfnamefont {A.}~\bibnamefont {Hearin}},
  \bibinfo {author} {\bibfnamefont {F.~C.}\ \bibnamefont {van~den Bosch}},
  \bibinfo {author} {\bibfnamefont {J.~U.}\ \bibnamefont {Lange}}, \ and\
  \bibinfo {author} {\bibfnamefont {A.~S.}\ \bibnamefont {Villarreal}},\
  }\href@noop {} {\bibfield  {journal} {\bibinfo  {journal} {Monthly Notices of
  the Royal Astronomical Society}\ }\textbf {\bibinfo {volume} {485}},\
  \bibinfo {pages} {1196} (\bibinfo {year} {2019})}\BibitemShut {NoStop}%
\bibitem [{\citenamefont {Hadzhiyska}\ \emph {et~al.}(2021)\citenamefont
  {Hadzhiyska}, \citenamefont {Liu}, \citenamefont {Somerville}, \citenamefont
  {Gabrielpillai}, \citenamefont {Bose}, \citenamefont {Eisenstein},\ and\
  \citenamefont {Hernquist}}]{hadzhiyska2021galaxy}%
  \BibitemOpen
  \bibfield  {author} {\bibinfo {author} {\bibfnamefont {B.}~\bibnamefont
  {Hadzhiyska}}, \bibinfo {author} {\bibfnamefont {S.}~\bibnamefont {Liu}},
  \bibinfo {author} {\bibfnamefont {R.~S.}\ \bibnamefont {Somerville}},
  \bibinfo {author} {\bibfnamefont {A.}~\bibnamefont {Gabrielpillai}}, \bibinfo
  {author} {\bibfnamefont {S.}~\bibnamefont {Bose}}, \bibinfo {author}
  {\bibfnamefont {D.}~\bibnamefont {Eisenstein}}, \ and\ \bibinfo {author}
  {\bibfnamefont {L.}~\bibnamefont {Hernquist}},\ }\href@noop {} {\bibfield
  {journal} {\bibinfo  {journal} {Monthly Notices of the Royal Astronomical
  Society}\ }\textbf {\bibinfo {volume} {508}},\ \bibinfo {pages} {698}
  (\bibinfo {year} {2021})}\BibitemShut {NoStop}%
\bibitem [{\citenamefont {Carlson}\ and\ \citenamefont
  {White}(2010)}]{carlson2010embedding}%
  \BibitemOpen
  \bibfield  {author} {\bibinfo {author} {\bibfnamefont {J.}~\bibnamefont
  {Carlson}}\ and\ \bibinfo {author} {\bibfnamefont {M.}~\bibnamefont
  {White}},\ }\href@noop {} {\bibfield  {journal} {\bibinfo  {journal} {The
  Astrophysical Journal Supplement Series}\ }\textbf {\bibinfo {volume}
  {190}},\ \bibinfo {pages} {311} (\bibinfo {year} {2010})}\BibitemShut
  {NoStop}%
\bibitem [{\citenamefont {Davis}\ \emph {et~al.}(1985)\citenamefont {Davis},
  \citenamefont {Efstathiou}, \citenamefont {Frenk},\ and\ \citenamefont
  {White}}]{davis1985evolution}%
  \BibitemOpen
  \bibfield  {author} {\bibinfo {author} {\bibfnamefont {M.}~\bibnamefont
  {Davis}}, \bibinfo {author} {\bibfnamefont {G.}~\bibnamefont {Efstathiou}},
  \bibinfo {author} {\bibfnamefont {C.~S.}\ \bibnamefont {Frenk}}, \ and\
  \bibinfo {author} {\bibfnamefont {S.~D.}\ \bibnamefont {White}},\ }\href@noop
  {} {\bibfield  {journal} {\bibinfo  {journal} {The Astrophysical Journal}\
  }\textbf {\bibinfo {volume} {292}},\ \bibinfo {pages} {371} (\bibinfo {year}
  {1985})}\BibitemShut {NoStop}%
\bibitem [{\citenamefont {Maksimova}\ \emph {et~al.}(2021)\citenamefont
  {Maksimova}, \citenamefont {Garrison}, \citenamefont {Eisenstein},
  \citenamefont {Hadzhiyska}, \citenamefont {Bose},\ and\ \citenamefont
  {Satterthwaite}}]{maksimova2021abacussummit}%
  \BibitemOpen
  \bibfield  {author} {\bibinfo {author} {\bibfnamefont {N.~A.}\ \bibnamefont
  {Maksimova}}, \bibinfo {author} {\bibfnamefont {L.~H.}\ \bibnamefont
  {Garrison}}, \bibinfo {author} {\bibfnamefont {D.~J.}\ \bibnamefont
  {Eisenstein}}, \bibinfo {author} {\bibfnamefont {B.}~\bibnamefont
  {Hadzhiyska}}, \bibinfo {author} {\bibfnamefont {S.}~\bibnamefont {Bose}}, \
  and\ \bibinfo {author} {\bibfnamefont {T.~P.}\ \bibnamefont
  {Satterthwaite}},\ }\href@noop {} {\bibfield  {journal} {\bibinfo  {journal}
  {Monthly Notices of the Royal Astronomical Society}\ }\textbf {\bibinfo
  {volume} {508}},\ \bibinfo {pages} {4017} (\bibinfo {year}
  {2021})}\BibitemShut {NoStop}%
\bibitem [{\citenamefont {Hadzhiyska}\ \emph {et~al.}(2022)\citenamefont
  {Hadzhiyska}, \citenamefont {Eisenstein}, \citenamefont {Bose}, \citenamefont
  {Garrison},\ and\ \citenamefont {Maksimova}}]{hadzhiyska2022compaso}%
  \BibitemOpen
  \bibfield  {author} {\bibinfo {author} {\bibfnamefont {B.}~\bibnamefont
  {Hadzhiyska}}, \bibinfo {author} {\bibfnamefont {D.}~\bibnamefont
  {Eisenstein}}, \bibinfo {author} {\bibfnamefont {S.}~\bibnamefont {Bose}},
  \bibinfo {author} {\bibfnamefont {L.~H.}\ \bibnamefont {Garrison}}, \ and\
  \bibinfo {author} {\bibfnamefont {N.}~\bibnamefont {Maksimova}},\ }\href@noop
  {} {\bibfield  {journal} {\bibinfo  {journal} {Monthly Notices of the Royal
  Astronomical Society}\ }\textbf {\bibinfo {volume} {509}},\ \bibinfo {pages}
  {501} (\bibinfo {year} {2022})}\BibitemShut {NoStop}%
\bibitem [{\citenamefont {Birdsall}\ and\ \citenamefont
  {Fuss}(1969)}]{birdsall1969clouds}%
  \BibitemOpen
  \bibfield  {author} {\bibinfo {author} {\bibfnamefont {C.~K.}\ \bibnamefont
  {Birdsall}}\ and\ \bibinfo {author} {\bibfnamefont {D.}~\bibnamefont
  {Fuss}},\ }\href@noop {} {\bibfield  {journal} {\bibinfo  {journal} {Journal
  of Computational Physics}\ }\textbf {\bibinfo {volume} {3}},\ \bibinfo
  {pages} {494} (\bibinfo {year} {1969})}\BibitemShut {NoStop}%
\bibitem [{\citenamefont {Anderson}\ \emph {et~al.}(2012)\citenamefont
  {Anderson}, \citenamefont {Aubourg}, \citenamefont {Bailey}, \citenamefont
  {Bizyaev}, \citenamefont {Blanton}, \citenamefont {Bolton}, \citenamefont
  {Brinkmann}, \citenamefont {Brownstein}, \citenamefont {Burden},
  \citenamefont {Cuesta} \emph {et~al.}}]{anderson2012clustering}%
  \BibitemOpen
  \bibfield  {author} {\bibinfo {author} {\bibfnamefont {L.}~\bibnamefont
  {Anderson}}, \bibinfo {author} {\bibfnamefont {E.}~\bibnamefont {Aubourg}},
  \bibinfo {author} {\bibfnamefont {S.}~\bibnamefont {Bailey}}, \bibinfo
  {author} {\bibfnamefont {D.}~\bibnamefont {Bizyaev}}, \bibinfo {author}
  {\bibfnamefont {M.}~\bibnamefont {Blanton}}, \bibinfo {author} {\bibfnamefont
  {A.~S.}\ \bibnamefont {Bolton}}, \bibinfo {author} {\bibfnamefont
  {J.}~\bibnamefont {Brinkmann}}, \bibinfo {author} {\bibfnamefont {J.~R.}\
  \bibnamefont {Brownstein}}, \bibinfo {author} {\bibfnamefont
  {A.}~\bibnamefont {Burden}}, \bibinfo {author} {\bibfnamefont {A.~J.}\
  \bibnamefont {Cuesta}},  \emph {et~al.},\ }\href@noop {} {\bibfield
  {journal} {\bibinfo  {journal} {Monthly Notices of the Royal Astronomical
  Society}\ }\textbf {\bibinfo {volume} {427}},\ \bibinfo {pages} {3435}
  (\bibinfo {year} {2012})}\BibitemShut {NoStop}%
\bibitem [{\citenamefont {Hahn}\ \emph
  {et~al.}(2022{\natexlab{b}})\citenamefont {Hahn}, \citenamefont {Abidi},
  \citenamefont {Eickenberg}, \citenamefont {Ho}, \citenamefont {Lemos},
  \citenamefont {Massara}, \citenamefont {Moradinezhad~Dizgah},\ and\
  \citenamefont {R{\'e}galdo-Saint~Blancard}}]{hahnsimbig}%
  \BibitemOpen
  \bibfield  {author} {\bibinfo {author} {\bibfnamefont {C.}~\bibnamefont
  {Hahn}}, \bibinfo {author} {\bibfnamefont {M.}~\bibnamefont {Abidi}},
  \bibinfo {author} {\bibfnamefont {M.}~\bibnamefont {Eickenberg}}, \bibinfo
  {author} {\bibfnamefont {S.}~\bibnamefont {Ho}}, \bibinfo {author}
  {\bibfnamefont {P.}~\bibnamefont {Lemos}}, \bibinfo {author} {\bibfnamefont
  {E.}~\bibnamefont {Massara}}, \bibinfo {author} {\bibfnamefont
  {A.}~\bibnamefont {Moradinezhad~Dizgah}}, \ and\ \bibinfo {author}
  {\bibfnamefont {B.}~\bibnamefont {R{\'e}galdo-Saint~Blancard}},\ }\href@noop
  {} {\bibfield  {journal} {\bibinfo  {journal} {Machine Learning for
  Astrophysics}\ ,\ \bibinfo {pages} {24}} (\bibinfo {year}
  {2022}{\natexlab{b}})}\BibitemShut {NoStop}%
\bibitem [{\citenamefont {Alzubaidi}\ \emph {et~al.}(2021)\citenamefont
  {Alzubaidi}, \citenamefont {Zhang}, \citenamefont {Humaidi}, \citenamefont
  {Al-Dujaili}, \citenamefont {Duan}, \citenamefont {Al-Shamma}, \citenamefont
  {Santamar{\'\i}a}, \citenamefont {Fadhel}, \citenamefont {Al-Amidie},\ and\
  \citenamefont {Farhan}}]{alzubaidi2021review}%
  \BibitemOpen
  \bibfield  {author} {\bibinfo {author} {\bibfnamefont {L.}~\bibnamefont
  {Alzubaidi}}, \bibinfo {author} {\bibfnamefont {J.}~\bibnamefont {Zhang}},
  \bibinfo {author} {\bibfnamefont {A.~J.}\ \bibnamefont {Humaidi}}, \bibinfo
  {author} {\bibfnamefont {A.}~\bibnamefont {Al-Dujaili}}, \bibinfo {author}
  {\bibfnamefont {Y.}~\bibnamefont {Duan}}, \bibinfo {author} {\bibfnamefont
  {O.}~\bibnamefont {Al-Shamma}}, \bibinfo {author} {\bibfnamefont
  {J.}~\bibnamefont {Santamar{\'\i}a}}, \bibinfo {author} {\bibfnamefont
  {M.~A.}\ \bibnamefont {Fadhel}}, \bibinfo {author} {\bibfnamefont
  {M.}~\bibnamefont {Al-Amidie}}, \ and\ \bibinfo {author} {\bibfnamefont
  {L.}~\bibnamefont {Farhan}},\ }\href@noop {} {\bibfield  {journal} {\bibinfo
  {journal} {Journal of big Data}\ }\textbf {\bibinfo {volume} {8}},\ \bibinfo
  {pages} {1} (\bibinfo {year} {2021})}\BibitemShut {NoStop}%
\bibitem [{\citenamefont {Santurkar}\ \emph {et~al.}(2018)\citenamefont
  {Santurkar}, \citenamefont {Tsipras}, \citenamefont {Ilyas},\ and\
  \citenamefont {Madry}}]{santurkar2018does}%
  \BibitemOpen
  \bibfield  {author} {\bibinfo {author} {\bibfnamefont {S.}~\bibnamefont
  {Santurkar}}, \bibinfo {author} {\bibfnamefont {D.}~\bibnamefont {Tsipras}},
  \bibinfo {author} {\bibfnamefont {A.}~\bibnamefont {Ilyas}}, \ and\ \bibinfo
  {author} {\bibfnamefont {A.}~\bibnamefont {Madry}},\ }\href@noop {}
  {\bibfield  {journal} {\bibinfo  {journal} {Advances in neural information
  processing systems}\ }\textbf {\bibinfo {volume} {31}} (\bibinfo {year}
  {2018})}\BibitemShut {NoStop}%
\bibitem [{\citenamefont {Smith}\ and\ \citenamefont
  {Topin}(2019)}]{smith2019super}%
  \BibitemOpen
  \bibfield  {author} {\bibinfo {author} {\bibfnamefont {L.~N.}\ \bibnamefont
  {Smith}}\ and\ \bibinfo {author} {\bibfnamefont {N.}~\bibnamefont {Topin}},\
  }in\ \href@noop {} {\emph {\bibinfo {booktitle} {Artificial intelligence and
  machine learning for multi-domain operations applications}}},\ Vol.\ \bibinfo
  {volume} {11006}\ (\bibinfo {organization} {SPIE},\ \bibinfo {year} {2019})\
  pp.\ \bibinfo {pages} {369--386}\BibitemShut {NoStop}%
\bibitem [{\citenamefont {Szegedy}\ \emph {et~al.}(2016)\citenamefont
  {Szegedy}, \citenamefont {Vanhoucke}, \citenamefont {Ioffe}, \citenamefont
  {Shlens},\ and\ \citenamefont {Wojna}}]{szegedy2016rethinking}%
  \BibitemOpen
  \bibfield  {author} {\bibinfo {author} {\bibfnamefont {C.}~\bibnamefont
  {Szegedy}}, \bibinfo {author} {\bibfnamefont {V.}~\bibnamefont {Vanhoucke}},
  \bibinfo {author} {\bibfnamefont {S.}~\bibnamefont {Ioffe}}, \bibinfo
  {author} {\bibfnamefont {J.}~\bibnamefont {Shlens}}, \ and\ \bibinfo {author}
  {\bibfnamefont {Z.}~\bibnamefont {Wojna}},\ }in\ \href@noop {} {\emph
  {\bibinfo {booktitle} {Proceedings of the IEEE conference on computer vision
  and pattern recognition}}}\ (\bibinfo {year} {2016})\ pp.\ \bibinfo {pages}
  {2818--2826}\BibitemShut {NoStop}%
\bibitem [{\citenamefont {Krizhevsky}\ \emph {et~al.}(2017)\citenamefont
  {Krizhevsky}, \citenamefont {Sutskever},\ and\ \citenamefont
  {Hinton}}]{krizhevsky2017imagenet}%
  \BibitemOpen
  \bibfield  {author} {\bibinfo {author} {\bibfnamefont {A.}~\bibnamefont
  {Krizhevsky}}, \bibinfo {author} {\bibfnamefont {I.}~\bibnamefont
  {Sutskever}}, \ and\ \bibinfo {author} {\bibfnamefont {G.~E.}\ \bibnamefont
  {Hinton}},\ }\href@noop {} {\bibfield  {journal} {\bibinfo  {journal}
  {Communications of the ACM}\ }\textbf {\bibinfo {volume} {60}},\ \bibinfo
  {pages} {84} (\bibinfo {year} {2017})}\BibitemShut {NoStop}%
\bibitem [{\citenamefont {Akiba}\ \emph {et~al.}(2019)\citenamefont {Akiba},
  \citenamefont {Sano}, \citenamefont {Yanase}, \citenamefont {Ohta},\ and\
  \citenamefont {Koyama}}]{optuna}%
  \BibitemOpen
  \bibfield  {author} {\bibinfo {author} {\bibfnamefont {T.}~\bibnamefont
  {Akiba}}, \bibinfo {author} {\bibfnamefont {S.}~\bibnamefont {Sano}},
  \bibinfo {author} {\bibfnamefont {T.}~\bibnamefont {Yanase}}, \bibinfo
  {author} {\bibfnamefont {T.}~\bibnamefont {Ohta}}, \ and\ \bibinfo {author}
  {\bibfnamefont {M.}~\bibnamefont {Koyama}},\ }in\ \href@noop {} {\emph
  {\bibinfo {booktitle} {Proceedings of the 25th ACM SIGKDD international
  conference on knowledge discovery \& data mining}}}\ (\bibinfo {year}
  {2019})\ pp.\ \bibinfo {pages} {2623--2631}\BibitemShut {NoStop}%
\bibitem [{\citenamefont {Maddox}\ \emph {et~al.}(2019)\citenamefont {Maddox},
  \citenamefont {Izmailov}, \citenamefont {Garipov}, \citenamefont {Vetrov},\
  and\ \citenamefont {Wilson}}]{maddox2019simple}%
  \BibitemOpen
  \bibfield  {author} {\bibinfo {author} {\bibfnamefont {W.~J.}\ \bibnamefont
  {Maddox}}, \bibinfo {author} {\bibfnamefont {P.}~\bibnamefont {Izmailov}},
  \bibinfo {author} {\bibfnamefont {T.}~\bibnamefont {Garipov}}, \bibinfo
  {author} {\bibfnamefont {D.~P.}\ \bibnamefont {Vetrov}}, \ and\ \bibinfo
  {author} {\bibfnamefont {A.~G.}\ \bibnamefont {Wilson}},\ }\href@noop {}
  {\bibfield  {journal} {\bibinfo  {journal} {Advances in neural information
  processing systems}\ }\textbf {\bibinfo {volume} {32}} (\bibinfo {year}
  {2019})}\BibitemShut {NoStop}%
\bibitem [{\citenamefont {Wilson}\ and\ \citenamefont
  {Izmailov}(2020)}]{wilson2020bayesian}%
  \BibitemOpen
  \bibfield  {author} {\bibinfo {author} {\bibfnamefont {A.~G.}\ \bibnamefont
  {Wilson}}\ and\ \bibinfo {author} {\bibfnamefont {P.}~\bibnamefont
  {Izmailov}},\ }\href@noop {} {\bibfield  {journal} {\bibinfo  {journal}
  {Advances in neural information processing systems}\ }\textbf {\bibinfo
  {volume} {33}},\ \bibinfo {pages} {4697} (\bibinfo {year}
  {2020})}\BibitemShut {NoStop}%
\bibitem [{\citenamefont {Cranmer}\ \emph {et~al.}(2021)\citenamefont
  {Cranmer}, \citenamefont {Tamayo}, \citenamefont {Rein}, \citenamefont
  {Battaglia}, \citenamefont {Hadden}, \citenamefont {Armitage}, \citenamefont
  {Ho},\ and\ \citenamefont {Spergel}}]{cranmer2021bayesian}%
  \BibitemOpen
  \bibfield  {author} {\bibinfo {author} {\bibfnamefont {M.}~\bibnamefont
  {Cranmer}}, \bibinfo {author} {\bibfnamefont {D.}~\bibnamefont {Tamayo}},
  \bibinfo {author} {\bibfnamefont {H.}~\bibnamefont {Rein}}, \bibinfo {author}
  {\bibfnamefont {P.}~\bibnamefont {Battaglia}}, \bibinfo {author}
  {\bibfnamefont {S.}~\bibnamefont {Hadden}}, \bibinfo {author} {\bibfnamefont
  {P.~J.}\ \bibnamefont {Armitage}}, \bibinfo {author} {\bibfnamefont
  {S.}~\bibnamefont {Ho}}, \ and\ \bibinfo {author} {\bibfnamefont {D.~N.}\
  \bibnamefont {Spergel}},\ }\href@noop {} {\bibfield  {journal} {\bibinfo
  {journal} {Proceedings of the National Academy of Sciences}\ }\textbf
  {\bibinfo {volume} {118}},\ \bibinfo {pages} {e2026053118} (\bibinfo {year}
  {2021})}\BibitemShut {NoStop}%
\bibitem [{\citenamefont {Lemos}\ \emph
  {et~al.}(2023{\natexlab{a}})\citenamefont {Lemos}, \citenamefont {Cranmer},
  \citenamefont {Abidi}, \citenamefont {Hahn}, \citenamefont {Eickenberg},
  \citenamefont {Massara}, \citenamefont {Yallup},\ and\ \citenamefont
  {Ho}}]{lemos2023robust}%
  \BibitemOpen
  \bibfield  {author} {\bibinfo {author} {\bibfnamefont {P.}~\bibnamefont
  {Lemos}}, \bibinfo {author} {\bibfnamefont {M.}~\bibnamefont {Cranmer}},
  \bibinfo {author} {\bibfnamefont {M.}~\bibnamefont {Abidi}}, \bibinfo
  {author} {\bibfnamefont {C.}~\bibnamefont {Hahn}}, \bibinfo {author}
  {\bibfnamefont {M.}~\bibnamefont {Eickenberg}}, \bibinfo {author}
  {\bibfnamefont {E.}~\bibnamefont {Massara}}, \bibinfo {author} {\bibfnamefont
  {D.}~\bibnamefont {Yallup}}, \ and\ \bibinfo {author} {\bibfnamefont
  {S.}~\bibnamefont {Ho}},\ }\href@noop {} {\bibfield  {journal} {\bibinfo
  {journal} {Machine Learning: Science and Technology}\ }\textbf {\bibinfo
  {volume} {4}},\ \bibinfo {pages} {01LT01} (\bibinfo {year}
  {2023}{\natexlab{a}})}\BibitemShut {NoStop}%
\bibitem [{\citenamefont {Rubin}(1984)}]{10.1214/aos/1176346785}%
  \BibitemOpen
  \bibfield  {author} {\bibinfo {author} {\bibfnamefont {D.~B.}\ \bibnamefont
  {Rubin}},\ }\href {\doibase 10.1214/aos/1176346785} {\bibfield  {journal}
  {\bibinfo  {journal} {The Annals of Statistics}\ }\textbf {\bibinfo {volume}
  {12}},\ \bibinfo {pages} {1151 } (\bibinfo {year} {1984})}\BibitemShut
  {NoStop}%
\bibitem [{\citenamefont {Pritchard}\ \emph {et~al.}(1999)\citenamefont
  {Pritchard}, \citenamefont {Seielstad}, \citenamefont {Perez-Lezaun},\ and\
  \citenamefont {Feldman}}]{pritchard1999population}%
  \BibitemOpen
  \bibfield  {author} {\bibinfo {author} {\bibfnamefont {J.~K.}\ \bibnamefont
  {Pritchard}}, \bibinfo {author} {\bibfnamefont {M.~T.}\ \bibnamefont
  {Seielstad}}, \bibinfo {author} {\bibfnamefont {A.}~\bibnamefont
  {Perez-Lezaun}}, \ and\ \bibinfo {author} {\bibfnamefont {M.~W.}\
  \bibnamefont {Feldman}},\ }\href@noop {} {\bibfield  {journal} {\bibinfo
  {journal} {Molecular biology and evolution}\ }\textbf {\bibinfo {volume}
  {16}},\ \bibinfo {pages} {1791} (\bibinfo {year} {1999})}\BibitemShut
  {NoStop}%
\bibitem [{\citenamefont {Beaumont}\ \emph {et~al.}(2002)\citenamefont
  {Beaumont}, \citenamefont {Zhang},\ and\ \citenamefont
  {Balding}}]{beaumont2002approximate}%
  \BibitemOpen
  \bibfield  {author} {\bibinfo {author} {\bibfnamefont {M.~A.}\ \bibnamefont
  {Beaumont}}, \bibinfo {author} {\bibfnamefont {W.}~\bibnamefont {Zhang}}, \
  and\ \bibinfo {author} {\bibfnamefont {D.~J.}\ \bibnamefont {Balding}},\
  }\href@noop {} {\bibfield  {journal} {\bibinfo  {journal} {Genetics}\
  }\textbf {\bibinfo {volume} {162}},\ \bibinfo {pages} {2025} (\bibinfo {year}
  {2002})}\BibitemShut {NoStop}%
\bibitem [{\citenamefont {Marjoram}\ \emph {et~al.}(2003)\citenamefont
  {Marjoram}, \citenamefont {Molitor}, \citenamefont {Plagnol},\ and\
  \citenamefont {Tavar{\'e}}}]{marjoram2003markov}%
  \BibitemOpen
  \bibfield  {author} {\bibinfo {author} {\bibfnamefont {P.}~\bibnamefont
  {Marjoram}}, \bibinfo {author} {\bibfnamefont {J.}~\bibnamefont {Molitor}},
  \bibinfo {author} {\bibfnamefont {V.}~\bibnamefont {Plagnol}}, \ and\
  \bibinfo {author} {\bibfnamefont {S.}~\bibnamefont {Tavar{\'e}}},\
  }\href@noop {} {\bibfield  {journal} {\bibinfo  {journal} {Proceedings of the
  National Academy of Sciences}\ }\textbf {\bibinfo {volume} {100}},\ \bibinfo
  {pages} {15324} (\bibinfo {year} {2003})}\BibitemShut {NoStop}%
\bibitem [{\citenamefont {Fearnhead}\ and\ \citenamefont
  {Prangle}(2012)}]{fearnhead2012constructing}%
  \BibitemOpen
  \bibfield  {author} {\bibinfo {author} {\bibfnamefont {P.}~\bibnamefont
  {Fearnhead}}\ and\ \bibinfo {author} {\bibfnamefont {D.}~\bibnamefont
  {Prangle}},\ }\href@noop {} {\bibfield  {journal} {\bibinfo  {journal}
  {Journal of the Royal Statistical Society: Series B (Statistical
  Methodology)}\ }\textbf {\bibinfo {volume} {74}},\ \bibinfo {pages} {419}
  (\bibinfo {year} {2012})}\BibitemShut {NoStop}%
\bibitem [{\citenamefont {Cranmer}\ \emph {et~al.}(2015)\citenamefont
  {Cranmer}, \citenamefont {Pavez},\ and\ \citenamefont
  {Louppe}}]{cranmer2015approximating}%
  \BibitemOpen
  \bibfield  {author} {\bibinfo {author} {\bibfnamefont {K.}~\bibnamefont
  {Cranmer}}, \bibinfo {author} {\bibfnamefont {J.}~\bibnamefont {Pavez}}, \
  and\ \bibinfo {author} {\bibfnamefont {G.}~\bibnamefont {Louppe}},\
  }\href@noop {} {\bibfield  {journal} {\bibinfo  {journal} {arXiv preprint
  arXiv:1506.02169}\ } (\bibinfo {year} {2015})}\BibitemShut {NoStop}%
\bibitem [{\citenamefont {Thomas}\ \emph {et~al.}(2022)\citenamefont {Thomas},
  \citenamefont {Dutta}, \citenamefont {Corander}, \citenamefont {Kaski},\ and\
  \citenamefont {Gutmann}}]{thomas2022likelihood}%
  \BibitemOpen
  \bibfield  {author} {\bibinfo {author} {\bibfnamefont {O.}~\bibnamefont
  {Thomas}}, \bibinfo {author} {\bibfnamefont {R.}~\bibnamefont {Dutta}},
  \bibinfo {author} {\bibfnamefont {J.}~\bibnamefont {Corander}}, \bibinfo
  {author} {\bibfnamefont {S.}~\bibnamefont {Kaski}}, \ and\ \bibinfo {author}
  {\bibfnamefont {M.~U.}\ \bibnamefont {Gutmann}},\ }\href@noop {} {\bibfield
  {journal} {\bibinfo  {journal} {Bayesian Analysis}\ }\textbf {\bibinfo
  {volume} {17}},\ \bibinfo {pages} {1} (\bibinfo {year} {2022})}\BibitemShut
  {NoStop}%
\bibitem [{\citenamefont {Hermans}\ \emph {et~al.}(2020)\citenamefont
  {Hermans}, \citenamefont {Begy},\ and\ \citenamefont
  {Louppe}}]{hermans2020likelihood}%
  \BibitemOpen
  \bibfield  {author} {\bibinfo {author} {\bibfnamefont {J.}~\bibnamefont
  {Hermans}}, \bibinfo {author} {\bibfnamefont {V.}~\bibnamefont {Begy}}, \
  and\ \bibinfo {author} {\bibfnamefont {G.}~\bibnamefont {Louppe}},\ }in\
  \href@noop {} {\emph {\bibinfo {booktitle} {International Conference on
  Machine Learning}}}\ (\bibinfo {organization} {PMLR},\ \bibinfo {year}
  {2020})\ pp.\ \bibinfo {pages} {4239--4248}\BibitemShut {NoStop}%
\bibitem [{\citenamefont {Durkan}\ \emph {et~al.}(2020)\citenamefont {Durkan},
  \citenamefont {Murray},\ and\ \citenamefont
  {Papamakarios}}]{durkan2020contrastive}%
  \BibitemOpen
  \bibfield  {author} {\bibinfo {author} {\bibfnamefont {C.}~\bibnamefont
  {Durkan}}, \bibinfo {author} {\bibfnamefont {I.}~\bibnamefont {Murray}}, \
  and\ \bibinfo {author} {\bibfnamefont {G.}~\bibnamefont {Papamakarios}},\
  }in\ \href@noop {} {\emph {\bibinfo {booktitle} {International Conference on
  Machine Learning}}}\ (\bibinfo {organization} {PMLR},\ \bibinfo {year}
  {2020})\ pp.\ \bibinfo {pages} {2771--2781}\BibitemShut {NoStop}%
\bibitem [{\citenamefont {Miller}\ \emph {et~al.}(2022)\citenamefont {Miller},
  \citenamefont {Weniger},\ and\ \citenamefont {Forr\'e}}]{Miller:2022haf}%
  \BibitemOpen
  \bibfield  {author} {\bibinfo {author} {\bibfnamefont {B.~K.}\ \bibnamefont
  {Miller}}, \bibinfo {author} {\bibfnamefont {C.}~\bibnamefont {Weniger}}, \
  and\ \bibinfo {author} {\bibfnamefont {P.}~\bibnamefont {Forr\'e}},\
  }\href@noop {} {\  (\bibinfo {year} {2022})},\ \Eprint
  {http://arxiv.org/abs/2210.06170} {arXiv:2210.06170 [stat.ML]} \BibitemShut
  {NoStop}%
\bibitem [{\citenamefont {Price}\ \emph {et~al.}(2018)\citenamefont {Price},
  \citenamefont {Drovandi}, \citenamefont {Lee},\ and\ \citenamefont
  {Nott}}]{price2018bayesian}%
  \BibitemOpen
  \bibfield  {author} {\bibinfo {author} {\bibfnamefont {L.~F.}\ \bibnamefont
  {Price}}, \bibinfo {author} {\bibfnamefont {C.~C.}\ \bibnamefont {Drovandi}},
  \bibinfo {author} {\bibfnamefont {A.}~\bibnamefont {Lee}}, \ and\ \bibinfo
  {author} {\bibfnamefont {D.~J.}\ \bibnamefont {Nott}},\ }\href@noop {}
  {\bibfield  {journal} {\bibinfo  {journal} {Journal of Computational and
  Graphical Statistics}\ }\textbf {\bibinfo {volume} {27}},\ \bibinfo {pages}
  {1} (\bibinfo {year} {2018})}\BibitemShut {NoStop}%
\bibitem [{\citenamefont {Papamakarios}\ \emph {et~al.}(2019)\citenamefont
  {Papamakarios}, \citenamefont {Sterratt},\ and\ \citenamefont
  {Murray}}]{papamakarios2019sequential}%
  \BibitemOpen
  \bibfield  {author} {\bibinfo {author} {\bibfnamefont {G.}~\bibnamefont
  {Papamakarios}}, \bibinfo {author} {\bibfnamefont {D.}~\bibnamefont
  {Sterratt}}, \ and\ \bibinfo {author} {\bibfnamefont {I.}~\bibnamefont
  {Murray}},\ }in\ \href@noop {} {\emph {\bibinfo {booktitle} {The 22nd
  International Conference on Artificial Intelligence and Statistics}}}\
  (\bibinfo {organization} {PMLR},\ \bibinfo {year} {2019})\ pp.\ \bibinfo
  {pages} {837--848}\BibitemShut {NoStop}%
\bibitem [{\citenamefont {Frazier}\ \emph {et~al.}(2022)\citenamefont
  {Frazier}, \citenamefont {Nott}, \citenamefont {Drovandi},\ and\
  \citenamefont {Kohn}}]{frazier2022bayesian}%
  \BibitemOpen
  \bibfield  {author} {\bibinfo {author} {\bibfnamefont {D.~T.}\ \bibnamefont
  {Frazier}}, \bibinfo {author} {\bibfnamefont {D.~J.}\ \bibnamefont {Nott}},
  \bibinfo {author} {\bibfnamefont {C.}~\bibnamefont {Drovandi}}, \ and\
  \bibinfo {author} {\bibfnamefont {R.}~\bibnamefont {Kohn}},\ }\href@noop {}
  {\bibfield  {journal} {\bibinfo  {journal} {Journal of the American
  Statistical Association}\ ,\ \bibinfo {pages} {1}} (\bibinfo {year}
  {2022})}\BibitemShut {NoStop}%
\bibitem [{\citenamefont {Sharrock}\ \emph {et~al.}(2022)\citenamefont
  {Sharrock}, \citenamefont {Simons}, \citenamefont {Liu},\ and\ \citenamefont
  {Beaumont}}]{sharrock2022sequential}%
  \BibitemOpen
  \bibfield  {author} {\bibinfo {author} {\bibfnamefont {L.}~\bibnamefont
  {Sharrock}}, \bibinfo {author} {\bibfnamefont {J.}~\bibnamefont {Simons}},
  \bibinfo {author} {\bibfnamefont {S.}~\bibnamefont {Liu}}, \ and\ \bibinfo
  {author} {\bibfnamefont {M.}~\bibnamefont {Beaumont}},\ }\href@noop {}
  {\bibfield  {journal} {\bibinfo  {journal} {arXiv preprint arXiv:2210.04872}\
  } (\bibinfo {year} {2022})}\BibitemShut {NoStop}%
\bibitem [{\citenamefont {Geffner}\ \emph {et~al.}(2022)\citenamefont
  {Geffner}, \citenamefont {Papamakarios},\ and\ \citenamefont
  {Mnih}}]{geffner2022score}%
  \BibitemOpen
  \bibfield  {author} {\bibinfo {author} {\bibfnamefont {T.}~\bibnamefont
  {Geffner}}, \bibinfo {author} {\bibfnamefont {G.}~\bibnamefont
  {Papamakarios}}, \ and\ \bibinfo {author} {\bibfnamefont {A.}~\bibnamefont
  {Mnih}},\ }\href@noop {} {\bibfield  {journal} {\bibinfo  {journal} {arXiv
  preprint arXiv:2209.14249}\ } (\bibinfo {year} {2022})}\BibitemShut {NoStop}%
\bibitem [{\citenamefont {Rezende}\ and\ \citenamefont
  {Mohamed}(2015)}]{rezende2015variational}%
  \BibitemOpen
  \bibfield  {author} {\bibinfo {author} {\bibfnamefont {D.}~\bibnamefont
  {Rezende}}\ and\ \bibinfo {author} {\bibfnamefont {S.}~\bibnamefont
  {Mohamed}},\ }in\ \href@noop {} {\emph {\bibinfo {booktitle} {International
  conference on machine learning}}}\ (\bibinfo {organization} {PMLR},\ \bibinfo
  {year} {2015})\ pp.\ \bibinfo {pages} {1530--1538}\BibitemShut {NoStop}%
\bibitem [{\citenamefont {Papamakarios}\ and\ \citenamefont
  {Murray}(2016)}]{papamakarios2016fast}%
  \BibitemOpen
  \bibfield  {author} {\bibinfo {author} {\bibfnamefont {G.}~\bibnamefont
  {Papamakarios}}\ and\ \bibinfo {author} {\bibfnamefont {I.}~\bibnamefont
  {Murray}},\ }\href@noop {} {\bibfield  {journal} {\bibinfo  {journal}
  {Advances in neural information processing systems}\ }\textbf {\bibinfo
  {volume} {29}} (\bibinfo {year} {2016})}\BibitemShut {NoStop}%
\bibitem [{\citenamefont {{Lueckmann}}\ \emph {et~al.}(2018)\citenamefont
  {{Lueckmann}}, \citenamefont {{Bassetto}}, \citenamefont {{Karaletsos}},\
  and\ \citenamefont {{Macke}}}]{2018arXiv180509294L}%
  \BibitemOpen
  \bibfield  {author} {\bibinfo {author} {\bibfnamefont {J.-M.}\ \bibnamefont
  {{Lueckmann}}}, \bibinfo {author} {\bibfnamefont {G.}~\bibnamefont
  {{Bassetto}}}, \bibinfo {author} {\bibfnamefont {T.}~\bibnamefont
  {{Karaletsos}}}, \ and\ \bibinfo {author} {\bibfnamefont {J.~H.}\
  \bibnamefont {{Macke}}},\ }\href@noop {} {\bibfield  {journal} {\bibinfo
  {journal} {arXiv e-prints}\ ,\ \bibinfo {eid} {arXiv:1805.09294}} (\bibinfo
  {year} {2018})},\ \Eprint {http://arxiv.org/abs/1805.09294} {arXiv:1805.09294
  [stat.ML]} \BibitemShut {NoStop}%
\bibitem [{\citenamefont {Lueckmann}\ \emph {et~al.}(2017)\citenamefont
  {Lueckmann}, \citenamefont {Goncalves}, \citenamefont {Bassetto},
  \citenamefont {{\"O}cal}, \citenamefont {Nonnenmacher},\ and\ \citenamefont
  {Macke}}]{lueckmann2017flexible}%
  \BibitemOpen
  \bibfield  {author} {\bibinfo {author} {\bibfnamefont {J.-M.}\ \bibnamefont
  {Lueckmann}}, \bibinfo {author} {\bibfnamefont {P.~J.}\ \bibnamefont
  {Goncalves}}, \bibinfo {author} {\bibfnamefont {G.}~\bibnamefont {Bassetto}},
  \bibinfo {author} {\bibfnamefont {K.}~\bibnamefont {{\"O}cal}}, \bibinfo
  {author} {\bibfnamefont {M.}~\bibnamefont {Nonnenmacher}}, \ and\ \bibinfo
  {author} {\bibfnamefont {J.~H.}\ \bibnamefont {Macke}},\ }\href@noop {}
  {\bibfield  {journal} {\bibinfo  {journal} {Advances in neural information
  processing systems}\ }\textbf {\bibinfo {volume} {30}} (\bibinfo {year}
  {2017})}\BibitemShut {NoStop}%
\bibitem [{\citenamefont {Greenberg}\ \emph {et~al.}(2019)\citenamefont
  {Greenberg}, \citenamefont {Nonnenmacher},\ and\ \citenamefont
  {Macke}}]{greenberg2019automatic}%
  \BibitemOpen
  \bibfield  {author} {\bibinfo {author} {\bibfnamefont {D.}~\bibnamefont
  {Greenberg}}, \bibinfo {author} {\bibfnamefont {M.}~\bibnamefont
  {Nonnenmacher}}, \ and\ \bibinfo {author} {\bibfnamefont {J.}~\bibnamefont
  {Macke}},\ }in\ \href@noop {} {\emph {\bibinfo {booktitle} {International
  Conference on Machine Learning}}}\ (\bibinfo {organization} {PMLR},\ \bibinfo
  {year} {2019})\ pp.\ \bibinfo {pages} {2404--2414}\BibitemShut {NoStop}%
\bibitem [{\citenamefont {Tejero-Cantero}\ \emph {et~al.}(2020)\citenamefont
  {Tejero-Cantero}, \citenamefont {Boelts}, \citenamefont {Deistler},
  \citenamefont {Lueckmann}, \citenamefont {Durkan}, \citenamefont
  {Gonçalves}, \citenamefont {Greenberg},\ and\ \citenamefont
  {Macke}}]{tejero-cantero2020sbi}%
  \BibitemOpen
  \bibfield  {author} {\bibinfo {author} {\bibfnamefont {A.}~\bibnamefont
  {Tejero-Cantero}}, \bibinfo {author} {\bibfnamefont {J.}~\bibnamefont
  {Boelts}}, \bibinfo {author} {\bibfnamefont {M.}~\bibnamefont {Deistler}},
  \bibinfo {author} {\bibfnamefont {J.-M.}\ \bibnamefont {Lueckmann}}, \bibinfo
  {author} {\bibfnamefont {C.}~\bibnamefont {Durkan}}, \bibinfo {author}
  {\bibfnamefont {P.~J.}\ \bibnamefont {Gonçalves}}, \bibinfo {author}
  {\bibfnamefont {D.~S.}\ \bibnamefont {Greenberg}}, \ and\ \bibinfo {author}
  {\bibfnamefont {J.~H.}\ \bibnamefont {Macke}},\ }\href {\doibase
  10.21105/joss.02505} {\bibfield  {journal} {\bibinfo  {journal} {Journal of
  Open Source Software}\ }\textbf {\bibinfo {volume} {5}},\ \bibinfo {pages}
  {2505} (\bibinfo {year} {2020})}\BibitemShut {NoStop}%
\bibitem [{\citenamefont {{Papamakarios}}\ \emph {et~al.}(2017)\citenamefont
  {{Papamakarios}}, \citenamefont {{Pavlakou}},\ and\ \citenamefont
  {{Murray}}}]{Papamakarios2017}%
  \BibitemOpen
  \bibfield  {author} {\bibinfo {author} {\bibfnamefont {G.}~\bibnamefont
  {{Papamakarios}}}, \bibinfo {author} {\bibfnamefont {T.}~\bibnamefont
  {{Pavlakou}}}, \ and\ \bibinfo {author} {\bibfnamefont {I.}~\bibnamefont
  {{Murray}}},\ }\href@noop {} {\bibfield  {journal} {\bibinfo  {journal}
  {arXiv e-prints}\ ,\ \bibinfo {eid} {arXiv:1705.07057}} (\bibinfo {year}
  {2017})},\ \Eprint {http://arxiv.org/abs/1705.07057} {arXiv:1705.07057
  [stat.ML]} \BibitemShut {NoStop}%
\bibitem [{\citenamefont {Durkan}\ \emph {et~al.}(2019)\citenamefont {Durkan},
  \citenamefont {Bekasov}, \citenamefont {Murray},\ and\ \citenamefont
  {Papamakarios}}]{durkan2019neural}%
  \BibitemOpen
  \bibfield  {author} {\bibinfo {author} {\bibfnamefont {C.}~\bibnamefont
  {Durkan}}, \bibinfo {author} {\bibfnamefont {A.}~\bibnamefont {Bekasov}},
  \bibinfo {author} {\bibfnamefont {I.}~\bibnamefont {Murray}}, \ and\ \bibinfo
  {author} {\bibfnamefont {G.}~\bibnamefont {Papamakarios}},\ }\href@noop {}
  {\bibfield  {journal} {\bibinfo  {journal} {Advances in neural information
  processing systems}\ }\textbf {\bibinfo {volume} {32}} (\bibinfo {year}
  {2019})}\BibitemShut {NoStop}%
\bibitem [{\citenamefont {Lakshminarayanan}\ \emph {et~al.}(2017)\citenamefont
  {Lakshminarayanan}, \citenamefont {Pritzel},\ and\ \citenamefont
  {Blundell}}]{lakshminarayanan2017simple}%
  \BibitemOpen
  \bibfield  {author} {\bibinfo {author} {\bibfnamefont {B.}~\bibnamefont
  {Lakshminarayanan}}, \bibinfo {author} {\bibfnamefont {A.}~\bibnamefont
  {Pritzel}}, \ and\ \bibinfo {author} {\bibfnamefont {C.}~\bibnamefont
  {Blundell}},\ }\href@noop {} {\bibfield  {journal} {\bibinfo  {journal}
  {Advances in neural information processing systems}\ }\textbf {\bibinfo
  {volume} {30}} (\bibinfo {year} {2017})}\BibitemShut {NoStop}%
\bibitem [{\citenamefont {Hermans}\ \emph {et~al.}(2022)\citenamefont
  {Hermans}, \citenamefont {Delaunoy}, \citenamefont {Rozet}, \citenamefont
  {Wehenkel}, \citenamefont {Begy},\ and\ \citenamefont
  {Louppe}}]{hermans2022trust}%
  \BibitemOpen
  \bibfield  {author} {\bibinfo {author} {\bibfnamefont {J.}~\bibnamefont
  {Hermans}}, \bibinfo {author} {\bibfnamefont {A.}~\bibnamefont {Delaunoy}},
  \bibinfo {author} {\bibfnamefont {F.}~\bibnamefont {Rozet}}, \bibinfo
  {author} {\bibfnamefont {A.}~\bibnamefont {Wehenkel}}, \bibinfo {author}
  {\bibfnamefont {V.}~\bibnamefont {Begy}}, \ and\ \bibinfo {author}
  {\bibfnamefont {G.}~\bibnamefont {Louppe}},\ }\href@noop {} {\bibfield
  {journal} {\bibinfo  {journal} {stat}\ }\textbf {\bibinfo {volume} {1050}},\
  \bibinfo {pages} {4} (\bibinfo {year} {2022})}\BibitemShut {NoStop}%
\bibitem [{\citenamefont {Lemos}\ \emph
  {et~al.}(2023{\natexlab{b}})\citenamefont {Lemos}, \citenamefont {Coogan},
  \citenamefont {Hezaveh},\ and\ \citenamefont
  {Perreault-Levasseur}}]{lemos2023sampling}%
  \BibitemOpen
  \bibfield  {author} {\bibinfo {author} {\bibfnamefont {P.}~\bibnamefont
  {Lemos}}, \bibinfo {author} {\bibfnamefont {A.}~\bibnamefont {Coogan}},
  \bibinfo {author} {\bibfnamefont {Y.}~\bibnamefont {Hezaveh}}, \ and\
  \bibinfo {author} {\bibfnamefont {L.}~\bibnamefont {Perreault-Levasseur}},\
  }\href@noop {} {\bibfield  {journal} {\bibinfo  {journal} {arXiv preprint
  arXiv:2302.03026}\ } (\bibinfo {year} {2023}{\natexlab{b}})}\BibitemShut
  {NoStop}%
\bibitem [{\citenamefont {Talts}\ \emph {et~al.}(2018)\citenamefont {Talts},
  \citenamefont {Betancourt}, \citenamefont {Simpson}, \citenamefont
  {Vehtari},\ and\ \citenamefont {Gelman}}]{talts2018validating}%
  \BibitemOpen
  \bibfield  {author} {\bibinfo {author} {\bibfnamefont {S.}~\bibnamefont
  {Talts}}, \bibinfo {author} {\bibfnamefont {M.}~\bibnamefont {Betancourt}},
  \bibinfo {author} {\bibfnamefont {D.}~\bibnamefont {Simpson}}, \bibinfo
  {author} {\bibfnamefont {A.}~\bibnamefont {Vehtari}}, \ and\ \bibinfo
  {author} {\bibfnamefont {A.}~\bibnamefont {Gelman}},\ }\href@noop {}
  {\bibfield  {journal} {\bibinfo  {journal} {arXiv preprint arXiv:1804.06788}\
  } (\bibinfo {year} {2018})}\BibitemShut {NoStop}%
\bibitem [{\citenamefont {Aghanim}\ \emph {et~al.}(2020)\citenamefont
  {Aghanim}, \citenamefont {Akrami}, \citenamefont {Ashdown}, \citenamefont
  {Aumont}, \citenamefont {Baccigalupi}, \citenamefont {Ballardini},
  \citenamefont {Banday}, \citenamefont {Barreiro}, \citenamefont {Bartolo},
  \citenamefont {Basak} \emph {et~al.}}]{aghanim2020planck}%
  \BibitemOpen
  \bibfield  {author} {\bibinfo {author} {\bibfnamefont {N.}~\bibnamefont
  {Aghanim}}, \bibinfo {author} {\bibfnamefont {Y.}~\bibnamefont {Akrami}},
  \bibinfo {author} {\bibfnamefont {M.}~\bibnamefont {Ashdown}}, \bibinfo
  {author} {\bibfnamefont {J.}~\bibnamefont {Aumont}}, \bibinfo {author}
  {\bibfnamefont {C.}~\bibnamefont {Baccigalupi}}, \bibinfo {author}
  {\bibfnamefont {M.}~\bibnamefont {Ballardini}}, \bibinfo {author}
  {\bibfnamefont {A.}~\bibnamefont {Banday}}, \bibinfo {author} {\bibfnamefont
  {R.}~\bibnamefont {Barreiro}}, \bibinfo {author} {\bibfnamefont
  {N.}~\bibnamefont {Bartolo}}, \bibinfo {author} {\bibfnamefont
  {S.}~\bibnamefont {Basak}},  \emph {et~al.},\ }\href@noop {} {\bibfield
  {journal} {\bibinfo  {journal} {Astronomy \& Astrophysics}\ }\textbf
  {\bibinfo {volume} {641}},\ \bibinfo {pages} {A6} (\bibinfo {year}
  {2020})}\BibitemShut {NoStop}%
\bibitem [{\citenamefont {{{\sc SimBIG}
  Collaboration}}(2023)}]{simbig2023compare}%
  \BibitemOpen
  \bibfield  {author} {\bibinfo {author} {\bibnamefont {{{\sc SimBIG}
  Collaboration}}},\ }\href@noop {} {\bibfield  {journal} {\bibinfo  {journal}
  {In preparation}\ } (\bibinfo {year} {2023})}\BibitemShut {NoStop}%
\bibitem [{\citenamefont {Collaboration}\ \emph
  {et~al.}(2016{\natexlab{a}})\citenamefont {Collaboration}, \citenamefont
  {Aghamousa}, \citenamefont {Aguilar}, \citenamefont {Ahlen}, \citenamefont
  {Alam}, \citenamefont {Allen}, \citenamefont {Prieto}, \citenamefont {Annis},
  \citenamefont {Bailey}, \citenamefont {Balland}, \citenamefont {Ballester},
  \citenamefont {Baltay}, \citenamefont {Beaufore}, \citenamefont {Bebek},
  \citenamefont {Beers}, \citenamefont {Bell}, \citenamefont {Bernal},
  \citenamefont {Besuner}, \citenamefont {Beutler}, \citenamefont {Blake},
  \citenamefont {Bleuler}, \citenamefont {Blomqvist}, \citenamefont {Blum},
  \citenamefont {Bolton}, \citenamefont {Briceno}, \citenamefont {Brooks},
  \citenamefont {Brownstein}, \citenamefont {{Buckley-Geer}}, \citenamefont
  {Burden}, \citenamefont {Burtin}, \citenamefont {Busca}, \citenamefont
  {Cahn}, \citenamefont {Cai}, \citenamefont {{Cardiel-Sas}}, \citenamefont
  {Carlberg}, \citenamefont {Carton}, \citenamefont {Casas}, \citenamefont
  {Castander}, \citenamefont {{Cervantes-Cota}}, \citenamefont {Claybaugh},
  \citenamefont {Close}, \citenamefont {Coker}, \citenamefont {Cole},
  \citenamefont {Comparat}, \citenamefont {Cooper}, \citenamefont {Cousinou},
  \citenamefont {Crocce}, \citenamefont {Cuby}, \citenamefont {Cunningham},
  \citenamefont {Davis}, \citenamefont {Dawson}, \citenamefont {{de la
  Macorra}}, \citenamefont {De~Vicente}, \citenamefont {Delubac}, \citenamefont
  {Derwent}, \citenamefont {Dey}, \citenamefont {Dhungana}, \citenamefont
  {Ding}, \citenamefont {Doel}, \citenamefont {Duan}, \citenamefont {Ealet},
  \citenamefont {Edelstein}, \citenamefont {Eftekharzadeh}, \citenamefont
  {Eisenstein}, \citenamefont {Elliott}, \citenamefont {Escoffier},
  \citenamefont {Evatt}, \citenamefont {Fagrelius}, \citenamefont {Fan},
  \citenamefont {Fanning}, \citenamefont {Farahi}, \citenamefont {Farihi},
  \citenamefont {Favole}, \citenamefont {Feng}, \citenamefont {Fernandez},
  \citenamefont {Findlay}, \citenamefont {Finkbeiner}, \citenamefont
  {Fitzpatrick}, \citenamefont {Flaugher}, \citenamefont {Flender},
  \citenamefont {{Font-Ribera}}, \citenamefont {{Forero-Romero}}, \citenamefont
  {Fosalba}, \citenamefont {Frenk}, \citenamefont {Fumagalli}, \citenamefont
  {Gaensicke}, \citenamefont {Gallo}, \citenamefont {{Garcia-Bellido}},
  \citenamefont {Gaztanaga}, \citenamefont {Fusillo}, \citenamefont {Gerard},
  \citenamefont {Gershkovich}, \citenamefont {Giannantonio}, \citenamefont
  {Gillet}, \citenamefont {{Gonzalez-de-Rivera}}, \citenamefont
  {{Gonzalez-Perez}}, \citenamefont {Gott}, \citenamefont {Graur},
  \citenamefont {Gutierrez}, \citenamefont {Guy}, \citenamefont {Habib},
  \citenamefont {Heetderks}, \citenamefont {Heetderks}, \citenamefont
  {Heitmann}, \citenamefont {Hellwing}, \citenamefont {Herrera}, \citenamefont
  {Ho}, \citenamefont {Holland}, \citenamefont {Honscheid}, \citenamefont
  {Huff}, \citenamefont {Hutchinson}, \citenamefont {Huterer}, \citenamefont
  {Hwang}, \citenamefont {Laguna}, \citenamefont {Ishikawa}, \citenamefont
  {Jacobs}, \citenamefont {Jeffrey}, \citenamefont {Jelinsky}, \citenamefont
  {Jennings}, \citenamefont {Jiang}, \citenamefont {Jimenez}, \citenamefont
  {Johnson}, \citenamefont {Joyce}, \citenamefont {Jullo}, \citenamefont
  {Juneau}, \citenamefont {Kama}, \citenamefont {Karcher}, \citenamefont
  {Karkar}, \citenamefont {Kehoe}, \citenamefont {Kennamer}, \citenamefont
  {Kent}, \citenamefont {Kilbinger}, \citenamefont {Kim}, \citenamefont
  {Kirkby}, \citenamefont {Kisner}, \citenamefont {Kitanidis}, \citenamefont
  {Kneib}, \citenamefont {Koposov}, \citenamefont {Kovacs}, \citenamefont
  {Koyama}, \citenamefont {Kremin}, \citenamefont {Kron}, \citenamefont
  {Kronig}, \citenamefont {{Kueter-Young}}, \citenamefont {Lacey},
  \citenamefont {Lafever}, \citenamefont {Lahav}, \citenamefont {Lambert},
  \citenamefont {Lampton}, \citenamefont {Landriau}, \citenamefont {Lang},
  \citenamefont {Lauer}, \citenamefont {Goff}, \citenamefont {Guillou},
  \citenamefont {Van~Suu}, \citenamefont {Lee}, \citenamefont {Lee},
  \citenamefont {Leitner}, \citenamefont {Lesser}, \citenamefont {Levi},
  \citenamefont {L'Huillier}, \citenamefont {Li}, \citenamefont {Liang},
  \citenamefont {Lin}, \citenamefont {Linder}, \citenamefont {Loebman},
  \citenamefont {Luki{\'c}}, \citenamefont {Ma}, \citenamefont {MacCrann},
  \citenamefont {Magneville}, \citenamefont {Makarem}, \citenamefont {Manera},
  \citenamefont {Manser}, \citenamefont {Marshall}, \citenamefont {Martini},
  \citenamefont {Massey}, \citenamefont {Matheson}, \citenamefont {McCauley},
  \citenamefont {McDonald}, \citenamefont {McGreer}, \citenamefont {Meisner},
  \citenamefont {Metcalfe}, \citenamefont {Miller}, \citenamefont {Miquel},
  \citenamefont {Moustakas}, \citenamefont {Myers}, \citenamefont {Naik},
  \citenamefont {Newman}, \citenamefont {Nichol}, \citenamefont {Nicola},
  \citenamefont {{da Costa}}, \citenamefont {Nie}, \citenamefont {Niz},
  \citenamefont {Norberg}, \citenamefont {Nord}, \citenamefont {Norman},
  \citenamefont {Nugent}, \citenamefont {O'Brien}, \citenamefont {Oh},
  \citenamefont {Olsen}, \citenamefont {Padilla}, \citenamefont {Padmanabhan},
  \citenamefont {Padmanabhan}, \citenamefont {{Palanque-Delabrouille}},
  \citenamefont {Palmese}, \citenamefont {Pappalardo}, \citenamefont
  {P{\^a}ris}, \citenamefont {Park}, \citenamefont {Patej}, \citenamefont
  {Peacock}, \citenamefont {Peiris}, \citenamefont {Peng}, \citenamefont
  {Percival}, \citenamefont {Perruchot}, \citenamefont {Pieri}, \citenamefont
  {Pogge}, \citenamefont {Pollack}, \citenamefont {Poppett}, \citenamefont
  {Prada}, \citenamefont {Prakash}, \citenamefont {Probst}, \citenamefont
  {Rabinowitz}, \citenamefont {Raichoor}, \citenamefont {Ree}, \citenamefont
  {Refregier}, \citenamefont {Regal}, \citenamefont {Reid}, \citenamefont
  {Reil}, \citenamefont {Rezaie}, \citenamefont {Rockosi}, \citenamefont {Roe},
  \citenamefont {Ronayette}, \citenamefont {Roodman}, \citenamefont {Ross},
  \citenamefont {Ross}, \citenamefont {Rossi}, \citenamefont {Rozo},
  \citenamefont {{Ruhlmann-Kleider}}, \citenamefont {Rykoff}, \citenamefont
  {Sabiu}, \citenamefont {Samushia}, \citenamefont {Sanchez}, \citenamefont
  {Sanchez}, \citenamefont {Schlegel}, \citenamefont {Schneider}, \citenamefont
  {Schubnell}, \citenamefont {Secroun}, \citenamefont {Seljak}, \citenamefont
  {Seo}, \citenamefont {Serrano}, \citenamefont {Shafieloo}, \citenamefont
  {Shan}, \citenamefont {Sharples}, \citenamefont {Sholl}, \citenamefont
  {Shourt}, \citenamefont {Silber}, \citenamefont {Silva}, \citenamefont
  {Sirk}, \citenamefont {Slosar}, \citenamefont {Smith}, \citenamefont {Smoot},
  \citenamefont {Som}, \citenamefont {Song}, \citenamefont {Sprayberry},
  \citenamefont {Staten}, \citenamefont {Stefanik}, \citenamefont {Tarle},
  \citenamefont {Tie}, \citenamefont {Tinker}, \citenamefont {Tojeiro},
  \citenamefont {Valdes}, \citenamefont {Valenzuela}, \citenamefont {Valluri},
  \citenamefont {{Vargas-Magana}}, \citenamefont {Verde}, \citenamefont
  {Walker}, \citenamefont {Wang}, \citenamefont {Wang}, \citenamefont {Weaver},
  \citenamefont {Weaverdyck}, \citenamefont {Wechsler}, \citenamefont
  {Weinberg}, \citenamefont {White}, \citenamefont {Yang}, \citenamefont
  {Yeche}, \citenamefont {Zhang}, \citenamefont {Zhao}, \citenamefont {Zheng},
  \citenamefont {Zhou}, \citenamefont {Zhou}, \citenamefont {Zhu},
  \citenamefont {Zou},\ and\ \citenamefont {Zu}}]{desicollaboration2016}%
  \BibitemOpen
  \bibfield  {author} {\bibinfo {author} {\bibfnamefont {D.}~\bibnamefont
  {Collaboration}}, \bibinfo {author} {\bibfnamefont {A.}~\bibnamefont
  {Aghamousa}}, \bibinfo {author} {\bibfnamefont {J.}~\bibnamefont {Aguilar}},
  \bibinfo {author} {\bibfnamefont {S.}~\bibnamefont {Ahlen}}, \bibinfo
  {author} {\bibfnamefont {S.}~\bibnamefont {Alam}}, \bibinfo {author}
  {\bibfnamefont {L.~E.}\ \bibnamefont {Allen}}, \bibinfo {author}
  {\bibfnamefont {C.~A.}\ \bibnamefont {Prieto}}, \bibinfo {author}
  {\bibfnamefont {J.}~\bibnamefont {Annis}}, \bibinfo {author} {\bibfnamefont
  {S.}~\bibnamefont {Bailey}}, \bibinfo {author} {\bibfnamefont
  {C.}~\bibnamefont {Balland}}, \bibinfo {author} {\bibfnamefont
  {O.}~\bibnamefont {Ballester}}, \bibinfo {author} {\bibfnamefont
  {C.}~\bibnamefont {Baltay}}, \bibinfo {author} {\bibfnamefont
  {L.}~\bibnamefont {Beaufore}}, \bibinfo {author} {\bibfnamefont
  {C.}~\bibnamefont {Bebek}}, \bibinfo {author} {\bibfnamefont {T.~C.}\
  \bibnamefont {Beers}}, \bibinfo {author} {\bibfnamefont {E.~F.}\ \bibnamefont
  {Bell}}, \bibinfo {author} {\bibfnamefont {J.~L.}\ \bibnamefont {Bernal}},
  \bibinfo {author} {\bibfnamefont {R.}~\bibnamefont {Besuner}}, \bibinfo
  {author} {\bibfnamefont {F.}~\bibnamefont {Beutler}}, \bibinfo {author}
  {\bibfnamefont {C.}~\bibnamefont {Blake}}, \bibinfo {author} {\bibfnamefont
  {H.}~\bibnamefont {Bleuler}}, \bibinfo {author} {\bibfnamefont
  {M.}~\bibnamefont {Blomqvist}}, \bibinfo {author} {\bibfnamefont
  {R.}~\bibnamefont {Blum}}, \bibinfo {author} {\bibfnamefont {A.~S.}\
  \bibnamefont {Bolton}}, \bibinfo {author} {\bibfnamefont {C.}~\bibnamefont
  {Briceno}}, \bibinfo {author} {\bibfnamefont {D.}~\bibnamefont {Brooks}},
  \bibinfo {author} {\bibfnamefont {J.~R.}\ \bibnamefont {Brownstein}},
  \bibinfo {author} {\bibfnamefont {E.}~\bibnamefont {{Buckley-Geer}}},
  \bibinfo {author} {\bibfnamefont {A.}~\bibnamefont {Burden}}, \bibinfo
  {author} {\bibfnamefont {E.}~\bibnamefont {Burtin}}, \bibinfo {author}
  {\bibfnamefont {N.~G.}\ \bibnamefont {Busca}}, \bibinfo {author}
  {\bibfnamefont {R.~N.}\ \bibnamefont {Cahn}}, \bibinfo {author}
  {\bibfnamefont {Y.-C.}\ \bibnamefont {Cai}}, \bibinfo {author} {\bibfnamefont
  {L.}~\bibnamefont {{Cardiel-Sas}}}, \bibinfo {author} {\bibfnamefont {R.~G.}\
  \bibnamefont {Carlberg}}, \bibinfo {author} {\bibfnamefont {P.-H.}\
  \bibnamefont {Carton}}, \bibinfo {author} {\bibfnamefont {R.}~\bibnamefont
  {Casas}}, \bibinfo {author} {\bibfnamefont {F.~J.}\ \bibnamefont
  {Castander}}, \bibinfo {author} {\bibfnamefont {J.~L.}\ \bibnamefont
  {{Cervantes-Cota}}}, \bibinfo {author} {\bibfnamefont {T.~M.}\ \bibnamefont
  {Claybaugh}}, \bibinfo {author} {\bibfnamefont {M.}~\bibnamefont {Close}},
  \bibinfo {author} {\bibfnamefont {C.~T.}\ \bibnamefont {Coker}}, \bibinfo
  {author} {\bibfnamefont {S.}~\bibnamefont {Cole}}, \bibinfo {author}
  {\bibfnamefont {J.}~\bibnamefont {Comparat}}, \bibinfo {author}
  {\bibfnamefont {A.~P.}\ \bibnamefont {Cooper}}, \bibinfo {author}
  {\bibfnamefont {M.-C.}\ \bibnamefont {Cousinou}}, \bibinfo {author}
  {\bibfnamefont {M.}~\bibnamefont {Crocce}}, \bibinfo {author} {\bibfnamefont
  {J.-G.}\ \bibnamefont {Cuby}}, \bibinfo {author} {\bibfnamefont {D.~P.}\
  \bibnamefont {Cunningham}}, \bibinfo {author} {\bibfnamefont {T.~M.}\
  \bibnamefont {Davis}}, \bibinfo {author} {\bibfnamefont {K.~S.}\ \bibnamefont
  {Dawson}}, \bibinfo {author} {\bibfnamefont {A.}~\bibnamefont {{de la
  Macorra}}}, \bibinfo {author} {\bibfnamefont {J.}~\bibnamefont {De~Vicente}},
  \bibinfo {author} {\bibfnamefont {T.}~\bibnamefont {Delubac}}, \bibinfo
  {author} {\bibfnamefont {M.}~\bibnamefont {Derwent}}, \bibinfo {author}
  {\bibfnamefont {A.}~\bibnamefont {Dey}}, \bibinfo {author} {\bibfnamefont
  {G.}~\bibnamefont {Dhungana}}, \bibinfo {author} {\bibfnamefont
  {Z.}~\bibnamefont {Ding}}, \bibinfo {author} {\bibfnamefont {P.}~\bibnamefont
  {Doel}}, \bibinfo {author} {\bibfnamefont {Y.~T.}\ \bibnamefont {Duan}},
  \bibinfo {author} {\bibfnamefont {A.}~\bibnamefont {Ealet}}, \bibinfo
  {author} {\bibfnamefont {J.}~\bibnamefont {Edelstein}}, \bibinfo {author}
  {\bibfnamefont {S.}~\bibnamefont {Eftekharzadeh}}, \bibinfo {author}
  {\bibfnamefont {D.~J.}\ \bibnamefont {Eisenstein}}, \bibinfo {author}
  {\bibfnamefont {A.}~\bibnamefont {Elliott}}, \bibinfo {author} {\bibfnamefont
  {S.}~\bibnamefont {Escoffier}}, \bibinfo {author} {\bibfnamefont
  {M.}~\bibnamefont {Evatt}}, \bibinfo {author} {\bibfnamefont
  {P.}~\bibnamefont {Fagrelius}}, \bibinfo {author} {\bibfnamefont
  {X.}~\bibnamefont {Fan}}, \bibinfo {author} {\bibfnamefont {K.}~\bibnamefont
  {Fanning}}, \bibinfo {author} {\bibfnamefont {A.}~\bibnamefont {Farahi}},
  \bibinfo {author} {\bibfnamefont {J.}~\bibnamefont {Farihi}}, \bibinfo
  {author} {\bibfnamefont {G.}~\bibnamefont {Favole}}, \bibinfo {author}
  {\bibfnamefont {Y.}~\bibnamefont {Feng}}, \bibinfo {author} {\bibfnamefont
  {E.}~\bibnamefont {Fernandez}}, \bibinfo {author} {\bibfnamefont {J.~R.}\
  \bibnamefont {Findlay}}, \bibinfo {author} {\bibfnamefont {D.~P.}\
  \bibnamefont {Finkbeiner}}, \bibinfo {author} {\bibfnamefont {M.~J.}\
  \bibnamefont {Fitzpatrick}}, \bibinfo {author} {\bibfnamefont
  {B.}~\bibnamefont {Flaugher}}, \bibinfo {author} {\bibfnamefont
  {S.}~\bibnamefont {Flender}}, \bibinfo {author} {\bibfnamefont
  {A.}~\bibnamefont {{Font-Ribera}}}, \bibinfo {author} {\bibfnamefont {J.~E.}\
  \bibnamefont {{Forero-Romero}}}, \bibinfo {author} {\bibfnamefont
  {P.}~\bibnamefont {Fosalba}}, \bibinfo {author} {\bibfnamefont {C.~S.}\
  \bibnamefont {Frenk}}, \bibinfo {author} {\bibfnamefont {M.}~\bibnamefont
  {Fumagalli}}, \bibinfo {author} {\bibfnamefont {B.~T.}\ \bibnamefont
  {Gaensicke}}, \bibinfo {author} {\bibfnamefont {G.}~\bibnamefont {Gallo}},
  \bibinfo {author} {\bibfnamefont {J.}~\bibnamefont {{Garcia-Bellido}}},
  \bibinfo {author} {\bibfnamefont {E.}~\bibnamefont {Gaztanaga}}, \bibinfo
  {author} {\bibfnamefont {N.~P.~G.}\ \bibnamefont {Fusillo}}, \bibinfo
  {author} {\bibfnamefont {T.}~\bibnamefont {Gerard}}, \bibinfo {author}
  {\bibfnamefont {I.}~\bibnamefont {Gershkovich}}, \bibinfo {author}
  {\bibfnamefont {T.}~\bibnamefont {Giannantonio}}, \bibinfo {author}
  {\bibfnamefont {D.}~\bibnamefont {Gillet}}, \bibinfo {author} {\bibfnamefont
  {G.}~\bibnamefont {{Gonzalez-de-Rivera}}}, \bibinfo {author} {\bibfnamefont
  {V.}~\bibnamefont {{Gonzalez-Perez}}}, \bibinfo {author} {\bibfnamefont
  {S.}~\bibnamefont {Gott}}, \bibinfo {author} {\bibfnamefont {O.}~\bibnamefont
  {Graur}}, \bibinfo {author} {\bibfnamefont {G.}~\bibnamefont {Gutierrez}},
  \bibinfo {author} {\bibfnamefont {J.}~\bibnamefont {Guy}}, \bibinfo {author}
  {\bibfnamefont {S.}~\bibnamefont {Habib}}, \bibinfo {author} {\bibfnamefont
  {H.}~\bibnamefont {Heetderks}}, \bibinfo {author} {\bibfnamefont
  {I.}~\bibnamefont {Heetderks}}, \bibinfo {author} {\bibfnamefont
  {K.}~\bibnamefont {Heitmann}}, \bibinfo {author} {\bibfnamefont {W.~A.}\
  \bibnamefont {Hellwing}}, \bibinfo {author} {\bibfnamefont {D.~A.}\
  \bibnamefont {Herrera}}, \bibinfo {author} {\bibfnamefont {S.}~\bibnamefont
  {Ho}}, \bibinfo {author} {\bibfnamefont {S.}~\bibnamefont {Holland}},
  \bibinfo {author} {\bibfnamefont {K.}~\bibnamefont {Honscheid}}, \bibinfo
  {author} {\bibfnamefont {E.}~\bibnamefont {Huff}}, \bibinfo {author}
  {\bibfnamefont {T.~A.}\ \bibnamefont {Hutchinson}}, \bibinfo {author}
  {\bibfnamefont {D.}~\bibnamefont {Huterer}}, \bibinfo {author} {\bibfnamefont
  {H.~S.}\ \bibnamefont {Hwang}}, \bibinfo {author} {\bibfnamefont {J.~M.~I.}\
  \bibnamefont {Laguna}}, \bibinfo {author} {\bibfnamefont {Y.}~\bibnamefont
  {Ishikawa}}, \bibinfo {author} {\bibfnamefont {D.}~\bibnamefont {Jacobs}},
  \bibinfo {author} {\bibfnamefont {N.}~\bibnamefont {Jeffrey}}, \bibinfo
  {author} {\bibfnamefont {P.}~\bibnamefont {Jelinsky}}, \bibinfo {author}
  {\bibfnamefont {E.}~\bibnamefont {Jennings}}, \bibinfo {author}
  {\bibfnamefont {L.}~\bibnamefont {Jiang}}, \bibinfo {author} {\bibfnamefont
  {J.}~\bibnamefont {Jimenez}}, \bibinfo {author} {\bibfnamefont
  {J.}~\bibnamefont {Johnson}}, \bibinfo {author} {\bibfnamefont
  {R.}~\bibnamefont {Joyce}}, \bibinfo {author} {\bibfnamefont
  {E.}~\bibnamefont {Jullo}}, \bibinfo {author} {\bibfnamefont
  {S.}~\bibnamefont {Juneau}}, \bibinfo {author} {\bibfnamefont
  {S.}~\bibnamefont {Kama}}, \bibinfo {author} {\bibfnamefont {A.}~\bibnamefont
  {Karcher}}, \bibinfo {author} {\bibfnamefont {S.}~\bibnamefont {Karkar}},
  \bibinfo {author} {\bibfnamefont {R.}~\bibnamefont {Kehoe}}, \bibinfo
  {author} {\bibfnamefont {N.}~\bibnamefont {Kennamer}}, \bibinfo {author}
  {\bibfnamefont {S.}~\bibnamefont {Kent}}, \bibinfo {author} {\bibfnamefont
  {M.}~\bibnamefont {Kilbinger}}, \bibinfo {author} {\bibfnamefont {A.~G.}\
  \bibnamefont {Kim}}, \bibinfo {author} {\bibfnamefont {D.}~\bibnamefont
  {Kirkby}}, \bibinfo {author} {\bibfnamefont {T.}~\bibnamefont {Kisner}},
  \bibinfo {author} {\bibfnamefont {E.}~\bibnamefont {Kitanidis}}, \bibinfo
  {author} {\bibfnamefont {J.-P.}\ \bibnamefont {Kneib}}, \bibinfo {author}
  {\bibfnamefont {S.}~\bibnamefont {Koposov}}, \bibinfo {author} {\bibfnamefont
  {E.}~\bibnamefont {Kovacs}}, \bibinfo {author} {\bibfnamefont
  {K.}~\bibnamefont {Koyama}}, \bibinfo {author} {\bibfnamefont
  {A.}~\bibnamefont {Kremin}}, \bibinfo {author} {\bibfnamefont
  {R.}~\bibnamefont {Kron}}, \bibinfo {author} {\bibfnamefont {L.}~\bibnamefont
  {Kronig}}, \bibinfo {author} {\bibfnamefont {A.}~\bibnamefont
  {{Kueter-Young}}}, \bibinfo {author} {\bibfnamefont {C.~G.}\ \bibnamefont
  {Lacey}}, \bibinfo {author} {\bibfnamefont {R.}~\bibnamefont {Lafever}},
  \bibinfo {author} {\bibfnamefont {O.}~\bibnamefont {Lahav}}, \bibinfo
  {author} {\bibfnamefont {A.}~\bibnamefont {Lambert}}, \bibinfo {author}
  {\bibfnamefont {M.}~\bibnamefont {Lampton}}, \bibinfo {author} {\bibfnamefont
  {M.}~\bibnamefont {Landriau}}, \bibinfo {author} {\bibfnamefont
  {D.}~\bibnamefont {Lang}}, \bibinfo {author} {\bibfnamefont {T.~R.}\
  \bibnamefont {Lauer}}, \bibinfo {author} {\bibfnamefont {J.-M.~L.}\
  \bibnamefont {Goff}}, \bibinfo {author} {\bibfnamefont {L.~L.}\ \bibnamefont
  {Guillou}}, \bibinfo {author} {\bibfnamefont {A.~L.}\ \bibnamefont
  {Van~Suu}}, \bibinfo {author} {\bibfnamefont {J.~H.}\ \bibnamefont {Lee}},
  \bibinfo {author} {\bibfnamefont {S.-J.}\ \bibnamefont {Lee}}, \bibinfo
  {author} {\bibfnamefont {D.}~\bibnamefont {Leitner}}, \bibinfo {author}
  {\bibfnamefont {M.}~\bibnamefont {Lesser}}, \bibinfo {author} {\bibfnamefont
  {M.~E.}\ \bibnamefont {Levi}}, \bibinfo {author} {\bibfnamefont
  {B.}~\bibnamefont {L'Huillier}}, \bibinfo {author} {\bibfnamefont
  {B.}~\bibnamefont {Li}}, \bibinfo {author} {\bibfnamefont {M.}~\bibnamefont
  {Liang}}, \bibinfo {author} {\bibfnamefont {H.}~\bibnamefont {Lin}}, \bibinfo
  {author} {\bibfnamefont {E.}~\bibnamefont {Linder}}, \bibinfo {author}
  {\bibfnamefont {S.~R.}\ \bibnamefont {Loebman}}, \bibinfo {author}
  {\bibfnamefont {Z.}~\bibnamefont {Luki{\'c}}}, \bibinfo {author}
  {\bibfnamefont {J.}~\bibnamefont {Ma}}, \bibinfo {author} {\bibfnamefont
  {N.}~\bibnamefont {MacCrann}}, \bibinfo {author} {\bibfnamefont
  {C.}~\bibnamefont {Magneville}}, \bibinfo {author} {\bibfnamefont
  {L.}~\bibnamefont {Makarem}}, \bibinfo {author} {\bibfnamefont
  {M.}~\bibnamefont {Manera}}, \bibinfo {author} {\bibfnamefont {C.~J.}\
  \bibnamefont {Manser}}, \bibinfo {author} {\bibfnamefont {R.}~\bibnamefont
  {Marshall}}, \bibinfo {author} {\bibfnamefont {P.}~\bibnamefont {Martini}},
  \bibinfo {author} {\bibfnamefont {R.}~\bibnamefont {Massey}}, \bibinfo
  {author} {\bibfnamefont {T.}~\bibnamefont {Matheson}}, \bibinfo {author}
  {\bibfnamefont {J.}~\bibnamefont {McCauley}}, \bibinfo {author}
  {\bibfnamefont {P.}~\bibnamefont {McDonald}}, \bibinfo {author}
  {\bibfnamefont {I.~D.}\ \bibnamefont {McGreer}}, \bibinfo {author}
  {\bibfnamefont {A.}~\bibnamefont {Meisner}}, \bibinfo {author} {\bibfnamefont
  {N.}~\bibnamefont {Metcalfe}}, \bibinfo {author} {\bibfnamefont {T.~N.}\
  \bibnamefont {Miller}}, \bibinfo {author} {\bibfnamefont {R.}~\bibnamefont
  {Miquel}}, \bibinfo {author} {\bibfnamefont {J.}~\bibnamefont {Moustakas}},
  \bibinfo {author} {\bibfnamefont {A.}~\bibnamefont {Myers}}, \bibinfo
  {author} {\bibfnamefont {M.}~\bibnamefont {Naik}}, \bibinfo {author}
  {\bibfnamefont {J.~A.}\ \bibnamefont {Newman}}, \bibinfo {author}
  {\bibfnamefont {R.~C.}\ \bibnamefont {Nichol}}, \bibinfo {author}
  {\bibfnamefont {A.}~\bibnamefont {Nicola}}, \bibinfo {author} {\bibfnamefont
  {L.~N.}\ \bibnamefont {{da Costa}}}, \bibinfo {author} {\bibfnamefont
  {J.}~\bibnamefont {Nie}}, \bibinfo {author} {\bibfnamefont {G.}~\bibnamefont
  {Niz}}, \bibinfo {author} {\bibfnamefont {P.}~\bibnamefont {Norberg}},
  \bibinfo {author} {\bibfnamefont {B.}~\bibnamefont {Nord}}, \bibinfo {author}
  {\bibfnamefont {D.}~\bibnamefont {Norman}}, \bibinfo {author} {\bibfnamefont
  {P.}~\bibnamefont {Nugent}}, \bibinfo {author} {\bibfnamefont
  {T.}~\bibnamefont {O'Brien}}, \bibinfo {author} {\bibfnamefont
  {M.}~\bibnamefont {Oh}}, \bibinfo {author} {\bibfnamefont {K.~A.~G.}\
  \bibnamefont {Olsen}}, \bibinfo {author} {\bibfnamefont {C.}~\bibnamefont
  {Padilla}}, \bibinfo {author} {\bibfnamefont {H.}~\bibnamefont
  {Padmanabhan}}, \bibinfo {author} {\bibfnamefont {N.}~\bibnamefont
  {Padmanabhan}}, \bibinfo {author} {\bibfnamefont {N.}~\bibnamefont
  {{Palanque-Delabrouille}}}, \bibinfo {author} {\bibfnamefont
  {A.}~\bibnamefont {Palmese}}, \bibinfo {author} {\bibfnamefont
  {D.}~\bibnamefont {Pappalardo}}, \bibinfo {author} {\bibfnamefont
  {I.}~\bibnamefont {P{\^a}ris}}, \bibinfo {author} {\bibfnamefont
  {C.}~\bibnamefont {Park}}, \bibinfo {author} {\bibfnamefont {A.}~\bibnamefont
  {Patej}}, \bibinfo {author} {\bibfnamefont {J.~A.}\ \bibnamefont {Peacock}},
  \bibinfo {author} {\bibfnamefont {H.~V.}\ \bibnamefont {Peiris}}, \bibinfo
  {author} {\bibfnamefont {X.}~\bibnamefont {Peng}}, \bibinfo {author}
  {\bibfnamefont {W.~J.}\ \bibnamefont {Percival}}, \bibinfo {author}
  {\bibfnamefont {S.}~\bibnamefont {Perruchot}}, \bibinfo {author}
  {\bibfnamefont {M.~M.}\ \bibnamefont {Pieri}}, \bibinfo {author}
  {\bibfnamefont {R.}~\bibnamefont {Pogge}}, \bibinfo {author} {\bibfnamefont
  {J.~E.}\ \bibnamefont {Pollack}}, \bibinfo {author} {\bibfnamefont
  {C.}~\bibnamefont {Poppett}}, \bibinfo {author} {\bibfnamefont
  {F.}~\bibnamefont {Prada}}, \bibinfo {author} {\bibfnamefont
  {A.}~\bibnamefont {Prakash}}, \bibinfo {author} {\bibfnamefont {R.~G.}\
  \bibnamefont {Probst}}, \bibinfo {author} {\bibfnamefont {D.}~\bibnamefont
  {Rabinowitz}}, \bibinfo {author} {\bibfnamefont {A.}~\bibnamefont
  {Raichoor}}, \bibinfo {author} {\bibfnamefont {C.~H.}\ \bibnamefont {Ree}},
  \bibinfo {author} {\bibfnamefont {A.}~\bibnamefont {Refregier}}, \bibinfo
  {author} {\bibfnamefont {X.}~\bibnamefont {Regal}}, \bibinfo {author}
  {\bibfnamefont {B.}~\bibnamefont {Reid}}, \bibinfo {author} {\bibfnamefont
  {K.}~\bibnamefont {Reil}}, \bibinfo {author} {\bibfnamefont {M.}~\bibnamefont
  {Rezaie}}, \bibinfo {author} {\bibfnamefont {C.~M.}\ \bibnamefont {Rockosi}},
  \bibinfo {author} {\bibfnamefont {N.}~\bibnamefont {Roe}}, \bibinfo {author}
  {\bibfnamefont {S.}~\bibnamefont {Ronayette}}, \bibinfo {author}
  {\bibfnamefont {A.}~\bibnamefont {Roodman}}, \bibinfo {author} {\bibfnamefont
  {A.~J.}\ \bibnamefont {Ross}}, \bibinfo {author} {\bibfnamefont {N.~P.}\
  \bibnamefont {Ross}}, \bibinfo {author} {\bibfnamefont {G.}~\bibnamefont
  {Rossi}}, \bibinfo {author} {\bibfnamefont {E.}~\bibnamefont {Rozo}},
  \bibinfo {author} {\bibfnamefont {V.}~\bibnamefont {{Ruhlmann-Kleider}}},
  \bibinfo {author} {\bibfnamefont {E.~S.}\ \bibnamefont {Rykoff}}, \bibinfo
  {author} {\bibfnamefont {C.}~\bibnamefont {Sabiu}}, \bibinfo {author}
  {\bibfnamefont {L.}~\bibnamefont {Samushia}}, \bibinfo {author}
  {\bibfnamefont {E.}~\bibnamefont {Sanchez}}, \bibinfo {author} {\bibfnamefont
  {J.}~\bibnamefont {Sanchez}}, \bibinfo {author} {\bibfnamefont {D.~J.}\
  \bibnamefont {Schlegel}}, \bibinfo {author} {\bibfnamefont {M.}~\bibnamefont
  {Schneider}}, \bibinfo {author} {\bibfnamefont {M.}~\bibnamefont
  {Schubnell}}, \bibinfo {author} {\bibfnamefont {A.}~\bibnamefont {Secroun}},
  \bibinfo {author} {\bibfnamefont {U.}~\bibnamefont {Seljak}}, \bibinfo
  {author} {\bibfnamefont {H.-J.}\ \bibnamefont {Seo}}, \bibinfo {author}
  {\bibfnamefont {S.}~\bibnamefont {Serrano}}, \bibinfo {author} {\bibfnamefont
  {A.}~\bibnamefont {Shafieloo}}, \bibinfo {author} {\bibfnamefont
  {H.}~\bibnamefont {Shan}}, \bibinfo {author} {\bibfnamefont {R.}~\bibnamefont
  {Sharples}}, \bibinfo {author} {\bibfnamefont {M.~J.}\ \bibnamefont {Sholl}},
  \bibinfo {author} {\bibfnamefont {W.~V.}\ \bibnamefont {Shourt}}, \bibinfo
  {author} {\bibfnamefont {J.~H.}\ \bibnamefont {Silber}}, \bibinfo {author}
  {\bibfnamefont {D.~R.}\ \bibnamefont {Silva}}, \bibinfo {author}
  {\bibfnamefont {M.~M.}\ \bibnamefont {Sirk}}, \bibinfo {author}
  {\bibfnamefont {A.}~\bibnamefont {Slosar}}, \bibinfo {author} {\bibfnamefont
  {A.}~\bibnamefont {Smith}}, \bibinfo {author} {\bibfnamefont {G.~F.}\
  \bibnamefont {Smoot}}, \bibinfo {author} {\bibfnamefont {D.}~\bibnamefont
  {Som}}, \bibinfo {author} {\bibfnamefont {Y.-S.}\ \bibnamefont {Song}},
  \bibinfo {author} {\bibfnamefont {D.}~\bibnamefont {Sprayberry}}, \bibinfo
  {author} {\bibfnamefont {R.}~\bibnamefont {Staten}}, \bibinfo {author}
  {\bibfnamefont {A.}~\bibnamefont {Stefanik}}, \bibinfo {author}
  {\bibfnamefont {G.}~\bibnamefont {Tarle}}, \bibinfo {author} {\bibfnamefont
  {S.~S.}\ \bibnamefont {Tie}}, \bibinfo {author} {\bibfnamefont {J.~L.}\
  \bibnamefont {Tinker}}, \bibinfo {author} {\bibfnamefont {R.}~\bibnamefont
  {Tojeiro}}, \bibinfo {author} {\bibfnamefont {F.}~\bibnamefont {Valdes}},
  \bibinfo {author} {\bibfnamefont {O.}~\bibnamefont {Valenzuela}}, \bibinfo
  {author} {\bibfnamefont {M.}~\bibnamefont {Valluri}}, \bibinfo {author}
  {\bibfnamefont {M.}~\bibnamefont {{Vargas-Magana}}}, \bibinfo {author}
  {\bibfnamefont {L.}~\bibnamefont {Verde}}, \bibinfo {author} {\bibfnamefont
  {A.~R.}\ \bibnamefont {Walker}}, \bibinfo {author} {\bibfnamefont
  {J.}~\bibnamefont {Wang}}, \bibinfo {author} {\bibfnamefont {Y.}~\bibnamefont
  {Wang}}, \bibinfo {author} {\bibfnamefont {B.~A.}\ \bibnamefont {Weaver}},
  \bibinfo {author} {\bibfnamefont {C.}~\bibnamefont {Weaverdyck}}, \bibinfo
  {author} {\bibfnamefont {R.~H.}\ \bibnamefont {Wechsler}}, \bibinfo {author}
  {\bibfnamefont {D.~H.}\ \bibnamefont {Weinberg}}, \bibinfo {author}
  {\bibfnamefont {M.}~\bibnamefont {White}}, \bibinfo {author} {\bibfnamefont
  {Q.}~\bibnamefont {Yang}}, \bibinfo {author} {\bibfnamefont {C.}~\bibnamefont
  {Yeche}}, \bibinfo {author} {\bibfnamefont {T.}~\bibnamefont {Zhang}},
  \bibinfo {author} {\bibfnamefont {G.-B.}\ \bibnamefont {Zhao}}, \bibinfo
  {author} {\bibfnamefont {Y.}~\bibnamefont {Zheng}}, \bibinfo {author}
  {\bibfnamefont {X.}~\bibnamefont {Zhou}}, \bibinfo {author} {\bibfnamefont
  {Z.}~\bibnamefont {Zhou}}, \bibinfo {author} {\bibfnamefont {Y.}~\bibnamefont
  {Zhu}}, \bibinfo {author} {\bibfnamefont {H.}~\bibnamefont {Zou}}, \ and\
  \bibinfo {author} {\bibfnamefont {Y.}~\bibnamefont {Zu}},\ }\href@noop {}
  {\bibfield  {journal} {\bibinfo  {journal} {arXiv:1611.00036 [astro-ph]}\ }
  (\bibinfo {year} {2016}{\natexlab{a}})},\ \Eprint
  {http://arxiv.org/abs/1611.00036} {arXiv:1611.00036 [astro-ph]} \BibitemShut
  {NoStop}%
\bibitem [{\citenamefont {Collaboration}\ \emph
  {et~al.}(2016{\natexlab{b}})\citenamefont {Collaboration}, \citenamefont
  {Aghamousa}, \citenamefont {Aguilar}, \citenamefont {Ahlen}, \citenamefont
  {Alam}, \citenamefont {Allen}, \citenamefont {Prieto}, \citenamefont {Annis},
  \citenamefont {Bailey}, \citenamefont {Balland}, \citenamefont {Ballester},
  \citenamefont {Baltay}, \citenamefont {Beaufore}, \citenamefont {Bebek},
  \citenamefont {Beers}, \citenamefont {Bell}, \citenamefont {Bernal},
  \citenamefont {Besuner}, \citenamefont {Beutler}, \citenamefont {Blake},
  \citenamefont {Bleuler}, \citenamefont {Blomqvist}, \citenamefont {Blum},
  \citenamefont {Bolton}, \citenamefont {Briceno}, \citenamefont {Brooks},
  \citenamefont {Brownstein}, \citenamefont {{Buckley-Geer}}, \citenamefont
  {Burden}, \citenamefont {Burtin}, \citenamefont {Busca}, \citenamefont
  {Cahn}, \citenamefont {Cai}, \citenamefont {{Cardiel-Sas}}, \citenamefont
  {Carlberg}, \citenamefont {Carton}, \citenamefont {Casas}, \citenamefont
  {Castander}, \citenamefont {{Cervantes-Cota}}, \citenamefont {Claybaugh},
  \citenamefont {Close}, \citenamefont {Coker}, \citenamefont {Cole},
  \citenamefont {Comparat}, \citenamefont {Cooper}, \citenamefont {Cousinou},
  \citenamefont {Crocce}, \citenamefont {Cuby}, \citenamefont {Cunningham},
  \citenamefont {Davis}, \citenamefont {Dawson}, \citenamefont {{de la
  Macorra}}, \citenamefont {De~Vicente}, \citenamefont {Delubac}, \citenamefont
  {Derwent}, \citenamefont {Dey}, \citenamefont {Dhungana}, \citenamefont
  {Ding}, \citenamefont {Doel}, \citenamefont {Duan}, \citenamefont {Ealet},
  \citenamefont {Edelstein}, \citenamefont {Eftekharzadeh}, \citenamefont
  {Eisenstein}, \citenamefont {Elliott}, \citenamefont {Escoffier},
  \citenamefont {Evatt}, \citenamefont {Fagrelius}, \citenamefont {Fan},
  \citenamefont {Fanning}, \citenamefont {Farahi}, \citenamefont {Farihi},
  \citenamefont {Favole}, \citenamefont {Feng}, \citenamefont {Fernandez},
  \citenamefont {Findlay}, \citenamefont {Finkbeiner}, \citenamefont
  {Fitzpatrick}, \citenamefont {Flaugher}, \citenamefont {Flender},
  \citenamefont {{Font-Ribera}}, \citenamefont {{Forero-Romero}}, \citenamefont
  {Fosalba}, \citenamefont {Frenk}, \citenamefont {Fumagalli}, \citenamefont
  {Gaensicke}, \citenamefont {Gallo}, \citenamefont {{Garcia-Bellido}},
  \citenamefont {Gaztanaga}, \citenamefont {Fusillo}, \citenamefont {Gerard},
  \citenamefont {Gershkovich}, \citenamefont {Giannantonio}, \citenamefont
  {Gillet}, \citenamefont {{Gonzalez-de-Rivera}}, \citenamefont
  {{Gonzalez-Perez}}, \citenamefont {Gott}, \citenamefont {Graur},
  \citenamefont {Gutierrez}, \citenamefont {Guy}, \citenamefont {Habib},
  \citenamefont {Heetderks}, \citenamefont {Heetderks}, \citenamefont
  {Heitmann}, \citenamefont {Hellwing}, \citenamefont {Herrera}, \citenamefont
  {Ho}, \citenamefont {Holland}, \citenamefont {Honscheid}, \citenamefont
  {Huff}, \citenamefont {Hutchinson}, \citenamefont {Huterer}, \citenamefont
  {Hwang}, \citenamefont {Laguna}, \citenamefont {Ishikawa}, \citenamefont
  {Jacobs}, \citenamefont {Jeffrey}, \citenamefont {Jelinsky}, \citenamefont
  {Jennings}, \citenamefont {Jiang}, \citenamefont {Jimenez}, \citenamefont
  {Johnson}, \citenamefont {Joyce}, \citenamefont {Jullo}, \citenamefont
  {Juneau}, \citenamefont {Kama}, \citenamefont {Karcher}, \citenamefont
  {Karkar}, \citenamefont {Kehoe}, \citenamefont {Kennamer}, \citenamefont
  {Kent}, \citenamefont {Kilbinger}, \citenamefont {Kim}, \citenamefont
  {Kirkby}, \citenamefont {Kisner}, \citenamefont {Kitanidis}, \citenamefont
  {Kneib}, \citenamefont {Koposov}, \citenamefont {Kovacs}, \citenamefont
  {Koyama}, \citenamefont {Kremin}, \citenamefont {Kron}, \citenamefont
  {Kronig}, \citenamefont {{Kueter-Young}}, \citenamefont {Lacey},
  \citenamefont {Lafever}, \citenamefont {Lahav}, \citenamefont {Lambert},
  \citenamefont {Lampton}, \citenamefont {Landriau}, \citenamefont {Lang},
  \citenamefont {Lauer}, \citenamefont {Goff}, \citenamefont {Guillou},
  \citenamefont {Van~Suu}, \citenamefont {Lee}, \citenamefont {Lee},
  \citenamefont {Leitner}, \citenamefont {Lesser}, \citenamefont {Levi},
  \citenamefont {L'Huillier}, \citenamefont {Li}, \citenamefont {Liang},
  \citenamefont {Lin}, \citenamefont {Linder}, \citenamefont {Loebman},
  \citenamefont {Luki{\'c}}, \citenamefont {Ma}, \citenamefont {MacCrann},
  \citenamefont {Magneville}, \citenamefont {Makarem}, \citenamefont {Manera},
  \citenamefont {Manser}, \citenamefont {Marshall}, \citenamefont {Martini},
  \citenamefont {Massey}, \citenamefont {Matheson}, \citenamefont {McCauley},
  \citenamefont {McDonald}, \citenamefont {McGreer}, \citenamefont {Meisner},
  \citenamefont {Metcalfe}, \citenamefont {Miller}, \citenamefont {Miquel},
  \citenamefont {Moustakas}, \citenamefont {Myers}, \citenamefont {Naik},
  \citenamefont {Newman}, \citenamefont {Nichol}, \citenamefont {Nicola},
  \citenamefont {{da Costa}}, \citenamefont {Nie}, \citenamefont {Niz},
  \citenamefont {Norberg}, \citenamefont {Nord}, \citenamefont {Norman},
  \citenamefont {Nugent}, \citenamefont {O'Brien}, \citenamefont {Oh},
  \citenamefont {Olsen}, \citenamefont {Padilla}, \citenamefont {Padmanabhan},
  \citenamefont {Padmanabhan}, \citenamefont {{Palanque-Delabrouille}},
  \citenamefont {Palmese}, \citenamefont {Pappalardo}, \citenamefont
  {P{\^a}ris}, \citenamefont {Park}, \citenamefont {Patej}, \citenamefont
  {Peacock}, \citenamefont {Peiris}, \citenamefont {Peng}, \citenamefont
  {Percival}, \citenamefont {Perruchot}, \citenamefont {Pieri}, \citenamefont
  {Pogge}, \citenamefont {Pollack}, \citenamefont {Poppett}, \citenamefont
  {Prada}, \citenamefont {Prakash}, \citenamefont {Probst}, \citenamefont
  {Rabinowitz}, \citenamefont {Raichoor}, \citenamefont {Ree}, \citenamefont
  {Refregier}, \citenamefont {Regal}, \citenamefont {Reid}, \citenamefont
  {Reil}, \citenamefont {Rezaie}, \citenamefont {Rockosi}, \citenamefont {Roe},
  \citenamefont {Ronayette}, \citenamefont {Roodman}, \citenamefont {Ross},
  \citenamefont {Ross}, \citenamefont {Rossi}, \citenamefont {Rozo},
  \citenamefont {{Ruhlmann-Kleider}}, \citenamefont {Rykoff}, \citenamefont
  {Sabiu}, \citenamefont {Samushia}, \citenamefont {Sanchez}, \citenamefont
  {Sanchez}, \citenamefont {Schlegel}, \citenamefont {Schneider}, \citenamefont
  {Schubnell}, \citenamefont {Secroun}, \citenamefont {Seljak}, \citenamefont
  {Seo}, \citenamefont {Serrano}, \citenamefont {Shafieloo}, \citenamefont
  {Shan}, \citenamefont {Sharples}, \citenamefont {Sholl}, \citenamefont
  {Shourt}, \citenamefont {Silber}, \citenamefont {Silva}, \citenamefont
  {Sirk}, \citenamefont {Slosar}, \citenamefont {Smith}, \citenamefont {Smoot},
  \citenamefont {Som}, \citenamefont {Song}, \citenamefont {Sprayberry},
  \citenamefont {Staten}, \citenamefont {Stefanik}, \citenamefont {Tarle},
  \citenamefont {Tie}, \citenamefont {Tinker}, \citenamefont {Tojeiro},
  \citenamefont {Valdes}, \citenamefont {Valenzuela}, \citenamefont {Valluri},
  \citenamefont {{Vargas-Magana}}, \citenamefont {Verde}, \citenamefont
  {Walker}, \citenamefont {Wang}, \citenamefont {Wang}, \citenamefont {Weaver},
  \citenamefont {Weaverdyck}, \citenamefont {Wechsler}, \citenamefont
  {Weinberg}, \citenamefont {White}, \citenamefont {Yang}, \citenamefont
  {Yeche}, \citenamefont {Zhang}, \citenamefont {Zhao}, \citenamefont {Zheng},
  \citenamefont {Zhou}, \citenamefont {Zhou}, \citenamefont {Zhu},
  \citenamefont {Zou},\ and\ \citenamefont {Zu}}]{desicollaboration2016a}%
  \BibitemOpen
  \bibfield  {author} {\bibinfo {author} {\bibfnamefont {D.}~\bibnamefont
  {Collaboration}}, \bibinfo {author} {\bibfnamefont {A.}~\bibnamefont
  {Aghamousa}}, \bibinfo {author} {\bibfnamefont {J.}~\bibnamefont {Aguilar}},
  \bibinfo {author} {\bibfnamefont {S.}~\bibnamefont {Ahlen}}, \bibinfo
  {author} {\bibfnamefont {S.}~\bibnamefont {Alam}}, \bibinfo {author}
  {\bibfnamefont {L.~E.}\ \bibnamefont {Allen}}, \bibinfo {author}
  {\bibfnamefont {C.~A.}\ \bibnamefont {Prieto}}, \bibinfo {author}
  {\bibfnamefont {J.}~\bibnamefont {Annis}}, \bibinfo {author} {\bibfnamefont
  {S.}~\bibnamefont {Bailey}}, \bibinfo {author} {\bibfnamefont
  {C.}~\bibnamefont {Balland}}, \bibinfo {author} {\bibfnamefont
  {O.}~\bibnamefont {Ballester}}, \bibinfo {author} {\bibfnamefont
  {C.}~\bibnamefont {Baltay}}, \bibinfo {author} {\bibfnamefont
  {L.}~\bibnamefont {Beaufore}}, \bibinfo {author} {\bibfnamefont
  {C.}~\bibnamefont {Bebek}}, \bibinfo {author} {\bibfnamefont {T.~C.}\
  \bibnamefont {Beers}}, \bibinfo {author} {\bibfnamefont {E.~F.}\ \bibnamefont
  {Bell}}, \bibinfo {author} {\bibfnamefont {J.~L.}\ \bibnamefont {Bernal}},
  \bibinfo {author} {\bibfnamefont {R.}~\bibnamefont {Besuner}}, \bibinfo
  {author} {\bibfnamefont {F.}~\bibnamefont {Beutler}}, \bibinfo {author}
  {\bibfnamefont {C.}~\bibnamefont {Blake}}, \bibinfo {author} {\bibfnamefont
  {H.}~\bibnamefont {Bleuler}}, \bibinfo {author} {\bibfnamefont
  {M.}~\bibnamefont {Blomqvist}}, \bibinfo {author} {\bibfnamefont
  {R.}~\bibnamefont {Blum}}, \bibinfo {author} {\bibfnamefont {A.~S.}\
  \bibnamefont {Bolton}}, \bibinfo {author} {\bibfnamefont {C.}~\bibnamefont
  {Briceno}}, \bibinfo {author} {\bibfnamefont {D.}~\bibnamefont {Brooks}},
  \bibinfo {author} {\bibfnamefont {J.~R.}\ \bibnamefont {Brownstein}},
  \bibinfo {author} {\bibfnamefont {E.}~\bibnamefont {{Buckley-Geer}}},
  \bibinfo {author} {\bibfnamefont {A.}~\bibnamefont {Burden}}, \bibinfo
  {author} {\bibfnamefont {E.}~\bibnamefont {Burtin}}, \bibinfo {author}
  {\bibfnamefont {N.~G.}\ \bibnamefont {Busca}}, \bibinfo {author}
  {\bibfnamefont {R.~N.}\ \bibnamefont {Cahn}}, \bibinfo {author}
  {\bibfnamefont {Y.-C.}\ \bibnamefont {Cai}}, \bibinfo {author} {\bibfnamefont
  {L.}~\bibnamefont {{Cardiel-Sas}}}, \bibinfo {author} {\bibfnamefont {R.~G.}\
  \bibnamefont {Carlberg}}, \bibinfo {author} {\bibfnamefont {P.-H.}\
  \bibnamefont {Carton}}, \bibinfo {author} {\bibfnamefont {R.}~\bibnamefont
  {Casas}}, \bibinfo {author} {\bibfnamefont {F.~J.}\ \bibnamefont
  {Castander}}, \bibinfo {author} {\bibfnamefont {J.~L.}\ \bibnamefont
  {{Cervantes-Cota}}}, \bibinfo {author} {\bibfnamefont {T.~M.}\ \bibnamefont
  {Claybaugh}}, \bibinfo {author} {\bibfnamefont {M.}~\bibnamefont {Close}},
  \bibinfo {author} {\bibfnamefont {C.~T.}\ \bibnamefont {Coker}}, \bibinfo
  {author} {\bibfnamefont {S.}~\bibnamefont {Cole}}, \bibinfo {author}
  {\bibfnamefont {J.}~\bibnamefont {Comparat}}, \bibinfo {author}
  {\bibfnamefont {A.~P.}\ \bibnamefont {Cooper}}, \bibinfo {author}
  {\bibfnamefont {M.-C.}\ \bibnamefont {Cousinou}}, \bibinfo {author}
  {\bibfnamefont {M.}~\bibnamefont {Crocce}}, \bibinfo {author} {\bibfnamefont
  {J.-G.}\ \bibnamefont {Cuby}}, \bibinfo {author} {\bibfnamefont {D.~P.}\
  \bibnamefont {Cunningham}}, \bibinfo {author} {\bibfnamefont {T.~M.}\
  \bibnamefont {Davis}}, \bibinfo {author} {\bibfnamefont {K.~S.}\ \bibnamefont
  {Dawson}}, \bibinfo {author} {\bibfnamefont {A.}~\bibnamefont {{de la
  Macorra}}}, \bibinfo {author} {\bibfnamefont {J.}~\bibnamefont {De~Vicente}},
  \bibinfo {author} {\bibfnamefont {T.}~\bibnamefont {Delubac}}, \bibinfo
  {author} {\bibfnamefont {M.}~\bibnamefont {Derwent}}, \bibinfo {author}
  {\bibfnamefont {A.}~\bibnamefont {Dey}}, \bibinfo {author} {\bibfnamefont
  {G.}~\bibnamefont {Dhungana}}, \bibinfo {author} {\bibfnamefont
  {Z.}~\bibnamefont {Ding}}, \bibinfo {author} {\bibfnamefont {P.}~\bibnamefont
  {Doel}}, \bibinfo {author} {\bibfnamefont {Y.~T.}\ \bibnamefont {Duan}},
  \bibinfo {author} {\bibfnamefont {A.}~\bibnamefont {Ealet}}, \bibinfo
  {author} {\bibfnamefont {J.}~\bibnamefont {Edelstein}}, \bibinfo {author}
  {\bibfnamefont {S.}~\bibnamefont {Eftekharzadeh}}, \bibinfo {author}
  {\bibfnamefont {D.~J.}\ \bibnamefont {Eisenstein}}, \bibinfo {author}
  {\bibfnamefont {A.}~\bibnamefont {Elliott}}, \bibinfo {author} {\bibfnamefont
  {S.}~\bibnamefont {Escoffier}}, \bibinfo {author} {\bibfnamefont
  {M.}~\bibnamefont {Evatt}}, \bibinfo {author} {\bibfnamefont
  {P.}~\bibnamefont {Fagrelius}}, \bibinfo {author} {\bibfnamefont
  {X.}~\bibnamefont {Fan}}, \bibinfo {author} {\bibfnamefont {K.}~\bibnamefont
  {Fanning}}, \bibinfo {author} {\bibfnamefont {A.}~\bibnamefont {Farahi}},
  \bibinfo {author} {\bibfnamefont {J.}~\bibnamefont {Farihi}}, \bibinfo
  {author} {\bibfnamefont {G.}~\bibnamefont {Favole}}, \bibinfo {author}
  {\bibfnamefont {Y.}~\bibnamefont {Feng}}, \bibinfo {author} {\bibfnamefont
  {E.}~\bibnamefont {Fernandez}}, \bibinfo {author} {\bibfnamefont {J.~R.}\
  \bibnamefont {Findlay}}, \bibinfo {author} {\bibfnamefont {D.~P.}\
  \bibnamefont {Finkbeiner}}, \bibinfo {author} {\bibfnamefont {M.~J.}\
  \bibnamefont {Fitzpatrick}}, \bibinfo {author} {\bibfnamefont
  {B.}~\bibnamefont {Flaugher}}, \bibinfo {author} {\bibfnamefont
  {S.}~\bibnamefont {Flender}}, \bibinfo {author} {\bibfnamefont
  {A.}~\bibnamefont {{Font-Ribera}}}, \bibinfo {author} {\bibfnamefont {J.~E.}\
  \bibnamefont {{Forero-Romero}}}, \bibinfo {author} {\bibfnamefont
  {P.}~\bibnamefont {Fosalba}}, \bibinfo {author} {\bibfnamefont {C.~S.}\
  \bibnamefont {Frenk}}, \bibinfo {author} {\bibfnamefont {M.}~\bibnamefont
  {Fumagalli}}, \bibinfo {author} {\bibfnamefont {B.~T.}\ \bibnamefont
  {Gaensicke}}, \bibinfo {author} {\bibfnamefont {G.}~\bibnamefont {Gallo}},
  \bibinfo {author} {\bibfnamefont {J.}~\bibnamefont {{Garcia-Bellido}}},
  \bibinfo {author} {\bibfnamefont {E.}~\bibnamefont {Gaztanaga}}, \bibinfo
  {author} {\bibfnamefont {N.~P.~G.}\ \bibnamefont {Fusillo}}, \bibinfo
  {author} {\bibfnamefont {T.}~\bibnamefont {Gerard}}, \bibinfo {author}
  {\bibfnamefont {I.}~\bibnamefont {Gershkovich}}, \bibinfo {author}
  {\bibfnamefont {T.}~\bibnamefont {Giannantonio}}, \bibinfo {author}
  {\bibfnamefont {D.}~\bibnamefont {Gillet}}, \bibinfo {author} {\bibfnamefont
  {G.}~\bibnamefont {{Gonzalez-de-Rivera}}}, \bibinfo {author} {\bibfnamefont
  {V.}~\bibnamefont {{Gonzalez-Perez}}}, \bibinfo {author} {\bibfnamefont
  {S.}~\bibnamefont {Gott}}, \bibinfo {author} {\bibfnamefont {O.}~\bibnamefont
  {Graur}}, \bibinfo {author} {\bibfnamefont {G.}~\bibnamefont {Gutierrez}},
  \bibinfo {author} {\bibfnamefont {J.}~\bibnamefont {Guy}}, \bibinfo {author}
  {\bibfnamefont {S.}~\bibnamefont {Habib}}, \bibinfo {author} {\bibfnamefont
  {H.}~\bibnamefont {Heetderks}}, \bibinfo {author} {\bibfnamefont
  {I.}~\bibnamefont {Heetderks}}, \bibinfo {author} {\bibfnamefont
  {K.}~\bibnamefont {Heitmann}}, \bibinfo {author} {\bibfnamefont {W.~A.}\
  \bibnamefont {Hellwing}}, \bibinfo {author} {\bibfnamefont {D.~A.}\
  \bibnamefont {Herrera}}, \bibinfo {author} {\bibfnamefont {S.}~\bibnamefont
  {Ho}}, \bibinfo {author} {\bibfnamefont {S.}~\bibnamefont {Holland}},
  \bibinfo {author} {\bibfnamefont {K.}~\bibnamefont {Honscheid}}, \bibinfo
  {author} {\bibfnamefont {E.}~\bibnamefont {Huff}}, \bibinfo {author}
  {\bibfnamefont {T.~A.}\ \bibnamefont {Hutchinson}}, \bibinfo {author}
  {\bibfnamefont {D.}~\bibnamefont {Huterer}}, \bibinfo {author} {\bibfnamefont
  {H.~S.}\ \bibnamefont {Hwang}}, \bibinfo {author} {\bibfnamefont {J.~M.~I.}\
  \bibnamefont {Laguna}}, \bibinfo {author} {\bibfnamefont {Y.}~\bibnamefont
  {Ishikawa}}, \bibinfo {author} {\bibfnamefont {D.}~\bibnamefont {Jacobs}},
  \bibinfo {author} {\bibfnamefont {N.}~\bibnamefont {Jeffrey}}, \bibinfo
  {author} {\bibfnamefont {P.}~\bibnamefont {Jelinsky}}, \bibinfo {author}
  {\bibfnamefont {E.}~\bibnamefont {Jennings}}, \bibinfo {author}
  {\bibfnamefont {L.}~\bibnamefont {Jiang}}, \bibinfo {author} {\bibfnamefont
  {J.}~\bibnamefont {Jimenez}}, \bibinfo {author} {\bibfnamefont
  {J.}~\bibnamefont {Johnson}}, \bibinfo {author} {\bibfnamefont
  {R.}~\bibnamefont {Joyce}}, \bibinfo {author} {\bibfnamefont
  {E.}~\bibnamefont {Jullo}}, \bibinfo {author} {\bibfnamefont
  {S.}~\bibnamefont {Juneau}}, \bibinfo {author} {\bibfnamefont
  {S.}~\bibnamefont {Kama}}, \bibinfo {author} {\bibfnamefont {A.}~\bibnamefont
  {Karcher}}, \bibinfo {author} {\bibfnamefont {S.}~\bibnamefont {Karkar}},
  \bibinfo {author} {\bibfnamefont {R.}~\bibnamefont {Kehoe}}, \bibinfo
  {author} {\bibfnamefont {N.}~\bibnamefont {Kennamer}}, \bibinfo {author}
  {\bibfnamefont {S.}~\bibnamefont {Kent}}, \bibinfo {author} {\bibfnamefont
  {M.}~\bibnamefont {Kilbinger}}, \bibinfo {author} {\bibfnamefont {A.~G.}\
  \bibnamefont {Kim}}, \bibinfo {author} {\bibfnamefont {D.}~\bibnamefont
  {Kirkby}}, \bibinfo {author} {\bibfnamefont {T.}~\bibnamefont {Kisner}},
  \bibinfo {author} {\bibfnamefont {E.}~\bibnamefont {Kitanidis}}, \bibinfo
  {author} {\bibfnamefont {J.-P.}\ \bibnamefont {Kneib}}, \bibinfo {author}
  {\bibfnamefont {S.}~\bibnamefont {Koposov}}, \bibinfo {author} {\bibfnamefont
  {E.}~\bibnamefont {Kovacs}}, \bibinfo {author} {\bibfnamefont
  {K.}~\bibnamefont {Koyama}}, \bibinfo {author} {\bibfnamefont
  {A.}~\bibnamefont {Kremin}}, \bibinfo {author} {\bibfnamefont
  {R.}~\bibnamefont {Kron}}, \bibinfo {author} {\bibfnamefont {L.}~\bibnamefont
  {Kronig}}, \bibinfo {author} {\bibfnamefont {A.}~\bibnamefont
  {{Kueter-Young}}}, \bibinfo {author} {\bibfnamefont {C.~G.}\ \bibnamefont
  {Lacey}}, \bibinfo {author} {\bibfnamefont {R.}~\bibnamefont {Lafever}},
  \bibinfo {author} {\bibfnamefont {O.}~\bibnamefont {Lahav}}, \bibinfo
  {author} {\bibfnamefont {A.}~\bibnamefont {Lambert}}, \bibinfo {author}
  {\bibfnamefont {M.}~\bibnamefont {Lampton}}, \bibinfo {author} {\bibfnamefont
  {M.}~\bibnamefont {Landriau}}, \bibinfo {author} {\bibfnamefont
  {D.}~\bibnamefont {Lang}}, \bibinfo {author} {\bibfnamefont {T.~R.}\
  \bibnamefont {Lauer}}, \bibinfo {author} {\bibfnamefont {J.-M.~L.}\
  \bibnamefont {Goff}}, \bibinfo {author} {\bibfnamefont {L.~L.}\ \bibnamefont
  {Guillou}}, \bibinfo {author} {\bibfnamefont {A.~L.}\ \bibnamefont
  {Van~Suu}}, \bibinfo {author} {\bibfnamefont {J.~H.}\ \bibnamefont {Lee}},
  \bibinfo {author} {\bibfnamefont {S.-J.}\ \bibnamefont {Lee}}, \bibinfo
  {author} {\bibfnamefont {D.}~\bibnamefont {Leitner}}, \bibinfo {author}
  {\bibfnamefont {M.}~\bibnamefont {Lesser}}, \bibinfo {author} {\bibfnamefont
  {M.~E.}\ \bibnamefont {Levi}}, \bibinfo {author} {\bibfnamefont
  {B.}~\bibnamefont {L'Huillier}}, \bibinfo {author} {\bibfnamefont
  {B.}~\bibnamefont {Li}}, \bibinfo {author} {\bibfnamefont {M.}~\bibnamefont
  {Liang}}, \bibinfo {author} {\bibfnamefont {H.}~\bibnamefont {Lin}}, \bibinfo
  {author} {\bibfnamefont {E.}~\bibnamefont {Linder}}, \bibinfo {author}
  {\bibfnamefont {S.~R.}\ \bibnamefont {Loebman}}, \bibinfo {author}
  {\bibfnamefont {Z.}~\bibnamefont {Luki{\'c}}}, \bibinfo {author}
  {\bibfnamefont {J.}~\bibnamefont {Ma}}, \bibinfo {author} {\bibfnamefont
  {N.}~\bibnamefont {MacCrann}}, \bibinfo {author} {\bibfnamefont
  {C.}~\bibnamefont {Magneville}}, \bibinfo {author} {\bibfnamefont
  {L.}~\bibnamefont {Makarem}}, \bibinfo {author} {\bibfnamefont
  {M.}~\bibnamefont {Manera}}, \bibinfo {author} {\bibfnamefont {C.~J.}\
  \bibnamefont {Manser}}, \bibinfo {author} {\bibfnamefont {R.}~\bibnamefont
  {Marshall}}, \bibinfo {author} {\bibfnamefont {P.}~\bibnamefont {Martini}},
  \bibinfo {author} {\bibfnamefont {R.}~\bibnamefont {Massey}}, \bibinfo
  {author} {\bibfnamefont {T.}~\bibnamefont {Matheson}}, \bibinfo {author}
  {\bibfnamefont {J.}~\bibnamefont {McCauley}}, \bibinfo {author}
  {\bibfnamefont {P.}~\bibnamefont {McDonald}}, \bibinfo {author}
  {\bibfnamefont {I.~D.}\ \bibnamefont {McGreer}}, \bibinfo {author}
  {\bibfnamefont {A.}~\bibnamefont {Meisner}}, \bibinfo {author} {\bibfnamefont
  {N.}~\bibnamefont {Metcalfe}}, \bibinfo {author} {\bibfnamefont {T.~N.}\
  \bibnamefont {Miller}}, \bibinfo {author} {\bibfnamefont {R.}~\bibnamefont
  {Miquel}}, \bibinfo {author} {\bibfnamefont {J.}~\bibnamefont {Moustakas}},
  \bibinfo {author} {\bibfnamefont {A.}~\bibnamefont {Myers}}, \bibinfo
  {author} {\bibfnamefont {M.}~\bibnamefont {Naik}}, \bibinfo {author}
  {\bibfnamefont {J.~A.}\ \bibnamefont {Newman}}, \bibinfo {author}
  {\bibfnamefont {R.~C.}\ \bibnamefont {Nichol}}, \bibinfo {author}
  {\bibfnamefont {A.}~\bibnamefont {Nicola}}, \bibinfo {author} {\bibfnamefont
  {L.~N.}\ \bibnamefont {{da Costa}}}, \bibinfo {author} {\bibfnamefont
  {J.}~\bibnamefont {Nie}}, \bibinfo {author} {\bibfnamefont {G.}~\bibnamefont
  {Niz}}, \bibinfo {author} {\bibfnamefont {P.}~\bibnamefont {Norberg}},
  \bibinfo {author} {\bibfnamefont {B.}~\bibnamefont {Nord}}, \bibinfo {author}
  {\bibfnamefont {D.}~\bibnamefont {Norman}}, \bibinfo {author} {\bibfnamefont
  {P.}~\bibnamefont {Nugent}}, \bibinfo {author} {\bibfnamefont
  {T.}~\bibnamefont {O'Brien}}, \bibinfo {author} {\bibfnamefont
  {M.}~\bibnamefont {Oh}}, \bibinfo {author} {\bibfnamefont {K.~A.~G.}\
  \bibnamefont {Olsen}}, \bibinfo {author} {\bibfnamefont {C.}~\bibnamefont
  {Padilla}}, \bibinfo {author} {\bibfnamefont {H.}~\bibnamefont
  {Padmanabhan}}, \bibinfo {author} {\bibfnamefont {N.}~\bibnamefont
  {Padmanabhan}}, \bibinfo {author} {\bibfnamefont {N.}~\bibnamefont
  {{Palanque-Delabrouille}}}, \bibinfo {author} {\bibfnamefont
  {A.}~\bibnamefont {Palmese}}, \bibinfo {author} {\bibfnamefont
  {D.}~\bibnamefont {Pappalardo}}, \bibinfo {author} {\bibfnamefont
  {I.}~\bibnamefont {P{\^a}ris}}, \bibinfo {author} {\bibfnamefont
  {C.}~\bibnamefont {Park}}, \bibinfo {author} {\bibfnamefont {A.}~\bibnamefont
  {Patej}}, \bibinfo {author} {\bibfnamefont {J.~A.}\ \bibnamefont {Peacock}},
  \bibinfo {author} {\bibfnamefont {H.~V.}\ \bibnamefont {Peiris}}, \bibinfo
  {author} {\bibfnamefont {X.}~\bibnamefont {Peng}}, \bibinfo {author}
  {\bibfnamefont {W.~J.}\ \bibnamefont {Percival}}, \bibinfo {author}
  {\bibfnamefont {S.}~\bibnamefont {Perruchot}}, \bibinfo {author}
  {\bibfnamefont {M.~M.}\ \bibnamefont {Pieri}}, \bibinfo {author}
  {\bibfnamefont {R.}~\bibnamefont {Pogge}}, \bibinfo {author} {\bibfnamefont
  {J.~E.}\ \bibnamefont {Pollack}}, \bibinfo {author} {\bibfnamefont
  {C.}~\bibnamefont {Poppett}}, \bibinfo {author} {\bibfnamefont
  {F.}~\bibnamefont {Prada}}, \bibinfo {author} {\bibfnamefont
  {A.}~\bibnamefont {Prakash}}, \bibinfo {author} {\bibfnamefont {R.~G.}\
  \bibnamefont {Probst}}, \bibinfo {author} {\bibfnamefont {D.}~\bibnamefont
  {Rabinowitz}}, \bibinfo {author} {\bibfnamefont {A.}~\bibnamefont
  {Raichoor}}, \bibinfo {author} {\bibfnamefont {C.~H.}\ \bibnamefont {Ree}},
  \bibinfo {author} {\bibfnamefont {A.}~\bibnamefont {Refregier}}, \bibinfo
  {author} {\bibfnamefont {X.}~\bibnamefont {Regal}}, \bibinfo {author}
  {\bibfnamefont {B.}~\bibnamefont {Reid}}, \bibinfo {author} {\bibfnamefont
  {K.}~\bibnamefont {Reil}}, \bibinfo {author} {\bibfnamefont {M.}~\bibnamefont
  {Rezaie}}, \bibinfo {author} {\bibfnamefont {C.~M.}\ \bibnamefont {Rockosi}},
  \bibinfo {author} {\bibfnamefont {N.}~\bibnamefont {Roe}}, \bibinfo {author}
  {\bibfnamefont {S.}~\bibnamefont {Ronayette}}, \bibinfo {author}
  {\bibfnamefont {A.}~\bibnamefont {Roodman}}, \bibinfo {author} {\bibfnamefont
  {A.~J.}\ \bibnamefont {Ross}}, \bibinfo {author} {\bibfnamefont {N.~P.}\
  \bibnamefont {Ross}}, \bibinfo {author} {\bibfnamefont {G.}~\bibnamefont
  {Rossi}}, \bibinfo {author} {\bibfnamefont {E.}~\bibnamefont {Rozo}},
  \bibinfo {author} {\bibfnamefont {V.}~\bibnamefont {{Ruhlmann-Kleider}}},
  \bibinfo {author} {\bibfnamefont {E.~S.}\ \bibnamefont {Rykoff}}, \bibinfo
  {author} {\bibfnamefont {C.}~\bibnamefont {Sabiu}}, \bibinfo {author}
  {\bibfnamefont {L.}~\bibnamefont {Samushia}}, \bibinfo {author}
  {\bibfnamefont {E.}~\bibnamefont {Sanchez}}, \bibinfo {author} {\bibfnamefont
  {J.}~\bibnamefont {Sanchez}}, \bibinfo {author} {\bibfnamefont {D.~J.}\
  \bibnamefont {Schlegel}}, \bibinfo {author} {\bibfnamefont {M.}~\bibnamefont
  {Schneider}}, \bibinfo {author} {\bibfnamefont {M.}~\bibnamefont
  {Schubnell}}, \bibinfo {author} {\bibfnamefont {A.}~\bibnamefont {Secroun}},
  \bibinfo {author} {\bibfnamefont {U.}~\bibnamefont {Seljak}}, \bibinfo
  {author} {\bibfnamefont {H.-J.}\ \bibnamefont {Seo}}, \bibinfo {author}
  {\bibfnamefont {S.}~\bibnamefont {Serrano}}, \bibinfo {author} {\bibfnamefont
  {A.}~\bibnamefont {Shafieloo}}, \bibinfo {author} {\bibfnamefont
  {H.}~\bibnamefont {Shan}}, \bibinfo {author} {\bibfnamefont {R.}~\bibnamefont
  {Sharples}}, \bibinfo {author} {\bibfnamefont {M.~J.}\ \bibnamefont {Sholl}},
  \bibinfo {author} {\bibfnamefont {W.~V.}\ \bibnamefont {Shourt}}, \bibinfo
  {author} {\bibfnamefont {J.~H.}\ \bibnamefont {Silber}}, \bibinfo {author}
  {\bibfnamefont {D.~R.}\ \bibnamefont {Silva}}, \bibinfo {author}
  {\bibfnamefont {M.~M.}\ \bibnamefont {Sirk}}, \bibinfo {author}
  {\bibfnamefont {A.}~\bibnamefont {Slosar}}, \bibinfo {author} {\bibfnamefont
  {A.}~\bibnamefont {Smith}}, \bibinfo {author} {\bibfnamefont {G.~F.}\
  \bibnamefont {Smoot}}, \bibinfo {author} {\bibfnamefont {D.}~\bibnamefont
  {Som}}, \bibinfo {author} {\bibfnamefont {Y.-S.}\ \bibnamefont {Song}},
  \bibinfo {author} {\bibfnamefont {D.}~\bibnamefont {Sprayberry}}, \bibinfo
  {author} {\bibfnamefont {R.}~\bibnamefont {Staten}}, \bibinfo {author}
  {\bibfnamefont {A.}~\bibnamefont {Stefanik}}, \bibinfo {author}
  {\bibfnamefont {G.}~\bibnamefont {Tarle}}, \bibinfo {author} {\bibfnamefont
  {S.~S.}\ \bibnamefont {Tie}}, \bibinfo {author} {\bibfnamefont {J.~L.}\
  \bibnamefont {Tinker}}, \bibinfo {author} {\bibfnamefont {R.}~\bibnamefont
  {Tojeiro}}, \bibinfo {author} {\bibfnamefont {F.}~\bibnamefont {Valdes}},
  \bibinfo {author} {\bibfnamefont {O.}~\bibnamefont {Valenzuela}}, \bibinfo
  {author} {\bibfnamefont {M.}~\bibnamefont {Valluri}}, \bibinfo {author}
  {\bibfnamefont {M.}~\bibnamefont {{Vargas-Magana}}}, \bibinfo {author}
  {\bibfnamefont {L.}~\bibnamefont {Verde}}, \bibinfo {author} {\bibfnamefont
  {A.~R.}\ \bibnamefont {Walker}}, \bibinfo {author} {\bibfnamefont
  {J.}~\bibnamefont {Wang}}, \bibinfo {author} {\bibfnamefont {Y.}~\bibnamefont
  {Wang}}, \bibinfo {author} {\bibfnamefont {B.~A.}\ \bibnamefont {Weaver}},
  \bibinfo {author} {\bibfnamefont {C.}~\bibnamefont {Weaverdyck}}, \bibinfo
  {author} {\bibfnamefont {R.~H.}\ \bibnamefont {Wechsler}}, \bibinfo {author}
  {\bibfnamefont {D.~H.}\ \bibnamefont {Weinberg}}, \bibinfo {author}
  {\bibfnamefont {M.}~\bibnamefont {White}}, \bibinfo {author} {\bibfnamefont
  {Q.}~\bibnamefont {Yang}}, \bibinfo {author} {\bibfnamefont {C.}~\bibnamefont
  {Yeche}}, \bibinfo {author} {\bibfnamefont {T.}~\bibnamefont {Zhang}},
  \bibinfo {author} {\bibfnamefont {G.-B.}\ \bibnamefont {Zhao}}, \bibinfo
  {author} {\bibfnamefont {Y.}~\bibnamefont {Zheng}}, \bibinfo {author}
  {\bibfnamefont {X.}~\bibnamefont {Zhou}}, \bibinfo {author} {\bibfnamefont
  {Z.}~\bibnamefont {Zhou}}, \bibinfo {author} {\bibfnamefont {Y.}~\bibnamefont
  {Zhu}}, \bibinfo {author} {\bibfnamefont {H.}~\bibnamefont {Zou}}, \ and\
  \bibinfo {author} {\bibfnamefont {Y.}~\bibnamefont {Zu}},\ }\href@noop {}
  {\bibfield  {journal} {\bibinfo  {journal} {arXiv:1611.00037 [astro-ph]}\ }
  (\bibinfo {year} {2016}{\natexlab{b}})},\ \Eprint
  {http://arxiv.org/abs/1611.00037} {arXiv:1611.00037 [astro-ph]} \BibitemShut
  {NoStop}%
\bibitem [{\citenamefont {Abareshi}\ \emph {et~al.}(2022)\citenamefont
  {Abareshi}, \citenamefont {Aguilar}, \citenamefont {Ahlen}, \citenamefont
  {Alam}, \citenamefont {Alexander}, \citenamefont {Alfarsy}, \citenamefont
  {Allen}, \citenamefont {Prieto}, \citenamefont {Alves}, \citenamefont
  {Ameel}, \citenamefont {Armengaud}, \citenamefont {Asorey}, \citenamefont
  {Aviles}, \citenamefont {Bailey}, \citenamefont {{Balaguera-Antol{\'i}nez}},
  \citenamefont {Ballester}, \citenamefont {Baltay}, \citenamefont {Bault},
  \citenamefont {Beltran}, \citenamefont {Benavides}, \citenamefont {BenZvi},
  \citenamefont {Berti}, \citenamefont {Besuner}, \citenamefont {Beutler},
  \citenamefont {Bianchi}, \citenamefont {Blake}, \citenamefont {Blanc},
  \citenamefont {Blum}, \citenamefont {Bolton}, \citenamefont {Bose},
  \citenamefont {Bramall}, \citenamefont {Brieden}, \citenamefont {Brodzeller},
  \citenamefont {Brooks}, \citenamefont {Brownewell}, \citenamefont
  {{Buckley-Geer}}, \citenamefont {Cahn}, \citenamefont {Cai}, \citenamefont
  {Canning}, \citenamefont {Rosell}, \citenamefont {Carton}, \citenamefont
  {Casas}, \citenamefont {Castander}, \citenamefont {{Cervantes-Cota}},
  \citenamefont {Chabanier}, \citenamefont {Chaussidon}, \citenamefont
  {Chuang}, \citenamefont {Circosta}, \citenamefont {Cole}, \citenamefont
  {Cooper}, \citenamefont {{da Costa}}, \citenamefont {Cousinou}, \citenamefont
  {Cuceu}, \citenamefont {Davis}, \citenamefont {Dawson}, \citenamefont {{de la
  Cruz-Noriega}}, \citenamefont {{de la Macorra}}, \citenamefont {{de Mattia}},
  \citenamefont {Della~Costa}, \citenamefont {Demmer}, \citenamefont {Derwent},
  \citenamefont {Dey}, \citenamefont {Dey}, \citenamefont {Dhungana},
  \citenamefont {Ding}, \citenamefont {Dobson}, \citenamefont {Doel},
  \citenamefont {{Donald-McCann}}, \citenamefont {Donaldson}, \citenamefont
  {Douglass}, \citenamefont {Duan}, \citenamefont {Dunlop}, \citenamefont
  {Edelstein}, \citenamefont {Eftekharzadeh}, \citenamefont {Eisenstein},
  \citenamefont {{Enriquez-Vargas}}, \citenamefont {Escoffier}, \citenamefont
  {Evatt}, \citenamefont {Fagrelius}, \citenamefont {Fan}, \citenamefont
  {Fanning}, \citenamefont {Fawcett}, \citenamefont {Ferraro}, \citenamefont
  {Ereza}, \citenamefont {Flaugher}, \citenamefont {{Font-Ribera}},
  \citenamefont {{Forero-Romero}}, \citenamefont {Frenk}, \citenamefont
  {Fromenteau}, \citenamefont {G{\"a}nsicke}, \citenamefont
  {{Garcia-Quintero}}, \citenamefont {Garrison}, \citenamefont {Gazta{\~n}aga},
  \citenamefont {Gerardi}, \citenamefont {{Gil-Mar{\'i}n}}, \citenamefont
  {Gontcho}, \citenamefont {{Gonzalez-Morales}}, \citenamefont
  {{Gonzalez-de-Rivera}}, \citenamefont {{Gonzalez-Perez}}, \citenamefont
  {Gordon}, \citenamefont {Graur}, \citenamefont {Green}, \citenamefont
  {Grove}, \citenamefont {Gruen}, \citenamefont {Gutierrez}, \citenamefont
  {Guy}, \citenamefont {Hahn}, \citenamefont {Harris}, \citenamefont {Herrera},
  \citenamefont {{Herrera-Alcantar}}, \citenamefont {Honscheid}, \citenamefont
  {Howlett}, \citenamefont {Huterer}, \citenamefont {Ir{\v s}i{\v c}},
  \citenamefont {Ishak}, \citenamefont {Jelinsky}, \citenamefont {Jiang},
  \citenamefont {Jimenez}, \citenamefont {Jing}, \citenamefont {Joyce},
  \citenamefont {Jullo}, \citenamefont {Juneau}, \citenamefont {Kara{\c
  c}ayl{\i}}, \citenamefont {Karamanis}, \citenamefont {Karcher}, \citenamefont
  {Karim}, \citenamefont {Kehoe}, \citenamefont {Kent}, \citenamefont {Kirkby},
  \citenamefont {Kisner}, \citenamefont {Kitaura}, \citenamefont {Koposov},
  \citenamefont {Kov{\'a}cs}, \citenamefont {Kremin}, \citenamefont
  {Krolewski}, \citenamefont {L'Huillier}, \citenamefont {Lahav}, \citenamefont
  {Lambert}, \citenamefont {Lamman}, \citenamefont {Lan}, \citenamefont
  {Landriau}, \citenamefont {Lane}, \citenamefont {Lang}, \citenamefont
  {Lange}, \citenamefont {Lasker}, \citenamefont {Guillou}, \citenamefont
  {Leauthaud}, \citenamefont {Van~Suu}, \citenamefont {Levi}, \citenamefont
  {Li}, \citenamefont {Magneville}, \citenamefont {Manera}, \citenamefont
  {Manser}, \citenamefont {Marshall}, \citenamefont {McCollam}, \citenamefont
  {McDonald}, \citenamefont {Meisner}, \citenamefont {Mezcua}, \citenamefont
  {Miller}, \citenamefont {Miquel}, \citenamefont {{Montero-Camacho}},
  \citenamefont {Moon}, \citenamefont {Martini}, \citenamefont
  {{Meneses-Rizo}}, \citenamefont {Moustakas}, \citenamefont {Mueller},
  \citenamefont {{Mu{\~n}oz-Guti{\'e}rrez}}, \citenamefont {Myers},
  \citenamefont {Nadathur}, \citenamefont {Najita}, \citenamefont {Napolitano},
  \citenamefont {Neilsen}, \citenamefont {Newman}, \citenamefont {Nie},
  \citenamefont {Ning}, \citenamefont {Niz}, \citenamefont {Norberg},
  \citenamefont {Noriega}, \citenamefont {O'Brien}, \citenamefont {Obuljen},
  \citenamefont {{Palanque-Delabrouille}}, \citenamefont {Palmese},
  \citenamefont {Zhiwei}, \citenamefont {Pappalardo}, \citenamefont {Peng},
  \citenamefont {Percival}, \citenamefont {Perruchot}, \citenamefont {Pogge},
  \citenamefont {Poppett}, \citenamefont {Porredon}, \citenamefont {Prada},
  \citenamefont {Prochaska}, \citenamefont {Pucha}, \citenamefont
  {{P{\'e}rez-Fern{\'a}ndez}}, \citenamefont {{P{\'e}rez-R{\'a}fols}},
  \citenamefont {Rabinowitz}, \citenamefont {Raichoor}, \citenamefont
  {{Ramirez-Solano}}, \citenamefont {{Ram{\'i}rez-P{\'e}rez}}, \citenamefont
  {Ravoux}, \citenamefont {Reil}, \citenamefont {Rezaie}, \citenamefont
  {Rocher}, \citenamefont {Rockosi}, \citenamefont {Roe}, \citenamefont
  {Roodman}, \citenamefont {Ross}, \citenamefont {Rossi}, \citenamefont
  {Ruggeri}, \citenamefont {{Ruhlmann-Kleider}}, \citenamefont {Sabiu},
  \citenamefont {Safonova}, \citenamefont {Said}, \citenamefont {Saintonge},
  \citenamefont {Catonga}, \citenamefont {Samushia}, \citenamefont {Sanchez},
  \citenamefont {Saulder}, \citenamefont {Schaan}, \citenamefont {Schlafly},
  \citenamefont {Schlegel}, \citenamefont {Schmoll}, \citenamefont {Scholte},
  \citenamefont {Schubnell}, \citenamefont {Secroun}, \citenamefont {Seo},
  \citenamefont {Serrano}, \citenamefont {Sharples}, \citenamefont {Sholl},
  \citenamefont {Silber}, \citenamefont {Silva}, \citenamefont {Sirk},
  \citenamefont {Siudek}, \citenamefont {Smith}, \citenamefont {Sprayberry},
  \citenamefont {Staten}, \citenamefont {Stupak}, \citenamefont {Tan},
  \citenamefont {Tarl{\'e}}, \citenamefont {Tie}, \citenamefont {Tojeiro},
  \citenamefont {{Ure{\~n}a-L{\'o}pez}}, \citenamefont {Valdes}, \citenamefont
  {Valenzuela}, \citenamefont {Valluri}, \citenamefont {{Vargas-Maga{\~n}a}},
  \citenamefont {Verde}, \citenamefont {Walther}, \citenamefont {Wang},
  \citenamefont {Wang}, \citenamefont {Weaver}, \citenamefont {Weaverdyck},
  \citenamefont {Wechsler}, \citenamefont {Wilson}, \citenamefont {Yang},
  \citenamefont {Yu}, \citenamefont {Yuan}, \citenamefont {Y{\`e}che},
  \citenamefont {Zhang}, \citenamefont {Zhang}, \citenamefont {Zhao},
  \citenamefont {Zhou}, \citenamefont {Zhou}, \citenamefont {Zou},
  \citenamefont {Zou}, \citenamefont {Zou},\ and\ \citenamefont
  {Zu}}]{abareshi2022}%
  \BibitemOpen
  \bibfield  {author} {\bibinfo {author} {\bibfnamefont {B.}~\bibnamefont
  {Abareshi}}, \bibinfo {author} {\bibfnamefont {J.}~\bibnamefont {Aguilar}},
  \bibinfo {author} {\bibfnamefont {S.}~\bibnamefont {Ahlen}}, \bibinfo
  {author} {\bibfnamefont {S.}~\bibnamefont {Alam}}, \bibinfo {author}
  {\bibfnamefont {D.~M.}\ \bibnamefont {Alexander}}, \bibinfo {author}
  {\bibfnamefont {R.}~\bibnamefont {Alfarsy}}, \bibinfo {author} {\bibfnamefont
  {L.}~\bibnamefont {Allen}}, \bibinfo {author} {\bibfnamefont {C.~A.}\
  \bibnamefont {Prieto}}, \bibinfo {author} {\bibfnamefont {O.}~\bibnamefont
  {Alves}}, \bibinfo {author} {\bibfnamefont {J.}~\bibnamefont {Ameel}},
  \bibinfo {author} {\bibfnamefont {E.}~\bibnamefont {Armengaud}}, \bibinfo
  {author} {\bibfnamefont {J.}~\bibnamefont {Asorey}}, \bibinfo {author}
  {\bibfnamefont {A.}~\bibnamefont {Aviles}}, \bibinfo {author} {\bibfnamefont
  {S.}~\bibnamefont {Bailey}}, \bibinfo {author} {\bibfnamefont
  {A.}~\bibnamefont {{Balaguera-Antol{\'i}nez}}}, \bibinfo {author}
  {\bibfnamefont {O.}~\bibnamefont {Ballester}}, \bibinfo {author}
  {\bibfnamefont {C.}~\bibnamefont {Baltay}}, \bibinfo {author} {\bibfnamefont
  {A.}~\bibnamefont {Bault}}, \bibinfo {author} {\bibfnamefont {S.~F.}\
  \bibnamefont {Beltran}}, \bibinfo {author} {\bibfnamefont {B.}~\bibnamefont
  {Benavides}}, \bibinfo {author} {\bibfnamefont {S.}~\bibnamefont {BenZvi}},
  \bibinfo {author} {\bibfnamefont {A.}~\bibnamefont {Berti}}, \bibinfo
  {author} {\bibfnamefont {R.}~\bibnamefont {Besuner}}, \bibinfo {author}
  {\bibfnamefont {F.}~\bibnamefont {Beutler}}, \bibinfo {author} {\bibfnamefont
  {D.}~\bibnamefont {Bianchi}}, \bibinfo {author} {\bibfnamefont
  {C.}~\bibnamefont {Blake}}, \bibinfo {author} {\bibfnamefont
  {P.}~\bibnamefont {Blanc}}, \bibinfo {author} {\bibfnamefont
  {R.}~\bibnamefont {Blum}}, \bibinfo {author} {\bibfnamefont {A.}~\bibnamefont
  {Bolton}}, \bibinfo {author} {\bibfnamefont {S.}~\bibnamefont {Bose}},
  \bibinfo {author} {\bibfnamefont {D.}~\bibnamefont {Bramall}}, \bibinfo
  {author} {\bibfnamefont {S.}~\bibnamefont {Brieden}}, \bibinfo {author}
  {\bibfnamefont {A.}~\bibnamefont {Brodzeller}}, \bibinfo {author}
  {\bibfnamefont {D.}~\bibnamefont {Brooks}}, \bibinfo {author} {\bibfnamefont
  {C.}~\bibnamefont {Brownewell}}, \bibinfo {author} {\bibfnamefont
  {E.}~\bibnamefont {{Buckley-Geer}}}, \bibinfo {author} {\bibfnamefont
  {R.~N.}\ \bibnamefont {Cahn}}, \bibinfo {author} {\bibfnamefont
  {Z.}~\bibnamefont {Cai}}, \bibinfo {author} {\bibfnamefont {R.}~\bibnamefont
  {Canning}}, \bibinfo {author} {\bibfnamefont {A.~C.}\ \bibnamefont {Rosell}},
  \bibinfo {author} {\bibfnamefont {P.}~\bibnamefont {Carton}}, \bibinfo
  {author} {\bibfnamefont {R.}~\bibnamefont {Casas}}, \bibinfo {author}
  {\bibfnamefont {F.~J.}\ \bibnamefont {Castander}}, \bibinfo {author}
  {\bibfnamefont {J.~L.}\ \bibnamefont {{Cervantes-Cota}}}, \bibinfo {author}
  {\bibfnamefont {S.}~\bibnamefont {Chabanier}}, \bibinfo {author}
  {\bibfnamefont {E.}~\bibnamefont {Chaussidon}}, \bibinfo {author}
  {\bibfnamefont {C.}~\bibnamefont {Chuang}}, \bibinfo {author} {\bibfnamefont
  {C.}~\bibnamefont {Circosta}}, \bibinfo {author} {\bibfnamefont
  {S.}~\bibnamefont {Cole}}, \bibinfo {author} {\bibfnamefont {A.~P.}\
  \bibnamefont {Cooper}}, \bibinfo {author} {\bibfnamefont {L.}~\bibnamefont
  {{da Costa}}}, \bibinfo {author} {\bibfnamefont {M.-C.}\ \bibnamefont
  {Cousinou}}, \bibinfo {author} {\bibfnamefont {A.}~\bibnamefont {Cuceu}},
  \bibinfo {author} {\bibfnamefont {T.~M.}\ \bibnamefont {Davis}}, \bibinfo
  {author} {\bibfnamefont {K.}~\bibnamefont {Dawson}}, \bibinfo {author}
  {\bibfnamefont {R.}~\bibnamefont {{de la Cruz-Noriega}}}, \bibinfo {author}
  {\bibfnamefont {A.}~\bibnamefont {{de la Macorra}}}, \bibinfo {author}
  {\bibfnamefont {A.}~\bibnamefont {{de Mattia}}}, \bibinfo {author}
  {\bibfnamefont {J.}~\bibnamefont {Della~Costa}}, \bibinfo {author}
  {\bibfnamefont {P.}~\bibnamefont {Demmer}}, \bibinfo {author} {\bibfnamefont
  {M.}~\bibnamefont {Derwent}}, \bibinfo {author} {\bibfnamefont
  {A.}~\bibnamefont {Dey}}, \bibinfo {author} {\bibfnamefont {B.}~\bibnamefont
  {Dey}}, \bibinfo {author} {\bibfnamefont {G.}~\bibnamefont {Dhungana}},
  \bibinfo {author} {\bibfnamefont {Z.}~\bibnamefont {Ding}}, \bibinfo {author}
  {\bibfnamefont {C.}~\bibnamefont {Dobson}}, \bibinfo {author} {\bibfnamefont
  {P.}~\bibnamefont {Doel}}, \bibinfo {author} {\bibfnamefont {J.}~\bibnamefont
  {{Donald-McCann}}}, \bibinfo {author} {\bibfnamefont {J.}~\bibnamefont
  {Donaldson}}, \bibinfo {author} {\bibfnamefont {K.}~\bibnamefont {Douglass}},
  \bibinfo {author} {\bibfnamefont {Y.}~\bibnamefont {Duan}}, \bibinfo {author}
  {\bibfnamefont {P.}~\bibnamefont {Dunlop}}, \bibinfo {author} {\bibfnamefont
  {J.}~\bibnamefont {Edelstein}}, \bibinfo {author} {\bibfnamefont
  {S.}~\bibnamefont {Eftekharzadeh}}, \bibinfo {author} {\bibfnamefont {D.~J.}\
  \bibnamefont {Eisenstein}}, \bibinfo {author} {\bibfnamefont
  {M.}~\bibnamefont {{Enriquez-Vargas}}}, \bibinfo {author} {\bibfnamefont
  {S.}~\bibnamefont {Escoffier}}, \bibinfo {author} {\bibfnamefont
  {M.}~\bibnamefont {Evatt}}, \bibinfo {author} {\bibfnamefont
  {P.}~\bibnamefont {Fagrelius}}, \bibinfo {author} {\bibfnamefont
  {X.}~\bibnamefont {Fan}}, \bibinfo {author} {\bibfnamefont {K.}~\bibnamefont
  {Fanning}}, \bibinfo {author} {\bibfnamefont {V.~A.}\ \bibnamefont
  {Fawcett}}, \bibinfo {author} {\bibfnamefont {S.}~\bibnamefont {Ferraro}},
  \bibinfo {author} {\bibfnamefont {J.}~\bibnamefont {Ereza}}, \bibinfo
  {author} {\bibfnamefont {B.}~\bibnamefont {Flaugher}}, \bibinfo {author}
  {\bibfnamefont {A.}~\bibnamefont {{Font-Ribera}}}, \bibinfo {author}
  {\bibfnamefont {J.~E.}\ \bibnamefont {{Forero-Romero}}}, \bibinfo {author}
  {\bibfnamefont {C.~S.}\ \bibnamefont {Frenk}}, \bibinfo {author}
  {\bibfnamefont {S.}~\bibnamefont {Fromenteau}}, \bibinfo {author}
  {\bibfnamefont {B.~T.}\ \bibnamefont {G{\"a}nsicke}}, \bibinfo {author}
  {\bibfnamefont {C.}~\bibnamefont {{Garcia-Quintero}}}, \bibinfo {author}
  {\bibfnamefont {L.}~\bibnamefont {Garrison}}, \bibinfo {author}
  {\bibfnamefont {E.}~\bibnamefont {Gazta{\~n}aga}}, \bibinfo {author}
  {\bibfnamefont {F.}~\bibnamefont {Gerardi}}, \bibinfo {author} {\bibfnamefont
  {H.}~\bibnamefont {{Gil-Mar{\'i}n}}}, \bibinfo {author} {\bibfnamefont
  {S.~G.~A.}\ \bibnamefont {Gontcho}}, \bibinfo {author} {\bibfnamefont
  {A.~X.}\ \bibnamefont {{Gonzalez-Morales}}}, \bibinfo {author} {\bibfnamefont
  {G.}~\bibnamefont {{Gonzalez-de-Rivera}}}, \bibinfo {author} {\bibfnamefont
  {V.}~\bibnamefont {{Gonzalez-Perez}}}, \bibinfo {author} {\bibfnamefont
  {C.}~\bibnamefont {Gordon}}, \bibinfo {author} {\bibfnamefont
  {O.}~\bibnamefont {Graur}}, \bibinfo {author} {\bibfnamefont
  {D.}~\bibnamefont {Green}}, \bibinfo {author} {\bibfnamefont
  {C.}~\bibnamefont {Grove}}, \bibinfo {author} {\bibfnamefont
  {D.}~\bibnamefont {Gruen}}, \bibinfo {author} {\bibfnamefont
  {G.}~\bibnamefont {Gutierrez}}, \bibinfo {author} {\bibfnamefont
  {J.}~\bibnamefont {Guy}}, \bibinfo {author} {\bibfnamefont {C.}~\bibnamefont
  {Hahn}}, \bibinfo {author} {\bibfnamefont {S.}~\bibnamefont {Harris}},
  \bibinfo {author} {\bibfnamefont {D.}~\bibnamefont {Herrera}}, \bibinfo
  {author} {\bibfnamefont {H.~K.}\ \bibnamefont {{Herrera-Alcantar}}}, \bibinfo
  {author} {\bibfnamefont {K.}~\bibnamefont {Honscheid}}, \bibinfo {author}
  {\bibfnamefont {C.}~\bibnamefont {Howlett}}, \bibinfo {author} {\bibfnamefont
  {D.}~\bibnamefont {Huterer}}, \bibinfo {author} {\bibfnamefont
  {V.}~\bibnamefont {Ir{\v s}i{\v c}}}, \bibinfo {author} {\bibfnamefont
  {M.}~\bibnamefont {Ishak}}, \bibinfo {author} {\bibfnamefont
  {P.}~\bibnamefont {Jelinsky}}, \bibinfo {author} {\bibfnamefont
  {L.}~\bibnamefont {Jiang}}, \bibinfo {author} {\bibfnamefont
  {J.}~\bibnamefont {Jimenez}}, \bibinfo {author} {\bibfnamefont {Y.~P.}\
  \bibnamefont {Jing}}, \bibinfo {author} {\bibfnamefont {R.}~\bibnamefont
  {Joyce}}, \bibinfo {author} {\bibfnamefont {E.}~\bibnamefont {Jullo}},
  \bibinfo {author} {\bibfnamefont {S.}~\bibnamefont {Juneau}}, \bibinfo
  {author} {\bibfnamefont {N.~G.}\ \bibnamefont {Kara{\c c}ayl{\i}}}, \bibinfo
  {author} {\bibfnamefont {M.}~\bibnamefont {Karamanis}}, \bibinfo {author}
  {\bibfnamefont {A.}~\bibnamefont {Karcher}}, \bibinfo {author} {\bibfnamefont
  {T.}~\bibnamefont {Karim}}, \bibinfo {author} {\bibfnamefont
  {R.}~\bibnamefont {Kehoe}}, \bibinfo {author} {\bibfnamefont
  {S.}~\bibnamefont {Kent}}, \bibinfo {author} {\bibfnamefont {D.}~\bibnamefont
  {Kirkby}}, \bibinfo {author} {\bibfnamefont {T.}~\bibnamefont {Kisner}},
  \bibinfo {author} {\bibfnamefont {F.}~\bibnamefont {Kitaura}}, \bibinfo
  {author} {\bibfnamefont {S.~E.}\ \bibnamefont {Koposov}}, \bibinfo {author}
  {\bibfnamefont {A.}~\bibnamefont {Kov{\'a}cs}}, \bibinfo {author}
  {\bibfnamefont {A.}~\bibnamefont {Kremin}}, \bibinfo {author} {\bibfnamefont
  {A.}~\bibnamefont {Krolewski}}, \bibinfo {author} {\bibfnamefont
  {B.}~\bibnamefont {L'Huillier}}, \bibinfo {author} {\bibfnamefont
  {O.}~\bibnamefont {Lahav}}, \bibinfo {author} {\bibfnamefont
  {A.}~\bibnamefont {Lambert}}, \bibinfo {author} {\bibfnamefont
  {C.}~\bibnamefont {Lamman}}, \bibinfo {author} {\bibfnamefont {T.-W.}\
  \bibnamefont {Lan}}, \bibinfo {author} {\bibfnamefont {M.}~\bibnamefont
  {Landriau}}, \bibinfo {author} {\bibfnamefont {S.}~\bibnamefont {Lane}},
  \bibinfo {author} {\bibfnamefont {D.}~\bibnamefont {Lang}}, \bibinfo {author}
  {\bibfnamefont {J.~U.}\ \bibnamefont {Lange}}, \bibinfo {author}
  {\bibfnamefont {J.}~\bibnamefont {Lasker}}, \bibinfo {author} {\bibfnamefont
  {L.~L.}\ \bibnamefont {Guillou}}, \bibinfo {author} {\bibfnamefont
  {A.}~\bibnamefont {Leauthaud}}, \bibinfo {author} {\bibfnamefont {A.~L.}\
  \bibnamefont {Van~Suu}}, \bibinfo {author} {\bibfnamefont {M.~E.}\
  \bibnamefont {Levi}}, \bibinfo {author} {\bibfnamefont {T.~S.}\ \bibnamefont
  {Li}}, \bibinfo {author} {\bibfnamefont {C.}~\bibnamefont {Magneville}},
  \bibinfo {author} {\bibfnamefont {M.}~\bibnamefont {Manera}}, \bibinfo
  {author} {\bibfnamefont {C.~J.}\ \bibnamefont {Manser}}, \bibinfo {author}
  {\bibfnamefont {B.}~\bibnamefont {Marshall}}, \bibinfo {author}
  {\bibfnamefont {W.}~\bibnamefont {McCollam}}, \bibinfo {author}
  {\bibfnamefont {P.}~\bibnamefont {McDonald}}, \bibinfo {author}
  {\bibfnamefont {A.~M.}\ \bibnamefont {Meisner}}, \bibinfo {author}
  {\bibfnamefont {J.~M.-F.~M.}\ \bibnamefont {Mezcua}}, \bibinfo {author}
  {\bibfnamefont {T.}~\bibnamefont {Miller}}, \bibinfo {author} {\bibfnamefont
  {R.}~\bibnamefont {Miquel}}, \bibinfo {author} {\bibfnamefont
  {P.}~\bibnamefont {{Montero-Camacho}}}, \bibinfo {author} {\bibfnamefont
  {J.}~\bibnamefont {Moon}}, \bibinfo {author} {\bibfnamefont {J.~P.}\
  \bibnamefont {Martini}}, \bibinfo {author} {\bibfnamefont {J.}~\bibnamefont
  {{Meneses-Rizo}}}, \bibinfo {author} {\bibfnamefont {J.}~\bibnamefont
  {Moustakas}}, \bibinfo {author} {\bibfnamefont {E.}~\bibnamefont {Mueller}},
  \bibinfo {author} {\bibfnamefont {A.}~\bibnamefont
  {{Mu{\~n}oz-Guti{\'e}rrez}}}, \bibinfo {author} {\bibfnamefont {A.~D.}\
  \bibnamefont {Myers}}, \bibinfo {author} {\bibfnamefont {S.}~\bibnamefont
  {Nadathur}}, \bibinfo {author} {\bibfnamefont {J.}~\bibnamefont {Najita}},
  \bibinfo {author} {\bibfnamefont {L.}~\bibnamefont {Napolitano}}, \bibinfo
  {author} {\bibfnamefont {E.}~\bibnamefont {Neilsen}}, \bibinfo {author}
  {\bibfnamefont {J.~A.}\ \bibnamefont {Newman}}, \bibinfo {author}
  {\bibfnamefont {J.~D.}\ \bibnamefont {Nie}}, \bibinfo {author} {\bibfnamefont
  {Y.}~\bibnamefont {Ning}}, \bibinfo {author} {\bibfnamefont {G.}~\bibnamefont
  {Niz}}, \bibinfo {author} {\bibfnamefont {P.}~\bibnamefont {Norberg}},
  \bibinfo {author} {\bibfnamefont {H.~E.}\ \bibnamefont {Noriega}}, \bibinfo
  {author} {\bibfnamefont {T.}~\bibnamefont {O'Brien}}, \bibinfo {author}
  {\bibfnamefont {A.}~\bibnamefont {Obuljen}}, \bibinfo {author} {\bibfnamefont
  {N.}~\bibnamefont {{Palanque-Delabrouille}}}, \bibinfo {author}
  {\bibfnamefont {A.}~\bibnamefont {Palmese}}, \bibinfo {author} {\bibfnamefont
  {P.}~\bibnamefont {Zhiwei}}, \bibinfo {author} {\bibfnamefont
  {D.}~\bibnamefont {Pappalardo}}, \bibinfo {author} {\bibfnamefont
  {X.}~\bibnamefont {Peng}}, \bibinfo {author} {\bibfnamefont {W.~J.}\
  \bibnamefont {Percival}}, \bibinfo {author} {\bibfnamefont {S.}~\bibnamefont
  {Perruchot}}, \bibinfo {author} {\bibfnamefont {R.}~\bibnamefont {Pogge}},
  \bibinfo {author} {\bibfnamefont {C.}~\bibnamefont {Poppett}}, \bibinfo
  {author} {\bibfnamefont {A.}~\bibnamefont {Porredon}}, \bibinfo {author}
  {\bibfnamefont {F.}~\bibnamefont {Prada}}, \bibinfo {author} {\bibfnamefont
  {J.}~\bibnamefont {Prochaska}}, \bibinfo {author} {\bibfnamefont
  {R.}~\bibnamefont {Pucha}}, \bibinfo {author} {\bibfnamefont
  {A.}~\bibnamefont {{P{\'e}rez-Fern{\'a}ndez}}}, \bibinfo {author}
  {\bibfnamefont {I.}~\bibnamefont {{P{\'e}rez-R{\'a}fols}}}, \bibinfo {author}
  {\bibfnamefont {D.}~\bibnamefont {Rabinowitz}}, \bibinfo {author}
  {\bibfnamefont {A.}~\bibnamefont {Raichoor}}, \bibinfo {author}
  {\bibfnamefont {S.}~\bibnamefont {{Ramirez-Solano}}}, \bibinfo {author}
  {\bibfnamefont {C.}~\bibnamefont {{Ram{\'i}rez-P{\'e}rez}}}, \bibinfo
  {author} {\bibfnamefont {C.}~\bibnamefont {Ravoux}}, \bibinfo {author}
  {\bibfnamefont {K.}~\bibnamefont {Reil}}, \bibinfo {author} {\bibfnamefont
  {M.}~\bibnamefont {Rezaie}}, \bibinfo {author} {\bibfnamefont
  {A.}~\bibnamefont {Rocher}}, \bibinfo {author} {\bibfnamefont
  {C.}~\bibnamefont {Rockosi}}, \bibinfo {author} {\bibfnamefont {N.~A.}\
  \bibnamefont {Roe}}, \bibinfo {author} {\bibfnamefont {A.}~\bibnamefont
  {Roodman}}, \bibinfo {author} {\bibfnamefont {A.~J.}\ \bibnamefont {Ross}},
  \bibinfo {author} {\bibfnamefont {G.}~\bibnamefont {Rossi}}, \bibinfo
  {author} {\bibfnamefont {R.}~\bibnamefont {Ruggeri}}, \bibinfo {author}
  {\bibfnamefont {V.}~\bibnamefont {{Ruhlmann-Kleider}}}, \bibinfo {author}
  {\bibfnamefont {C.~G.}\ \bibnamefont {Sabiu}}, \bibinfo {author}
  {\bibfnamefont {S.}~\bibnamefont {Safonova}}, \bibinfo {author}
  {\bibfnamefont {K.}~\bibnamefont {Said}}, \bibinfo {author} {\bibfnamefont
  {A.}~\bibnamefont {Saintonge}}, \bibinfo {author} {\bibfnamefont {J.~S.}\
  \bibnamefont {Catonga}}, \bibinfo {author} {\bibfnamefont {L.}~\bibnamefont
  {Samushia}}, \bibinfo {author} {\bibfnamefont {E.}~\bibnamefont {Sanchez}},
  \bibinfo {author} {\bibfnamefont {C.}~\bibnamefont {Saulder}}, \bibinfo
  {author} {\bibfnamefont {E.}~\bibnamefont {Schaan}}, \bibinfo {author}
  {\bibfnamefont {E.}~\bibnamefont {Schlafly}}, \bibinfo {author}
  {\bibfnamefont {D.}~\bibnamefont {Schlegel}}, \bibinfo {author}
  {\bibfnamefont {J.}~\bibnamefont {Schmoll}}, \bibinfo {author} {\bibfnamefont
  {D.}~\bibnamefont {Scholte}}, \bibinfo {author} {\bibfnamefont
  {M.}~\bibnamefont {Schubnell}}, \bibinfo {author} {\bibfnamefont
  {A.}~\bibnamefont {Secroun}}, \bibinfo {author} {\bibfnamefont
  {H.}~\bibnamefont {Seo}}, \bibinfo {author} {\bibfnamefont {S.}~\bibnamefont
  {Serrano}}, \bibinfo {author} {\bibfnamefont {R.~M.}\ \bibnamefont
  {Sharples}}, \bibinfo {author} {\bibfnamefont {M.~J.}\ \bibnamefont {Sholl}},
  \bibinfo {author} {\bibfnamefont {J.~H.}\ \bibnamefont {Silber}}, \bibinfo
  {author} {\bibfnamefont {D.~R.}\ \bibnamefont {Silva}}, \bibinfo {author}
  {\bibfnamefont {M.}~\bibnamefont {Sirk}}, \bibinfo {author} {\bibfnamefont
  {M.}~\bibnamefont {Siudek}}, \bibinfo {author} {\bibfnamefont
  {A.}~\bibnamefont {Smith}}, \bibinfo {author} {\bibfnamefont
  {D.}~\bibnamefont {Sprayberry}}, \bibinfo {author} {\bibfnamefont
  {R.}~\bibnamefont {Staten}}, \bibinfo {author} {\bibfnamefont
  {B.}~\bibnamefont {Stupak}}, \bibinfo {author} {\bibfnamefont
  {T.}~\bibnamefont {Tan}}, \bibinfo {author} {\bibfnamefont {G.}~\bibnamefont
  {Tarl{\'e}}}, \bibinfo {author} {\bibfnamefont {S.~S.}\ \bibnamefont {Tie}},
  \bibinfo {author} {\bibfnamefont {R.}~\bibnamefont {Tojeiro}}, \bibinfo
  {author} {\bibfnamefont {L.~A.}\ \bibnamefont {{Ure{\~n}a-L{\'o}pez}}},
  \bibinfo {author} {\bibfnamefont {F.}~\bibnamefont {Valdes}}, \bibinfo
  {author} {\bibfnamefont {O.}~\bibnamefont {Valenzuela}}, \bibinfo {author}
  {\bibfnamefont {M.}~\bibnamefont {Valluri}}, \bibinfo {author} {\bibfnamefont
  {M.}~\bibnamefont {{Vargas-Maga{\~n}a}}}, \bibinfo {author} {\bibfnamefont
  {L.}~\bibnamefont {Verde}}, \bibinfo {author} {\bibfnamefont
  {M.}~\bibnamefont {Walther}}, \bibinfo {author} {\bibfnamefont
  {B.}~\bibnamefont {Wang}}, \bibinfo {author} {\bibfnamefont {M.~S.}\
  \bibnamefont {Wang}}, \bibinfo {author} {\bibfnamefont {B.~A.}\ \bibnamefont
  {Weaver}}, \bibinfo {author} {\bibfnamefont {C.}~\bibnamefont {Weaverdyck}},
  \bibinfo {author} {\bibfnamefont {R.}~\bibnamefont {Wechsler}}, \bibinfo
  {author} {\bibfnamefont {M.~J.}\ \bibnamefont {Wilson}}, \bibinfo {author}
  {\bibfnamefont {J.}~\bibnamefont {Yang}}, \bibinfo {author} {\bibfnamefont
  {Y.}~\bibnamefont {Yu}}, \bibinfo {author} {\bibfnamefont {S.}~\bibnamefont
  {Yuan}}, \bibinfo {author} {\bibfnamefont {C.}~\bibnamefont {Y{\`e}che}},
  \bibinfo {author} {\bibfnamefont {H.}~\bibnamefont {Zhang}}, \bibinfo
  {author} {\bibfnamefont {K.}~\bibnamefont {Zhang}}, \bibinfo {author}
  {\bibfnamefont {C.}~\bibnamefont {Zhao}}, \bibinfo {author} {\bibfnamefont
  {R.}~\bibnamefont {Zhou}}, \bibinfo {author} {\bibfnamefont {Z.}~\bibnamefont
  {Zhou}}, \bibinfo {author} {\bibfnamefont {H.}~\bibnamefont {Zou}}, \bibinfo
  {author} {\bibfnamefont {J.}~\bibnamefont {Zou}}, \bibinfo {author}
  {\bibfnamefont {S.}~\bibnamefont {Zou}}, \ and\ \bibinfo {author}
  {\bibfnamefont {Y.}~\bibnamefont {Zu}},\ }\href {\doibase
  10.48550/arXiv.2205.10939} {\enquote {\bibinfo {title} {Overview of the
  {{Instrumentation}} for the {{Dark Energy Spectroscopic Instrument}}},}\ }
  (\bibinfo {year} {2022}),\ \Eprint {http://arxiv.org/abs/2205.10939}
  {arXiv:2205.10939 [astro-ph]} \BibitemShut {NoStop}%
\bibitem [{\citenamefont {Takada}\ \emph {et~al.}(2014)\citenamefont {Takada},
  \citenamefont {Ellis}, \citenamefont {Chiba}, \citenamefont {Greene},
  \citenamefont {Aihara}, \citenamefont {Arimoto}, \citenamefont {Bundy},
  \citenamefont {Cohen}, \citenamefont {Dor{\'e}}, \citenamefont {Graves},
  \citenamefont {Gunn}, \citenamefont {Heckman}, \citenamefont {Hirata},
  \citenamefont {Ho}, \citenamefont {Kneib}, \citenamefont {Le~F{\`e}vre},
  \citenamefont {Lin}, \citenamefont {More}, \citenamefont {Murayama},
  \citenamefont {Nagao}, \citenamefont {Ouchi}, \citenamefont {Seiffert},
  \citenamefont {Silverman}, \citenamefont {Sodr{\'e}}, \citenamefont
  {Spergel}, \citenamefont {Strauss}, \citenamefont {Sugai}, \citenamefont
  {Suto}, \citenamefont {Takami},\ and\ \citenamefont {Wyse}}]{takada2014}%
  \BibitemOpen
  \bibfield  {author} {\bibinfo {author} {\bibfnamefont {M.}~\bibnamefont
  {Takada}}, \bibinfo {author} {\bibfnamefont {R.~S.}\ \bibnamefont {Ellis}},
  \bibinfo {author} {\bibfnamefont {M.}~\bibnamefont {Chiba}}, \bibinfo
  {author} {\bibfnamefont {J.~E.}\ \bibnamefont {Greene}}, \bibinfo {author}
  {\bibfnamefont {H.}~\bibnamefont {Aihara}}, \bibinfo {author} {\bibfnamefont
  {N.}~\bibnamefont {Arimoto}}, \bibinfo {author} {\bibfnamefont
  {K.}~\bibnamefont {Bundy}}, \bibinfo {author} {\bibfnamefont
  {J.}~\bibnamefont {Cohen}}, \bibinfo {author} {\bibfnamefont
  {O.}~\bibnamefont {Dor{\'e}}}, \bibinfo {author} {\bibfnamefont
  {G.}~\bibnamefont {Graves}}, \bibinfo {author} {\bibfnamefont {J.~E.}\
  \bibnamefont {Gunn}}, \bibinfo {author} {\bibfnamefont {T.}~\bibnamefont
  {Heckman}}, \bibinfo {author} {\bibfnamefont {C.~M.}\ \bibnamefont {Hirata}},
  \bibinfo {author} {\bibfnamefont {P.}~\bibnamefont {Ho}}, \bibinfo {author}
  {\bibfnamefont {J.-P.}\ \bibnamefont {Kneib}}, \bibinfo {author}
  {\bibfnamefont {O.}~\bibnamefont {Le~F{\`e}vre}}, \bibinfo {author}
  {\bibfnamefont {L.}~\bibnamefont {Lin}}, \bibinfo {author} {\bibfnamefont
  {S.}~\bibnamefont {More}}, \bibinfo {author} {\bibfnamefont {H.}~\bibnamefont
  {Murayama}}, \bibinfo {author} {\bibfnamefont {T.}~\bibnamefont {Nagao}},
  \bibinfo {author} {\bibfnamefont {M.}~\bibnamefont {Ouchi}}, \bibinfo
  {author} {\bibfnamefont {M.}~\bibnamefont {Seiffert}}, \bibinfo {author}
  {\bibfnamefont {J.~D.}\ \bibnamefont {Silverman}}, \bibinfo {author}
  {\bibfnamefont {L.}~\bibnamefont {Sodr{\'e}}}, \bibinfo {author}
  {\bibfnamefont {D.~N.}\ \bibnamefont {Spergel}}, \bibinfo {author}
  {\bibfnamefont {M.~A.}\ \bibnamefont {Strauss}}, \bibinfo {author}
  {\bibfnamefont {H.}~\bibnamefont {Sugai}}, \bibinfo {author} {\bibfnamefont
  {Y.}~\bibnamefont {Suto}}, \bibinfo {author} {\bibfnamefont {H.}~\bibnamefont
  {Takami}}, \ and\ \bibinfo {author} {\bibfnamefont {R.}~\bibnamefont
  {Wyse}},\ }\href {\doibase 10.1093/pasj/pst019} {\bibfield  {journal}
  {\bibinfo  {journal} {Publications of the Astronomical Society of Japan}\
  }\textbf {\bibinfo {volume} {66}},\ \bibinfo {pages} {R1} (\bibinfo {year}
  {2014})}\BibitemShut {NoStop}%
\bibitem [{\citenamefont {Tamura}\ \emph {et~al.}(2016)\citenamefont {Tamura},
  \citenamefont {Takato}, \citenamefont {Shimono}, \citenamefont {Moritani},
  \citenamefont {Yabe}, \citenamefont {Ishizuka}, \citenamefont {Ueda},
  \citenamefont {Kamata}, \citenamefont {Aghazarian}, \citenamefont {Arnouts},
  \citenamefont {Barban}, \citenamefont {Barkhouser}, \citenamefont {Borges},
  \citenamefont {Braun}, \citenamefont {Carr}, \citenamefont {Chabaud},
  \citenamefont {Chang}, \citenamefont {Chen}, \citenamefont {Chiba},
  \citenamefont {Chou}, \citenamefont {Chu}, \citenamefont {Cohen},
  \citenamefont {{de Almeida}}, \citenamefont {{de Oliveira}}, \citenamefont
  {{de Oliveira}}, \citenamefont {Dekany}, \citenamefont {Dohlen},
  \citenamefont {{dos Santos}}, \citenamefont {{dos Santos}}, \citenamefont
  {Ellis}, \citenamefont {Fabricius}, \citenamefont {Ferrand}, \citenamefont
  {Ferreira}, \citenamefont {Golebiowski}, \citenamefont {Greene},
  \citenamefont {Gross}, \citenamefont {Gunn}, \citenamefont {Hammond},
  \citenamefont {Harding}, \citenamefont {Hart}, \citenamefont {Heckman},
  \citenamefont {Hirata}, \citenamefont {Ho}, \citenamefont {Hope},
  \citenamefont {Hovland}, \citenamefont {Hsu}, \citenamefont {Hu},
  \citenamefont {Huang}, \citenamefont {Jaquet}, \citenamefont {Jing},
  \citenamefont {Karr}, \citenamefont {Kimura}, \citenamefont {King},
  \citenamefont {Komatsu}, \citenamefont {Le~Brun}, \citenamefont
  {Le~F{\`e}vre}, \citenamefont {Le~Fur}, \citenamefont {Le~Mignant},
  \citenamefont {Ling}, \citenamefont {Loomis}, \citenamefont {Lupton},
  \citenamefont {Madec}, \citenamefont {Mao}, \citenamefont {Marrara},
  \citenamefont {{Mendes de Oliveira}}, \citenamefont {Minowa}, \citenamefont
  {Morantz}, \citenamefont {Murayama}, \citenamefont {Murray}, \citenamefont
  {Ohyama}, \citenamefont {Orndorff}, \citenamefont {Pascal}, \citenamefont
  {Pereira}, \citenamefont {Reiley}, \citenamefont {Reinecke}, \citenamefont
  {Ritter}, \citenamefont {Roberts}, \citenamefont {Schwochert}, \citenamefont
  {Seiffert}, \citenamefont {Smee}, \citenamefont {Sodre}, \citenamefont
  {Spergel}, \citenamefont {Steinkraus}, \citenamefont {Strauss}, \citenamefont
  {Surace}, \citenamefont {Suto}, \citenamefont {Suzuki}, \citenamefont
  {Swinbank}, \citenamefont {Tait}, \citenamefont {Takada}, \citenamefont
  {Tamura}, \citenamefont {Tanaka}, \citenamefont {Tresse}, \citenamefont
  {Verducci}, \citenamefont {Vibert}, \citenamefont {Vidal}, \citenamefont
  {Wang}, \citenamefont {Wen}, \citenamefont {Yan},\ and\ \citenamefont
  {Yasuda}}]{tamura2016}%
  \BibitemOpen
  \bibfield  {author} {\bibinfo {author} {\bibfnamefont {N.}~\bibnamefont
  {Tamura}}, \bibinfo {author} {\bibfnamefont {N.}~\bibnamefont {Takato}},
  \bibinfo {author} {\bibfnamefont {A.}~\bibnamefont {Shimono}}, \bibinfo
  {author} {\bibfnamefont {Y.}~\bibnamefont {Moritani}}, \bibinfo {author}
  {\bibfnamefont {K.}~\bibnamefont {Yabe}}, \bibinfo {author} {\bibfnamefont
  {Y.}~\bibnamefont {Ishizuka}}, \bibinfo {author} {\bibfnamefont
  {A.}~\bibnamefont {Ueda}}, \bibinfo {author} {\bibfnamefont {Y.}~\bibnamefont
  {Kamata}}, \bibinfo {author} {\bibfnamefont {H.}~\bibnamefont {Aghazarian}},
  \bibinfo {author} {\bibfnamefont {S.}~\bibnamefont {Arnouts}}, \bibinfo
  {author} {\bibfnamefont {G.}~\bibnamefont {Barban}}, \bibinfo {author}
  {\bibfnamefont {R.~H.}\ \bibnamefont {Barkhouser}}, \bibinfo {author}
  {\bibfnamefont {R.~C.}\ \bibnamefont {Borges}}, \bibinfo {author}
  {\bibfnamefont {D.~F.}\ \bibnamefont {Braun}}, \bibinfo {author}
  {\bibfnamefont {M.~A.}\ \bibnamefont {Carr}}, \bibinfo {author}
  {\bibfnamefont {P.-Y.}\ \bibnamefont {Chabaud}}, \bibinfo {author}
  {\bibfnamefont {Y.-C.}\ \bibnamefont {Chang}}, \bibinfo {author}
  {\bibfnamefont {H.-Y.}\ \bibnamefont {Chen}}, \bibinfo {author}
  {\bibfnamefont {M.}~\bibnamefont {Chiba}}, \bibinfo {author} {\bibfnamefont
  {R.~C.~Y.}\ \bibnamefont {Chou}}, \bibinfo {author} {\bibfnamefont {Y.-H.}\
  \bibnamefont {Chu}}, \bibinfo {author} {\bibfnamefont {J.}~\bibnamefont
  {Cohen}}, \bibinfo {author} {\bibfnamefont {R.~P.}\ \bibnamefont {{de
  Almeida}}}, \bibinfo {author} {\bibfnamefont {A.~C.}\ \bibnamefont {{de
  Oliveira}}}, \bibinfo {author} {\bibfnamefont {L.~S.}\ \bibnamefont {{de
  Oliveira}}}, \bibinfo {author} {\bibfnamefont {R.~G.}\ \bibnamefont
  {Dekany}}, \bibinfo {author} {\bibfnamefont {K.}~\bibnamefont {Dohlen}},
  \bibinfo {author} {\bibfnamefont {J.~B.}\ \bibnamefont {{dos Santos}}},
  \bibinfo {author} {\bibfnamefont {L.~H.}\ \bibnamefont {{dos Santos}}},
  \bibinfo {author} {\bibfnamefont {R.}~\bibnamefont {Ellis}}, \bibinfo
  {author} {\bibfnamefont {M.}~\bibnamefont {Fabricius}}, \bibinfo {author}
  {\bibfnamefont {D.}~\bibnamefont {Ferrand}}, \bibinfo {author} {\bibfnamefont
  {D.}~\bibnamefont {Ferreira}}, \bibinfo {author} {\bibfnamefont
  {M.}~\bibnamefont {Golebiowski}}, \bibinfo {author} {\bibfnamefont {J.~E.}\
  \bibnamefont {Greene}}, \bibinfo {author} {\bibfnamefont {J.}~\bibnamefont
  {Gross}}, \bibinfo {author} {\bibfnamefont {J.~E.}\ \bibnamefont {Gunn}},
  \bibinfo {author} {\bibfnamefont {R.}~\bibnamefont {Hammond}}, \bibinfo
  {author} {\bibfnamefont {A.}~\bibnamefont {Harding}}, \bibinfo {author}
  {\bibfnamefont {M.}~\bibnamefont {Hart}}, \bibinfo {author} {\bibfnamefont
  {T.~M.}\ \bibnamefont {Heckman}}, \bibinfo {author} {\bibfnamefont {C.~M.}\
  \bibnamefont {Hirata}}, \bibinfo {author} {\bibfnamefont {P.}~\bibnamefont
  {Ho}}, \bibinfo {author} {\bibfnamefont {S.~C.}\ \bibnamefont {Hope}},
  \bibinfo {author} {\bibfnamefont {L.}~\bibnamefont {Hovland}}, \bibinfo
  {author} {\bibfnamefont {S.-F.}\ \bibnamefont {Hsu}}, \bibinfo {author}
  {\bibfnamefont {Y.-S.}\ \bibnamefont {Hu}}, \bibinfo {author} {\bibfnamefont
  {P.-J.}\ \bibnamefont {Huang}}, \bibinfo {author} {\bibfnamefont
  {M.}~\bibnamefont {Jaquet}}, \bibinfo {author} {\bibfnamefont
  {Y.}~\bibnamefont {Jing}}, \bibinfo {author} {\bibfnamefont {J.}~\bibnamefont
  {Karr}}, \bibinfo {author} {\bibfnamefont {M.}~\bibnamefont {Kimura}},
  \bibinfo {author} {\bibfnamefont {M.~E.}\ \bibnamefont {King}}, \bibinfo
  {author} {\bibfnamefont {E.}~\bibnamefont {Komatsu}}, \bibinfo {author}
  {\bibfnamefont {V.}~\bibnamefont {Le~Brun}}, \bibinfo {author} {\bibfnamefont
  {O.}~\bibnamefont {Le~F{\`e}vre}}, \bibinfo {author} {\bibfnamefont
  {A.}~\bibnamefont {Le~Fur}}, \bibinfo {author} {\bibfnamefont
  {D.}~\bibnamefont {Le~Mignant}}, \bibinfo {author} {\bibfnamefont {H.-H.}\
  \bibnamefont {Ling}}, \bibinfo {author} {\bibfnamefont {C.~P.}\ \bibnamefont
  {Loomis}}, \bibinfo {author} {\bibfnamefont {R.~H.}\ \bibnamefont {Lupton}},
  \bibinfo {author} {\bibfnamefont {F.}~\bibnamefont {Madec}}, \bibinfo
  {author} {\bibfnamefont {P.}~\bibnamefont {Mao}}, \bibinfo {author}
  {\bibfnamefont {L.~S.}\ \bibnamefont {Marrara}}, \bibinfo {author}
  {\bibfnamefont {C.}~\bibnamefont {{Mendes de Oliveira}}}, \bibinfo {author}
  {\bibfnamefont {Y.}~\bibnamefont {Minowa}}, \bibinfo {author} {\bibfnamefont
  {C.}~\bibnamefont {Morantz}}, \bibinfo {author} {\bibfnamefont
  {H.}~\bibnamefont {Murayama}}, \bibinfo {author} {\bibfnamefont {G.~J.}\
  \bibnamefont {Murray}}, \bibinfo {author} {\bibfnamefont {Y.}~\bibnamefont
  {Ohyama}}, \bibinfo {author} {\bibfnamefont {J.}~\bibnamefont {Orndorff}},
  \bibinfo {author} {\bibfnamefont {S.}~\bibnamefont {Pascal}}, \bibinfo
  {author} {\bibfnamefont {J.~M.}\ \bibnamefont {Pereira}}, \bibinfo {author}
  {\bibfnamefont {D.}~\bibnamefont {Reiley}}, \bibinfo {author} {\bibfnamefont
  {M.}~\bibnamefont {Reinecke}}, \bibinfo {author} {\bibfnamefont
  {A.}~\bibnamefont {Ritter}}, \bibinfo {author} {\bibfnamefont
  {M.}~\bibnamefont {Roberts}}, \bibinfo {author} {\bibfnamefont {M.~A.}\
  \bibnamefont {Schwochert}}, \bibinfo {author} {\bibfnamefont {M.~D.}\
  \bibnamefont {Seiffert}}, \bibinfo {author} {\bibfnamefont {S.~A.}\
  \bibnamefont {Smee}}, \bibinfo {author} {\bibfnamefont {L.}~\bibnamefont
  {Sodre}}, \bibinfo {author} {\bibfnamefont {D.~N.}\ \bibnamefont {Spergel}},
  \bibinfo {author} {\bibfnamefont {A.~J.}\ \bibnamefont {Steinkraus}},
  \bibinfo {author} {\bibfnamefont {M.~A.}\ \bibnamefont {Strauss}}, \bibinfo
  {author} {\bibfnamefont {C.}~\bibnamefont {Surace}}, \bibinfo {author}
  {\bibfnamefont {Y.}~\bibnamefont {Suto}}, \bibinfo {author} {\bibfnamefont
  {N.}~\bibnamefont {Suzuki}}, \bibinfo {author} {\bibfnamefont
  {J.}~\bibnamefont {Swinbank}}, \bibinfo {author} {\bibfnamefont {P.~J.}\
  \bibnamefont {Tait}}, \bibinfo {author} {\bibfnamefont {M.}~\bibnamefont
  {Takada}}, \bibinfo {author} {\bibfnamefont {T.}~\bibnamefont {Tamura}},
  \bibinfo {author} {\bibfnamefont {Y.}~\bibnamefont {Tanaka}}, \bibinfo
  {author} {\bibfnamefont {L.}~\bibnamefont {Tresse}}, \bibinfo {author}
  {\bibfnamefont {O.}~\bibnamefont {Verducci}}, \bibinfo {author}
  {\bibfnamefont {D.}~\bibnamefont {Vibert}}, \bibinfo {author} {\bibfnamefont
  {C.}~\bibnamefont {Vidal}}, \bibinfo {author} {\bibfnamefont {S.-Y.}\
  \bibnamefont {Wang}}, \bibinfo {author} {\bibfnamefont {C.-Y.}\ \bibnamefont
  {Wen}}, \bibinfo {author} {\bibfnamefont {C.-H.}\ \bibnamefont {Yan}}, \ and\
  \bibinfo {author} {\bibfnamefont {N.}~\bibnamefont {Yasuda}},\ }in\ \href
  {\doibase 10.1117/12.2232103} {\emph {\bibinfo {booktitle} {Ground-Based and
  {{Airborne Instrumentation}} for {{Astronomy VI}}}}},\ Vol.\ \bibinfo
  {volume} {9908}\ (\bibinfo {address} {{eprint: arXiv:1608.01075}},\ \bibinfo
  {year} {2016})\ p.\ \bibinfo {pages} {99081M}\BibitemShut {NoStop}%
\bibitem [{\citenamefont {Laureijs}\ \emph {et~al.}(2011)\citenamefont
  {Laureijs}, \citenamefont {Amiaux}, \citenamefont {Arduini}, \citenamefont
  {Augu{\`e}res}, \citenamefont {Brinchmann}, \citenamefont {Cole},
  \citenamefont {Cropper}, \citenamefont {Dabin}, \citenamefont {Duvet},
  \citenamefont {Ealet}, \citenamefont {Garilli}, \citenamefont {Gondoin},
  \citenamefont {Guzzo}, \citenamefont {Hoar}, \citenamefont {Hoekstra},
  \citenamefont {Holmes}, \citenamefont {Kitching}, \citenamefont {Maciaszek},
  \citenamefont {Mellier}, \citenamefont {Pasian}, \citenamefont {Percival},
  \citenamefont {Rhodes}, \citenamefont {Saavedra~Criado}, \citenamefont
  {Sauvage}, \citenamefont {Scaramella}, \citenamefont {Valenziano},
  \citenamefont {Warren}, \citenamefont {Bender}, \citenamefont {Castander},
  \citenamefont {Cimatti}, \citenamefont {Le~F{\`e}vre}, \citenamefont
  {{Kurki-Suonio}}, \citenamefont {Levi}, \citenamefont {Lilje}, \citenamefont
  {Meylan}, \citenamefont {Nichol}, \citenamefont {Pedersen}, \citenamefont
  {Popa}, \citenamefont {Rebolo~Lopez}, \citenamefont {Rix}, \citenamefont
  {Rottgering}, \citenamefont {Zeilinger}, \citenamefont {Grupp}, \citenamefont
  {Hudelot}, \citenamefont {Massey}, \citenamefont {Meneghetti}, \citenamefont
  {Miller}, \citenamefont {Paltani}, \citenamefont {{Paulin-Henriksson}},
  \citenamefont {Pires}, \citenamefont {Saxton}, \citenamefont {Schrabback},
  \citenamefont {Seidel}, \citenamefont {Walsh}, \citenamefont {Aghanim},
  \citenamefont {Amendola}, \citenamefont {Bartlett}, \citenamefont
  {Baccigalupi}, \citenamefont {Beaulieu}, \citenamefont {Benabed},
  \citenamefont {Cuby}, \citenamefont {Elbaz}, \citenamefont {Fosalba},
  \citenamefont {Gavazzi}, \citenamefont {Helmi}, \citenamefont {Hook},
  \citenamefont {Irwin}, \citenamefont {Kneib}, \citenamefont {Kunz},
  \citenamefont {Mannucci}, \citenamefont {Moscardini}, \citenamefont {Tao},
  \citenamefont {Teyssier}, \citenamefont {Weller}, \citenamefont {Zamorani},
  \citenamefont {Zapatero~Osorio}, \citenamefont {Boulade}, \citenamefont
  {Foumond}, \citenamefont {Di~Giorgio}, \citenamefont {Guttridge},
  \citenamefont {James}, \citenamefont {Kemp}, \citenamefont {Martignac},
  \citenamefont {Spencer}, \citenamefont {Walton}, \citenamefont
  {Bl{\"u}mchen}, \citenamefont {Bonoli}, \citenamefont {Bortoletto},
  \citenamefont {Cerna}, \citenamefont {Corcione}, \citenamefont {Fabron},
  \citenamefont {Jahnke}, \citenamefont {Ligori}, \citenamefont {Madrid},
  \citenamefont {Martin}, \citenamefont {Morgante}, \citenamefont {Pamplona},
  \citenamefont {Prieto}, \citenamefont {Riva}, \citenamefont {Toledo},
  \citenamefont {Trifoglio}, \citenamefont {Zerbi}, \citenamefont {Abdalla},
  \citenamefont {Douspis}, \citenamefont {Grenet}, \citenamefont {Borgani},
  \citenamefont {Bouwens}, \citenamefont {Courbin}, \citenamefont {Delouis},
  \citenamefont {Dubath}, \citenamefont {Fontana}, \citenamefont {Frailis},
  \citenamefont {Grazian}, \citenamefont {Koppenh{\"o}fer}, \citenamefont
  {Mansutti}, \citenamefont {Melchior}, \citenamefont {Mignoli}, \citenamefont
  {Mohr}, \citenamefont {Neissner}, \citenamefont {Noddle}, \citenamefont
  {Poncet}, \citenamefont {Scodeggio}, \citenamefont {Serrano}, \citenamefont
  {Shane}, \citenamefont {Starck}, \citenamefont {Surace}, \citenamefont
  {Taylor}, \citenamefont {{Verdoes-Kleijn}}, \citenamefont {Vuerli},
  \citenamefont {Williams}, \citenamefont {Zacchei}, \citenamefont {Altieri},
  \citenamefont {Escudero~Sanz}, \citenamefont {Kohley}, \citenamefont
  {Oosterbroek}, \citenamefont {Astier}, \citenamefont {Bacon}, \citenamefont
  {Bardelli}, \citenamefont {Baugh}, \citenamefont {Bellagamba}, \citenamefont
  {Benoist}, \citenamefont {Bianchi}, \citenamefont {Biviano}, \citenamefont
  {Branchini}, \citenamefont {Carbone}, \citenamefont {Cardone}, \citenamefont
  {Clements}, \citenamefont {Colombi}, \citenamefont {Conselice}, \citenamefont
  {Cresci}, \citenamefont {Deacon}, \citenamefont {Dunlop}, \citenamefont
  {Fedeli}, \citenamefont {Fontanot}, \citenamefont {Franzetti}, \citenamefont
  {Giocoli}, \citenamefont {{Garcia-Bellido}}, \citenamefont {Gow},
  \citenamefont {Heavens}, \citenamefont {Hewett}, \citenamefont {Heymans},
  \citenamefont {Holland}, \citenamefont {Huang}, \citenamefont {Ilbert},
  \citenamefont {Joachimi}, \citenamefont {Jennins}, \citenamefont {Kerins},
  \citenamefont {Kiessling}, \citenamefont {Kirk}, \citenamefont {Kotak},
  \citenamefont {Krause}, \citenamefont {Lahav}, \citenamefont {{van Leeuwen}},
  \citenamefont {Lesgourgues}, \citenamefont {Lombardi}, \citenamefont
  {Magliocchetti}, \citenamefont {Maguire}, \citenamefont {Majerotto},
  \citenamefont {Maoli}, \citenamefont {Marulli}, \citenamefont {Maurogordato},
  \citenamefont {McCracken}, \citenamefont {McLure}, \citenamefont
  {Melchiorri}, \citenamefont {Merson}, \citenamefont {Moresco}, \citenamefont
  {Nonino}, \citenamefont {Norberg}, \citenamefont {Peacock}, \citenamefont
  {Pello}, \citenamefont {Penny}, \citenamefont {Pettorino}, \citenamefont
  {Di~Porto}, \citenamefont {Pozzetti}, \citenamefont {Quercellini},
  \citenamefont {Radovich}, \citenamefont {Rassat}, \citenamefont {Roche},
  \citenamefont {Ronayette}, \citenamefont {Rossetti}, \citenamefont
  {Sartoris}, \citenamefont {Schneider}, \citenamefont {Semboloni},
  \citenamefont {Serjeant}, \citenamefont {Simpson}, \citenamefont {Skordis},
  \citenamefont {Smadja}, \citenamefont {Smartt}, \citenamefont {Spano},
  \citenamefont {Spiro}, \citenamefont {Sullivan}, \citenamefont {Tilquin},
  \citenamefont {Trotta}, \citenamefont {Verde}, \citenamefont {Wang},
  \citenamefont {Williger}, \citenamefont {Zhao}, \citenamefont {Zoubian},\
  and\ \citenamefont {Zucca}}]{laureijs2011}%
  \BibitemOpen
  \bibfield  {author} {\bibinfo {author} {\bibfnamefont {R.}~\bibnamefont
  {Laureijs}}, \bibinfo {author} {\bibfnamefont {J.}~\bibnamefont {Amiaux}},
  \bibinfo {author} {\bibfnamefont {S.}~\bibnamefont {Arduini}}, \bibinfo
  {author} {\bibfnamefont {J.-L.}\ \bibnamefont {Augu{\`e}res}}, \bibinfo
  {author} {\bibfnamefont {J.}~\bibnamefont {Brinchmann}}, \bibinfo {author}
  {\bibfnamefont {R.}~\bibnamefont {Cole}}, \bibinfo {author} {\bibfnamefont
  {M.}~\bibnamefont {Cropper}}, \bibinfo {author} {\bibfnamefont
  {C.}~\bibnamefont {Dabin}}, \bibinfo {author} {\bibfnamefont
  {L.}~\bibnamefont {Duvet}}, \bibinfo {author} {\bibfnamefont
  {A.}~\bibnamefont {Ealet}}, \bibinfo {author} {\bibfnamefont
  {B.}~\bibnamefont {Garilli}}, \bibinfo {author} {\bibfnamefont
  {P.}~\bibnamefont {Gondoin}}, \bibinfo {author} {\bibfnamefont
  {L.}~\bibnamefont {Guzzo}}, \bibinfo {author} {\bibfnamefont
  {J.}~\bibnamefont {Hoar}}, \bibinfo {author} {\bibfnamefont {H.}~\bibnamefont
  {Hoekstra}}, \bibinfo {author} {\bibfnamefont {R.}~\bibnamefont {Holmes}},
  \bibinfo {author} {\bibfnamefont {T.}~\bibnamefont {Kitching}}, \bibinfo
  {author} {\bibfnamefont {T.}~\bibnamefont {Maciaszek}}, \bibinfo {author}
  {\bibfnamefont {Y.}~\bibnamefont {Mellier}}, \bibinfo {author} {\bibfnamefont
  {F.}~\bibnamefont {Pasian}}, \bibinfo {author} {\bibfnamefont
  {W.}~\bibnamefont {Percival}}, \bibinfo {author} {\bibfnamefont
  {J.}~\bibnamefont {Rhodes}}, \bibinfo {author} {\bibfnamefont
  {G.}~\bibnamefont {Saavedra~Criado}}, \bibinfo {author} {\bibfnamefont
  {M.}~\bibnamefont {Sauvage}}, \bibinfo {author} {\bibfnamefont
  {R.}~\bibnamefont {Scaramella}}, \bibinfo {author} {\bibfnamefont
  {L.}~\bibnamefont {Valenziano}}, \bibinfo {author} {\bibfnamefont
  {S.}~\bibnamefont {Warren}}, \bibinfo {author} {\bibfnamefont
  {R.}~\bibnamefont {Bender}}, \bibinfo {author} {\bibfnamefont
  {F.}~\bibnamefont {Castander}}, \bibinfo {author} {\bibfnamefont
  {A.}~\bibnamefont {Cimatti}}, \bibinfo {author} {\bibfnamefont
  {O.}~\bibnamefont {Le~F{\`e}vre}}, \bibinfo {author} {\bibfnamefont
  {H.}~\bibnamefont {{Kurki-Suonio}}}, \bibinfo {author} {\bibfnamefont
  {M.}~\bibnamefont {Levi}}, \bibinfo {author} {\bibfnamefont {P.}~\bibnamefont
  {Lilje}}, \bibinfo {author} {\bibfnamefont {G.}~\bibnamefont {Meylan}},
  \bibinfo {author} {\bibfnamefont {R.}~\bibnamefont {Nichol}}, \bibinfo
  {author} {\bibfnamefont {K.}~\bibnamefont {Pedersen}}, \bibinfo {author}
  {\bibfnamefont {V.}~\bibnamefont {Popa}}, \bibinfo {author} {\bibfnamefont
  {R.}~\bibnamefont {Rebolo~Lopez}}, \bibinfo {author} {\bibfnamefont {H.-W.}\
  \bibnamefont {Rix}}, \bibinfo {author} {\bibfnamefont {H.}~\bibnamefont
  {Rottgering}}, \bibinfo {author} {\bibfnamefont {W.}~\bibnamefont
  {Zeilinger}}, \bibinfo {author} {\bibfnamefont {F.}~\bibnamefont {Grupp}},
  \bibinfo {author} {\bibfnamefont {P.}~\bibnamefont {Hudelot}}, \bibinfo
  {author} {\bibfnamefont {R.}~\bibnamefont {Massey}}, \bibinfo {author}
  {\bibfnamefont {M.}~\bibnamefont {Meneghetti}}, \bibinfo {author}
  {\bibfnamefont {L.}~\bibnamefont {Miller}}, \bibinfo {author} {\bibfnamefont
  {S.}~\bibnamefont {Paltani}}, \bibinfo {author} {\bibfnamefont
  {S.}~\bibnamefont {{Paulin-Henriksson}}}, \bibinfo {author} {\bibfnamefont
  {S.}~\bibnamefont {Pires}}, \bibinfo {author} {\bibfnamefont
  {C.}~\bibnamefont {Saxton}}, \bibinfo {author} {\bibfnamefont
  {T.}~\bibnamefont {Schrabback}}, \bibinfo {author} {\bibfnamefont
  {G.}~\bibnamefont {Seidel}}, \bibinfo {author} {\bibfnamefont
  {J.}~\bibnamefont {Walsh}}, \bibinfo {author} {\bibfnamefont
  {N.}~\bibnamefont {Aghanim}}, \bibinfo {author} {\bibfnamefont
  {L.}~\bibnamefont {Amendola}}, \bibinfo {author} {\bibfnamefont
  {J.}~\bibnamefont {Bartlett}}, \bibinfo {author} {\bibfnamefont
  {C.}~\bibnamefont {Baccigalupi}}, \bibinfo {author} {\bibfnamefont {J.-P.}\
  \bibnamefont {Beaulieu}}, \bibinfo {author} {\bibfnamefont {K.}~\bibnamefont
  {Benabed}}, \bibinfo {author} {\bibfnamefont {J.-G.}\ \bibnamefont {Cuby}},
  \bibinfo {author} {\bibfnamefont {D.}~\bibnamefont {Elbaz}}, \bibinfo
  {author} {\bibfnamefont {P.}~\bibnamefont {Fosalba}}, \bibinfo {author}
  {\bibfnamefont {G.}~\bibnamefont {Gavazzi}}, \bibinfo {author} {\bibfnamefont
  {A.}~\bibnamefont {Helmi}}, \bibinfo {author} {\bibfnamefont
  {I.}~\bibnamefont {Hook}}, \bibinfo {author} {\bibfnamefont {M.}~\bibnamefont
  {Irwin}}, \bibinfo {author} {\bibfnamefont {J.-P.}\ \bibnamefont {Kneib}},
  \bibinfo {author} {\bibfnamefont {M.}~\bibnamefont {Kunz}}, \bibinfo {author}
  {\bibfnamefont {F.}~\bibnamefont {Mannucci}}, \bibinfo {author}
  {\bibfnamefont {L.}~\bibnamefont {Moscardini}}, \bibinfo {author}
  {\bibfnamefont {C.}~\bibnamefont {Tao}}, \bibinfo {author} {\bibfnamefont
  {R.}~\bibnamefont {Teyssier}}, \bibinfo {author} {\bibfnamefont
  {J.}~\bibnamefont {Weller}}, \bibinfo {author} {\bibfnamefont
  {G.}~\bibnamefont {Zamorani}}, \bibinfo {author} {\bibfnamefont {M.~R.}\
  \bibnamefont {Zapatero~Osorio}}, \bibinfo {author} {\bibfnamefont
  {O.}~\bibnamefont {Boulade}}, \bibinfo {author} {\bibfnamefont {J.~J.}\
  \bibnamefont {Foumond}}, \bibinfo {author} {\bibfnamefont {A.}~\bibnamefont
  {Di~Giorgio}}, \bibinfo {author} {\bibfnamefont {P.}~\bibnamefont
  {Guttridge}}, \bibinfo {author} {\bibfnamefont {A.}~\bibnamefont {James}},
  \bibinfo {author} {\bibfnamefont {M.}~\bibnamefont {Kemp}}, \bibinfo {author}
  {\bibfnamefont {J.}~\bibnamefont {Martignac}}, \bibinfo {author}
  {\bibfnamefont {A.}~\bibnamefont {Spencer}}, \bibinfo {author} {\bibfnamefont
  {D.}~\bibnamefont {Walton}}, \bibinfo {author} {\bibfnamefont
  {T.}~\bibnamefont {Bl{\"u}mchen}}, \bibinfo {author} {\bibfnamefont
  {C.}~\bibnamefont {Bonoli}}, \bibinfo {author} {\bibfnamefont
  {F.}~\bibnamefont {Bortoletto}}, \bibinfo {author} {\bibfnamefont
  {C.}~\bibnamefont {Cerna}}, \bibinfo {author} {\bibfnamefont
  {L.}~\bibnamefont {Corcione}}, \bibinfo {author} {\bibfnamefont
  {C.}~\bibnamefont {Fabron}}, \bibinfo {author} {\bibfnamefont
  {K.}~\bibnamefont {Jahnke}}, \bibinfo {author} {\bibfnamefont
  {S.}~\bibnamefont {Ligori}}, \bibinfo {author} {\bibfnamefont
  {F.}~\bibnamefont {Madrid}}, \bibinfo {author} {\bibfnamefont
  {L.}~\bibnamefont {Martin}}, \bibinfo {author} {\bibfnamefont
  {G.}~\bibnamefont {Morgante}}, \bibinfo {author} {\bibfnamefont
  {T.}~\bibnamefont {Pamplona}}, \bibinfo {author} {\bibfnamefont
  {E.}~\bibnamefont {Prieto}}, \bibinfo {author} {\bibfnamefont
  {M.}~\bibnamefont {Riva}}, \bibinfo {author} {\bibfnamefont {R.}~\bibnamefont
  {Toledo}}, \bibinfo {author} {\bibfnamefont {M.}~\bibnamefont {Trifoglio}},
  \bibinfo {author} {\bibfnamefont {F.}~\bibnamefont {Zerbi}}, \bibinfo
  {author} {\bibfnamefont {F.}~\bibnamefont {Abdalla}}, \bibinfo {author}
  {\bibfnamefont {M.}~\bibnamefont {Douspis}}, \bibinfo {author} {\bibfnamefont
  {C.}~\bibnamefont {Grenet}}, \bibinfo {author} {\bibfnamefont
  {S.}~\bibnamefont {Borgani}}, \bibinfo {author} {\bibfnamefont
  {R.}~\bibnamefont {Bouwens}}, \bibinfo {author} {\bibfnamefont
  {F.}~\bibnamefont {Courbin}}, \bibinfo {author} {\bibfnamefont {J.-M.}\
  \bibnamefont {Delouis}}, \bibinfo {author} {\bibfnamefont {P.}~\bibnamefont
  {Dubath}}, \bibinfo {author} {\bibfnamefont {A.}~\bibnamefont {Fontana}},
  \bibinfo {author} {\bibfnamefont {M.}~\bibnamefont {Frailis}}, \bibinfo
  {author} {\bibfnamefont {A.}~\bibnamefont {Grazian}}, \bibinfo {author}
  {\bibfnamefont {J.}~\bibnamefont {Koppenh{\"o}fer}}, \bibinfo {author}
  {\bibfnamefont {O.}~\bibnamefont {Mansutti}}, \bibinfo {author}
  {\bibfnamefont {M.}~\bibnamefont {Melchior}}, \bibinfo {author}
  {\bibfnamefont {M.}~\bibnamefont {Mignoli}}, \bibinfo {author} {\bibfnamefont
  {J.}~\bibnamefont {Mohr}}, \bibinfo {author} {\bibfnamefont {C.}~\bibnamefont
  {Neissner}}, \bibinfo {author} {\bibfnamefont {K.}~\bibnamefont {Noddle}},
  \bibinfo {author} {\bibfnamefont {M.}~\bibnamefont {Poncet}}, \bibinfo
  {author} {\bibfnamefont {M.}~\bibnamefont {Scodeggio}}, \bibinfo {author}
  {\bibfnamefont {S.}~\bibnamefont {Serrano}}, \bibinfo {author} {\bibfnamefont
  {N.}~\bibnamefont {Shane}}, \bibinfo {author} {\bibfnamefont {J.-L.}\
  \bibnamefont {Starck}}, \bibinfo {author} {\bibfnamefont {C.}~\bibnamefont
  {Surace}}, \bibinfo {author} {\bibfnamefont {A.}~\bibnamefont {Taylor}},
  \bibinfo {author} {\bibfnamefont {G.}~\bibnamefont {{Verdoes-Kleijn}}},
  \bibinfo {author} {\bibfnamefont {C.}~\bibnamefont {Vuerli}}, \bibinfo
  {author} {\bibfnamefont {O.~R.}\ \bibnamefont {Williams}}, \bibinfo {author}
  {\bibfnamefont {A.}~\bibnamefont {Zacchei}}, \bibinfo {author} {\bibfnamefont
  {B.}~\bibnamefont {Altieri}}, \bibinfo {author} {\bibfnamefont
  {I.}~\bibnamefont {Escudero~Sanz}}, \bibinfo {author} {\bibfnamefont
  {R.}~\bibnamefont {Kohley}}, \bibinfo {author} {\bibfnamefont
  {T.}~\bibnamefont {Oosterbroek}}, \bibinfo {author} {\bibfnamefont
  {P.}~\bibnamefont {Astier}}, \bibinfo {author} {\bibfnamefont
  {D.}~\bibnamefont {Bacon}}, \bibinfo {author} {\bibfnamefont
  {S.}~\bibnamefont {Bardelli}}, \bibinfo {author} {\bibfnamefont
  {C.}~\bibnamefont {Baugh}}, \bibinfo {author} {\bibfnamefont
  {F.}~\bibnamefont {Bellagamba}}, \bibinfo {author} {\bibfnamefont
  {C.}~\bibnamefont {Benoist}}, \bibinfo {author} {\bibfnamefont
  {D.}~\bibnamefont {Bianchi}}, \bibinfo {author} {\bibfnamefont
  {A.}~\bibnamefont {Biviano}}, \bibinfo {author} {\bibfnamefont
  {E.}~\bibnamefont {Branchini}}, \bibinfo {author} {\bibfnamefont
  {C.}~\bibnamefont {Carbone}}, \bibinfo {author} {\bibfnamefont
  {V.}~\bibnamefont {Cardone}}, \bibinfo {author} {\bibfnamefont
  {D.}~\bibnamefont {Clements}}, \bibinfo {author} {\bibfnamefont
  {S.}~\bibnamefont {Colombi}}, \bibinfo {author} {\bibfnamefont
  {C.}~\bibnamefont {Conselice}}, \bibinfo {author} {\bibfnamefont
  {G.}~\bibnamefont {Cresci}}, \bibinfo {author} {\bibfnamefont
  {N.}~\bibnamefont {Deacon}}, \bibinfo {author} {\bibfnamefont
  {J.}~\bibnamefont {Dunlop}}, \bibinfo {author} {\bibfnamefont
  {C.}~\bibnamefont {Fedeli}}, \bibinfo {author} {\bibfnamefont
  {F.}~\bibnamefont {Fontanot}}, \bibinfo {author} {\bibfnamefont
  {P.}~\bibnamefont {Franzetti}}, \bibinfo {author} {\bibfnamefont
  {C.}~\bibnamefont {Giocoli}}, \bibinfo {author} {\bibfnamefont
  {J.}~\bibnamefont {{Garcia-Bellido}}}, \bibinfo {author} {\bibfnamefont
  {J.}~\bibnamefont {Gow}}, \bibinfo {author} {\bibfnamefont {A.}~\bibnamefont
  {Heavens}}, \bibinfo {author} {\bibfnamefont {P.}~\bibnamefont {Hewett}},
  \bibinfo {author} {\bibfnamefont {C.}~\bibnamefont {Heymans}}, \bibinfo
  {author} {\bibfnamefont {A.}~\bibnamefont {Holland}}, \bibinfo {author}
  {\bibfnamefont {Z.}~\bibnamefont {Huang}}, \bibinfo {author} {\bibfnamefont
  {O.}~\bibnamefont {Ilbert}}, \bibinfo {author} {\bibfnamefont
  {B.}~\bibnamefont {Joachimi}}, \bibinfo {author} {\bibfnamefont
  {E.}~\bibnamefont {Jennins}}, \bibinfo {author} {\bibfnamefont
  {E.}~\bibnamefont {Kerins}}, \bibinfo {author} {\bibfnamefont
  {A.}~\bibnamefont {Kiessling}}, \bibinfo {author} {\bibfnamefont
  {D.}~\bibnamefont {Kirk}}, \bibinfo {author} {\bibfnamefont {R.}~\bibnamefont
  {Kotak}}, \bibinfo {author} {\bibfnamefont {O.}~\bibnamefont {Krause}},
  \bibinfo {author} {\bibfnamefont {O.}~\bibnamefont {Lahav}}, \bibinfo
  {author} {\bibfnamefont {F.}~\bibnamefont {{van Leeuwen}}}, \bibinfo {author}
  {\bibfnamefont {J.}~\bibnamefont {Lesgourgues}}, \bibinfo {author}
  {\bibfnamefont {M.}~\bibnamefont {Lombardi}}, \bibinfo {author}
  {\bibfnamefont {M.}~\bibnamefont {Magliocchetti}}, \bibinfo {author}
  {\bibfnamefont {K.}~\bibnamefont {Maguire}}, \bibinfo {author} {\bibfnamefont
  {E.}~\bibnamefont {Majerotto}}, \bibinfo {author} {\bibfnamefont
  {R.}~\bibnamefont {Maoli}}, \bibinfo {author} {\bibfnamefont
  {F.}~\bibnamefont {Marulli}}, \bibinfo {author} {\bibfnamefont
  {S.}~\bibnamefont {Maurogordato}}, \bibinfo {author} {\bibfnamefont
  {H.}~\bibnamefont {McCracken}}, \bibinfo {author} {\bibfnamefont
  {R.}~\bibnamefont {McLure}}, \bibinfo {author} {\bibfnamefont
  {A.}~\bibnamefont {Melchiorri}}, \bibinfo {author} {\bibfnamefont
  {A.}~\bibnamefont {Merson}}, \bibinfo {author} {\bibfnamefont
  {M.}~\bibnamefont {Moresco}}, \bibinfo {author} {\bibfnamefont
  {M.}~\bibnamefont {Nonino}}, \bibinfo {author} {\bibfnamefont
  {P.}~\bibnamefont {Norberg}}, \bibinfo {author} {\bibfnamefont
  {J.}~\bibnamefont {Peacock}}, \bibinfo {author} {\bibfnamefont
  {R.}~\bibnamefont {Pello}}, \bibinfo {author} {\bibfnamefont
  {M.}~\bibnamefont {Penny}}, \bibinfo {author} {\bibfnamefont
  {V.}~\bibnamefont {Pettorino}}, \bibinfo {author} {\bibfnamefont
  {C.}~\bibnamefont {Di~Porto}}, \bibinfo {author} {\bibfnamefont
  {L.}~\bibnamefont {Pozzetti}}, \bibinfo {author} {\bibfnamefont
  {C.}~\bibnamefont {Quercellini}}, \bibinfo {author} {\bibfnamefont
  {M.}~\bibnamefont {Radovich}}, \bibinfo {author} {\bibfnamefont
  {A.}~\bibnamefont {Rassat}}, \bibinfo {author} {\bibfnamefont
  {N.}~\bibnamefont {Roche}}, \bibinfo {author} {\bibfnamefont
  {S.}~\bibnamefont {Ronayette}}, \bibinfo {author} {\bibfnamefont
  {E.}~\bibnamefont {Rossetti}}, \bibinfo {author} {\bibfnamefont
  {B.}~\bibnamefont {Sartoris}}, \bibinfo {author} {\bibfnamefont
  {P.}~\bibnamefont {Schneider}}, \bibinfo {author} {\bibfnamefont
  {E.}~\bibnamefont {Semboloni}}, \bibinfo {author} {\bibfnamefont
  {S.}~\bibnamefont {Serjeant}}, \bibinfo {author} {\bibfnamefont
  {F.}~\bibnamefont {Simpson}}, \bibinfo {author} {\bibfnamefont
  {C.}~\bibnamefont {Skordis}}, \bibinfo {author} {\bibfnamefont
  {G.}~\bibnamefont {Smadja}}, \bibinfo {author} {\bibfnamefont
  {S.}~\bibnamefont {Smartt}}, \bibinfo {author} {\bibfnamefont
  {P.}~\bibnamefont {Spano}}, \bibinfo {author} {\bibfnamefont
  {S.}~\bibnamefont {Spiro}}, \bibinfo {author} {\bibfnamefont
  {M.}~\bibnamefont {Sullivan}}, \bibinfo {author} {\bibfnamefont
  {A.}~\bibnamefont {Tilquin}}, \bibinfo {author} {\bibfnamefont
  {R.}~\bibnamefont {Trotta}}, \bibinfo {author} {\bibfnamefont
  {L.}~\bibnamefont {Verde}}, \bibinfo {author} {\bibfnamefont
  {Y.}~\bibnamefont {Wang}}, \bibinfo {author} {\bibfnamefont {G.}~\bibnamefont
  {Williger}}, \bibinfo {author} {\bibfnamefont {G.}~\bibnamefont {Zhao}},
  \bibinfo {author} {\bibfnamefont {J.}~\bibnamefont {Zoubian}}, \ and\
  \bibinfo {author} {\bibfnamefont {E.}~\bibnamefont {Zucca}},\ }\href@noop {}
  {\bibfield  {journal} {\bibinfo  {journal} {arXiv e-prints}\ ,\ \bibinfo
  {pages} {arXiv:1110.3193}} (\bibinfo {year} {2011})}\BibitemShut {NoStop}%
\bibitem [{\citenamefont {Spergel}\ \emph {et~al.}(2015)\citenamefont
  {Spergel}, \citenamefont {Gehrels}, \citenamefont {Baltay}, \citenamefont
  {Bennett}, \citenamefont {Breckinridge}, \citenamefont {Donahue},
  \citenamefont {Dressler}, \citenamefont {Gaudi}, \citenamefont {Greene},
  \citenamefont {Guyon}, \citenamefont {Hirata}, \citenamefont {Kalirai},
  \citenamefont {Kasdin}, \citenamefont {Macintosh}, \citenamefont {Moos},
  \citenamefont {Perlmutter}, \citenamefont {Postman}, \citenamefont
  {Rauscher}, \citenamefont {Rhodes}, \citenamefont {Wang}, \citenamefont
  {Weinberg}, \citenamefont {Benford}, \citenamefont {Hudson}, \citenamefont
  {Jeong}, \citenamefont {Mellier}, \citenamefont {Traub}, \citenamefont
  {Yamada}, \citenamefont {Capak}, \citenamefont {Colbert}, \citenamefont
  {Masters}, \citenamefont {Penny}, \citenamefont {Savransky}, \citenamefont
  {Stern}, \citenamefont {Zimmerman}, \citenamefont {Barry}, \citenamefont
  {Bartusek}, \citenamefont {Carpenter}, \citenamefont {Cheng}, \citenamefont
  {Content}, \citenamefont {Dekens}, \citenamefont {Demers}, \citenamefont
  {Grady}, \citenamefont {Jackson}, \citenamefont {Kuan}, \citenamefont {Kruk},
  \citenamefont {Melton}, \citenamefont {Nemati}, \citenamefont {Parvin},
  \citenamefont {Poberezhskiy}, \citenamefont {Peddie}, \citenamefont {Ruffa},
  \citenamefont {Wallace}, \citenamefont {Whipple}, \citenamefont {Wollack},\
  and\ \citenamefont {Zhao}}]{spergel2015}%
  \BibitemOpen
  \bibfield  {author} {\bibinfo {author} {\bibfnamefont {D.}~\bibnamefont
  {Spergel}}, \bibinfo {author} {\bibfnamefont {N.}~\bibnamefont {Gehrels}},
  \bibinfo {author} {\bibfnamefont {C.}~\bibnamefont {Baltay}}, \bibinfo
  {author} {\bibfnamefont {D.}~\bibnamefont {Bennett}}, \bibinfo {author}
  {\bibfnamefont {J.}~\bibnamefont {Breckinridge}}, \bibinfo {author}
  {\bibfnamefont {M.}~\bibnamefont {Donahue}}, \bibinfo {author} {\bibfnamefont
  {A.}~\bibnamefont {Dressler}}, \bibinfo {author} {\bibfnamefont {B.~S.}\
  \bibnamefont {Gaudi}}, \bibinfo {author} {\bibfnamefont {T.}~\bibnamefont
  {Greene}}, \bibinfo {author} {\bibfnamefont {O.}~\bibnamefont {Guyon}},
  \bibinfo {author} {\bibfnamefont {C.}~\bibnamefont {Hirata}}, \bibinfo
  {author} {\bibfnamefont {J.}~\bibnamefont {Kalirai}}, \bibinfo {author}
  {\bibfnamefont {N.~J.}\ \bibnamefont {Kasdin}}, \bibinfo {author}
  {\bibfnamefont {B.}~\bibnamefont {Macintosh}}, \bibinfo {author}
  {\bibfnamefont {W.}~\bibnamefont {Moos}}, \bibinfo {author} {\bibfnamefont
  {S.}~\bibnamefont {Perlmutter}}, \bibinfo {author} {\bibfnamefont
  {M.}~\bibnamefont {Postman}}, \bibinfo {author} {\bibfnamefont
  {B.}~\bibnamefont {Rauscher}}, \bibinfo {author} {\bibfnamefont
  {J.}~\bibnamefont {Rhodes}}, \bibinfo {author} {\bibfnamefont
  {Y.}~\bibnamefont {Wang}}, \bibinfo {author} {\bibfnamefont {D.}~\bibnamefont
  {Weinberg}}, \bibinfo {author} {\bibfnamefont {D.}~\bibnamefont {Benford}},
  \bibinfo {author} {\bibfnamefont {M.}~\bibnamefont {Hudson}}, \bibinfo
  {author} {\bibfnamefont {W.~S.}\ \bibnamefont {Jeong}}, \bibinfo {author}
  {\bibfnamefont {Y.}~\bibnamefont {Mellier}}, \bibinfo {author} {\bibfnamefont
  {W.}~\bibnamefont {Traub}}, \bibinfo {author} {\bibfnamefont
  {T.}~\bibnamefont {Yamada}}, \bibinfo {author} {\bibfnamefont
  {P.}~\bibnamefont {Capak}}, \bibinfo {author} {\bibfnamefont
  {J.}~\bibnamefont {Colbert}}, \bibinfo {author} {\bibfnamefont
  {D.}~\bibnamefont {Masters}}, \bibinfo {author} {\bibfnamefont
  {M.}~\bibnamefont {Penny}}, \bibinfo {author} {\bibfnamefont
  {D.}~\bibnamefont {Savransky}}, \bibinfo {author} {\bibfnamefont
  {D.}~\bibnamefont {Stern}}, \bibinfo {author} {\bibfnamefont
  {N.}~\bibnamefont {Zimmerman}}, \bibinfo {author} {\bibfnamefont
  {R.}~\bibnamefont {Barry}}, \bibinfo {author} {\bibfnamefont
  {L.}~\bibnamefont {Bartusek}}, \bibinfo {author} {\bibfnamefont
  {K.}~\bibnamefont {Carpenter}}, \bibinfo {author} {\bibfnamefont
  {E.}~\bibnamefont {Cheng}}, \bibinfo {author} {\bibfnamefont
  {D.}~\bibnamefont {Content}}, \bibinfo {author} {\bibfnamefont
  {F.}~\bibnamefont {Dekens}}, \bibinfo {author} {\bibfnamefont
  {R.}~\bibnamefont {Demers}}, \bibinfo {author} {\bibfnamefont
  {K.}~\bibnamefont {Grady}}, \bibinfo {author} {\bibfnamefont
  {C.}~\bibnamefont {Jackson}}, \bibinfo {author} {\bibfnamefont
  {G.}~\bibnamefont {Kuan}}, \bibinfo {author} {\bibfnamefont {J.}~\bibnamefont
  {Kruk}}, \bibinfo {author} {\bibfnamefont {M.}~\bibnamefont {Melton}},
  \bibinfo {author} {\bibfnamefont {B.}~\bibnamefont {Nemati}}, \bibinfo
  {author} {\bibfnamefont {B.}~\bibnamefont {Parvin}}, \bibinfo {author}
  {\bibfnamefont {I.}~\bibnamefont {Poberezhskiy}}, \bibinfo {author}
  {\bibfnamefont {C.}~\bibnamefont {Peddie}}, \bibinfo {author} {\bibfnamefont
  {J.}~\bibnamefont {Ruffa}}, \bibinfo {author} {\bibfnamefont {J.~K.}\
  \bibnamefont {Wallace}}, \bibinfo {author} {\bibfnamefont {A.}~\bibnamefont
  {Whipple}}, \bibinfo {author} {\bibfnamefont {E.}~\bibnamefont {Wollack}}, \
  and\ \bibinfo {author} {\bibfnamefont {F.}~\bibnamefont {Zhao}},\ }\href@noop
  {} {\enquote {\bibinfo {title} {Wide-{{Field InfrarRed Survey
  Telescope-Astrophysics Focused Telescope Assets WFIRST-AFTA}} 2015
  {{Report}}},}\ } (\bibinfo {year} {2015})\BibitemShut {NoStop}%
\bibitem [{\citenamefont {Wang}\ \emph
  {et~al.}(2022{\natexlab{b}})\citenamefont {Wang}, \citenamefont {Zhai},
  \citenamefont {Alavi}, \citenamefont {Massara}, \citenamefont {Pisani},
  \citenamefont {Benson}, \citenamefont {Hirata}, \citenamefont {Samushia},
  \citenamefont {Weinberg}, \citenamefont {Colbert}, \citenamefont {Dor{\'e}},
  \citenamefont {Eifler}, \citenamefont {Heinrich}, \citenamefont {Ho},
  \citenamefont {Krause}, \citenamefont {Padmanabhan}, \citenamefont
  {Spergel},\ and\ \citenamefont {Teplitz}}]{wang2022a}%
  \BibitemOpen
  \bibfield  {author} {\bibinfo {author} {\bibfnamefont {Y.}~\bibnamefont
  {Wang}}, \bibinfo {author} {\bibfnamefont {Z.}~\bibnamefont {Zhai}}, \bibinfo
  {author} {\bibfnamefont {A.}~\bibnamefont {Alavi}}, \bibinfo {author}
  {\bibfnamefont {E.}~\bibnamefont {Massara}}, \bibinfo {author} {\bibfnamefont
  {A.}~\bibnamefont {Pisani}}, \bibinfo {author} {\bibfnamefont
  {A.}~\bibnamefont {Benson}}, \bibinfo {author} {\bibfnamefont {C.~M.}\
  \bibnamefont {Hirata}}, \bibinfo {author} {\bibfnamefont {L.}~\bibnamefont
  {Samushia}}, \bibinfo {author} {\bibfnamefont {D.~H.}\ \bibnamefont
  {Weinberg}}, \bibinfo {author} {\bibfnamefont {J.}~\bibnamefont {Colbert}},
  \bibinfo {author} {\bibfnamefont {O.}~\bibnamefont {Dor{\'e}}}, \bibinfo
  {author} {\bibfnamefont {T.}~\bibnamefont {Eifler}}, \bibinfo {author}
  {\bibfnamefont {C.}~\bibnamefont {Heinrich}}, \bibinfo {author}
  {\bibfnamefont {S.}~\bibnamefont {Ho}}, \bibinfo {author} {\bibfnamefont
  {E.}~\bibnamefont {Krause}}, \bibinfo {author} {\bibfnamefont
  {N.}~\bibnamefont {Padmanabhan}}, \bibinfo {author} {\bibfnamefont
  {D.}~\bibnamefont {Spergel}}, \ and\ \bibinfo {author} {\bibfnamefont
  {H.~I.}\ \bibnamefont {Teplitz}},\ }\href {\doibase 10.3847/1538-4357/ac4973}
  {\bibfield  {journal} {\bibinfo  {journal} {The Astrophysical Journal}\
  }\textbf {\bibinfo {volume} {928}},\ \bibinfo {pages} {1} (\bibinfo {year}
  {2022}{\natexlab{b}})}\BibitemShut {NoStop}%
\end{thebibliography}%

\end{document}